\documentclass[a4paper,fleqn,usenatbib]{mnras}
\usepackage[T1]{fontenc}
\usepackage{newtxtext,newtxmath}
\usepackage{ae,aecompl}
\usepackage{graphicx}
\usepackage{amsmath}	
\usepackage{hyperref}
\usepackage{microtype}
\usepackage{verbatim}
\usepackage{graphicx}
\usepackage{subfig}
\usepackage{url}
\usepackage{physics}
\usepackage{subcaption}
\usepackage{natbib,hyperref}

\title[Timing analysis of two variable ULXs]{Pulsations and constraints on time lags in two distinct ULXs: NGC 5204 X--1 and NGC 55 ULX--1}

\author[Rajath Sathyaprakash et al.]{Rajath Sathyaprakash$^{1}$\thanks{E-mail: rajath.sathyaprakash@gmail.com} Matthew. J. Middleton$^{2}$ and Chris Done$^{3}$
\\
$^{1}$Scuola Universitaria Superiore IUSS Pavia, Palazzo del Broletto, piazza della Vittoria 15, I-27100 Pavia, Italy\\
$^{2}$Department of Physics and Astronomy, University of Southampton, Southampton SO17 1BJ\\
$^{3}$Centre for Extragalactic Astronomy, Durham University, Dept of Physics, South Road, Durham DH1 3LE, UK\\
}

\date{Accepted XXX. Received October 4 2026; in original form ZZZ}

\pubyear{2026}

\begin{document}

\label{firstpage}
\pagerange{\pageref{firstpage}--\pageref{lastpage}}
\maketitle

\begin{abstract}
The nature of ultraluminous X--ray sources (ULXs) is shrouded in mystery as they're believed to be point sources with luminosities above the classical Eddington limit for a 10$M_{\odot}$ black hole (BH), but generally fit into the classification tree of super--Eddington X--ray binaries. Studies of reverberation lags and lag--energy spectra provide some insight into the nature of viscous turbulence in optically thick, super-Eddington discs, and are useful in conjunction with modelling multi--wavelength spectra and covariance spectra. In this study, we focus on the timing properties of two classically distinct ULXs: NGC 5204 X--1 and NGC 55 ULX--1, both of which are known to emit relativistic winds, as observed by modelling emission line spectra with photo--ionisation models such as SPEX, but are thought to be viewed at slightly different inclinations due to differences in the hardness--ratio of their X--ray continuum. Here, we perform an extensive timing study of the two ULXs to reveal new constraints on the accretion disc size scale and structure. We also perform an accelerated search for pulsations (including the correction for an orbital modulation of the signal) to discover significant spin--periodicities in both NGC 5204 X--1 and NGC 55 ULX--1  (at $> 3\sigma$ significance).  
\end{abstract}


\section{Introduction}

Among the plethora of binary systems with different classifications and evolutionary states as observed with recent deep X--ray observations and all sky monitoring, include gamma--ray bursts (\citealt{gehrels2004}; \citealt{berger2009}; \citealt{pavanzo2015}; \citealt{gehrels2012}), kilonovae (\citealt{metzger2017}; \citealt{kilonovap2}; \citealt{kilonovap3}), tidal disruption events (\citealt{rees1988}; \citealt{burrows2011}; \citealt{cenko2012}; \citealt{pasham2023}), cataclysmic variables \citealt{cataclysmic_var} and ultraluminous X--ray sources (\citealt{lasota2023}; ULXs). The definition of ULXs is somewhat empirical since they are only (or used to be) defined by their apparent luminosities to be in excess of the Eddington limit for a 10 $M_{\odot}$ black hole (i.e. $>3 \times 10^{39}$ erg s$^{-1}$; for detailed reviews, see \citealt{king_lasota_middleton_2023}; \citealt{pinto_walton_2023}; \citealt{kaaret_feng_roberts_2017}). Recent results from observations with several X--ray and UV/optical facilities suggest that they belong to a unique class of young high mass X--ray binaries undergoing a prolonged phase of super--Eddington accretion, likely owing to mass transfer from a nearby companion star (\citealt{king_2008}; \citealt{Zezas_2002}; \citealt{Misra_2024}; \citealt{sathyaprakash2026}). For example, the discovery of pulsating ULXs from as many as seven ULXs indicate that a sizeable fraction of the population could harbour neutron stars accreting well above the Eddington luminosity (\citealt{bachetti_2014}; \citealt{fuerst2016}; \citealt{israel_2017b,israel_2017a}; \citealt{carpano_2018}; \citealt{sathyaprakash_2019}; \citealt{rodriguez_castillo_2020}).  
\newline
\newline
Moreover, studies of broadband CCD and high--resolution RGS spectra of several ULXs reveal that the structure of the accretion disc in these sources is expected to deviate from a thin Shakura--Sunyaev disc (\citealt{shakura_sunyaev_1973}) in regions where the Eddington--limit is exceeded locally (\citealt{Soria_2015}). In particular, the disc is expected to become geometrically thick with strong winds driven by radiation pressure (\citealt{Done_2007}). Indeed, the X--ray continuum of several ULXs can be described by at--least two thermal components or a multi--colour disc and a thermal comptonisation model in the 0.3--10 keV energy range. {\it{Historically}}, the temperature of the soft thermal component (0.3--1.0 keV) was used to derive a BH mass for the ULX assuming a standard (sub--Eddington) thin disc truncated at the inner most stable orbit (ISCO), implying masses for ULXs to be in the range of intermediate mass black holes (IMBHs; \citealt{colbert_mushotzky_1999}). However, the absence of a power--law component at energies above 10 keV, as predicted for Galactic sub--Eddington X--ray binaries in the soft state has argued against this hypothesis, especially given the appearance of a downturn in ULX CCD spectra above 10 keV in the {\it{NuSTAR}} observations (\citealt{roberts_2015}). ULX spectra are broadly classified into four main states: supersoft (SS), soft ultraluminous (SUL), hard ultraluminous (HUL) and a broadened disc (BD). This classification scheme (as detailed in \citealt{sutton_2014}) separates SUL and HUL spectra based on the unabsorbed X--ray flux and the slope of the power--law used to fit the spectral component between 2--10 keV, while supersoft ULXs are identified by a pre--dominance of the soft disc component (with little X--ray flux above 2 keV; \citealt{kong_2003}). Moreover, broadened disc spectra suggest that ULX accretion discs could become advection dominated (\citealt{watarai_2001}) during some phases of mass transfer, especially close to the Eddington threshold, when photons are swept radially with the accretion flow rather than vertically, because the timescale for the vertical diffusion of photons exceeds the radial viscous diffusion timescale of the in--falling material. A key diagnostic for such advection-dominated discs is the correlation between the integrated disc luminosity $L_{X}$ and temperature $T$, which is expected to follow $L_{X} \propto T^{2}$ (\citealt{watarai_2001}) rather than $L_{X} \propto T^{4}$ for thin discs. For mass accretion rates much higher than Eddington, the disc flux and temperature could become anti--correlated since an increase in mass transfer rate at the outer disc radius leads to an increase in the spherisation radius and wind optical depth, and hence a decrease in the temperature of thermal photons from the wind photosphere. A full range of ULX spectral shapes can be found in \citet{king2023} and \citet{barra_2022}.
\newline
\newline
ULX CCD spectra also indicate strong residuals at energies below $\sim 1$ keV, which were resolved into narrow blue--shifted absorption and emission lines (\citealt{Pinto_2016}; \citealt{Kosec_2018}) that confirm the presence of multi--phase winds with variable ionisation states and velocities (\citealt{Pinto_kosec_2023}). The confirmation of disc winds in ULX spectra imply that their super--Eddington luminosities can partly be explained due to geometrical beaming of hard X--rays from the inner regions of the accretion flow (\citealt{king_2008}). However, in the case of magnetised NS ULXs the winds could be suppressed if the magnetospheric radius of the NS exceeds the spherisation radius, implying that another mechanism can explain their extreme luminosities, potentially via the reduction of electron scattering opacity in the presence of a strong magnetic field (\citealt{chashkina_2019}; \citealt{mushtukov_2019}). Timing analyses complement broadband X--ray spectroscopy in providing further insights into the nature of super-Eddington accretion flows and on the properties of the compact object (for e.g. its mass and spin). The detection of pulsations and super--orbital periodicities enables constraints on the formation channels of NS ULXs via binary population synthesis models (\citealt{teamcompas}), in addition to the spin properties of the NS (see for e.g. Table 1 of \citealt{king_2020}). 
\newline
\newline
Possible signatures of quasi--periodic oscillations (QPOs) have been found in some observations of ULXs with softer spectra at frequencies ranging from $\sim$ 10--100 mHz and accompanied by flat-topped noise (\citealt{de_marco_2013}), although their physical origin remains unclear. Some of the proposed models to explain the observed QPO properties include instabilities of a radiation--pressure supported super--Eddington accretion flow at radii smaller than the spherisation radius or the Lense--Thirring precession of such accretion flows (\citealt{de_marco_2013}; \citealt{middleton_2019}; \citealt{atapin_2019}; \citealt{pasham_2019}). Historically, QPOs in ULXs were thought to be analogous to type--C QPOs observed in the hard--intermediate state of BH binaries, hence providing a proxy for the BH mass. However, the detection of mHz QPOs in pulsating NS ULXs such as M82 X--2 and M51 X--7, with observed frequencies, widths and fractional--rms stable over many different observations, suggests mechanisms different from those of BH X--ray binaries are required to explain the observed QPO properties (\citealt{Feng_Roberto_2011}; \citealt{imbrogno_2024}). Aside from Lense--Thirring precession of the accretion flow, beats between the Keplerian frequency at the magnetospheric radius $R_{\text{M}}$ and the spin frequency of the NS were proposed as a possible explanation. However, this apparently would require the inner disc radius to be very close to the co--rotation radius, which is expected to lead to large changes in X--ray luminosity, which is un--observed in the particular case of M51 X--7. Moreover, it was noted that any model proposed to explain the QPOs in this source would need to explain the correlation between the strength of the QPO signal with the pulsations and vice--versa.   
\newline
\newline
In some ULXs that show sufficiently high fractional variability $F_{\text{var}}$ and count--rate, soft band time delays were detected: see for e.g. NGC 5408 X--1; \citealt{de_marco_2013}; NGC 55 ULX--1; \citealt{pinto_soria_2017} and NGC 1313 X--1; \citealt{kara_2020}. Such lags are usually explained by reprocessing of photons emitted from the inner regions of the disc by distant outflowing material or via the Compton scattering of photons from an accretion envelope surrounding the NS. Hence, such lags can provide constraints on the spherisation radius or the magnetospheric radius of the accretion disc surrounding the compact object, which then sets limits on the mass accretion rate $\dot{m}$ and its magnetic field geometry $B_{\text{dipolar}}$ (\citealt{mushtukov_2019}). A few particular sources that display signatures of many different spectral states, accompanied by recurrent variability, are ideal to study the structure of super--Eddington discs from multiple viewing angles. In this work, we perform a search for pulsations, quasi--periodic oscillations (QPOs) and time delays in the recent {\it{XMM-Newton}} observations of two archetypal and variable ULXs: NGC 5204 X--1 and NGC 55 ULX--1. We provide specific details of the two sources in the sub--sections below.   
\begin{figure}
\centering
\hspace{-0.36cm}
\includegraphics[scale=0.25]{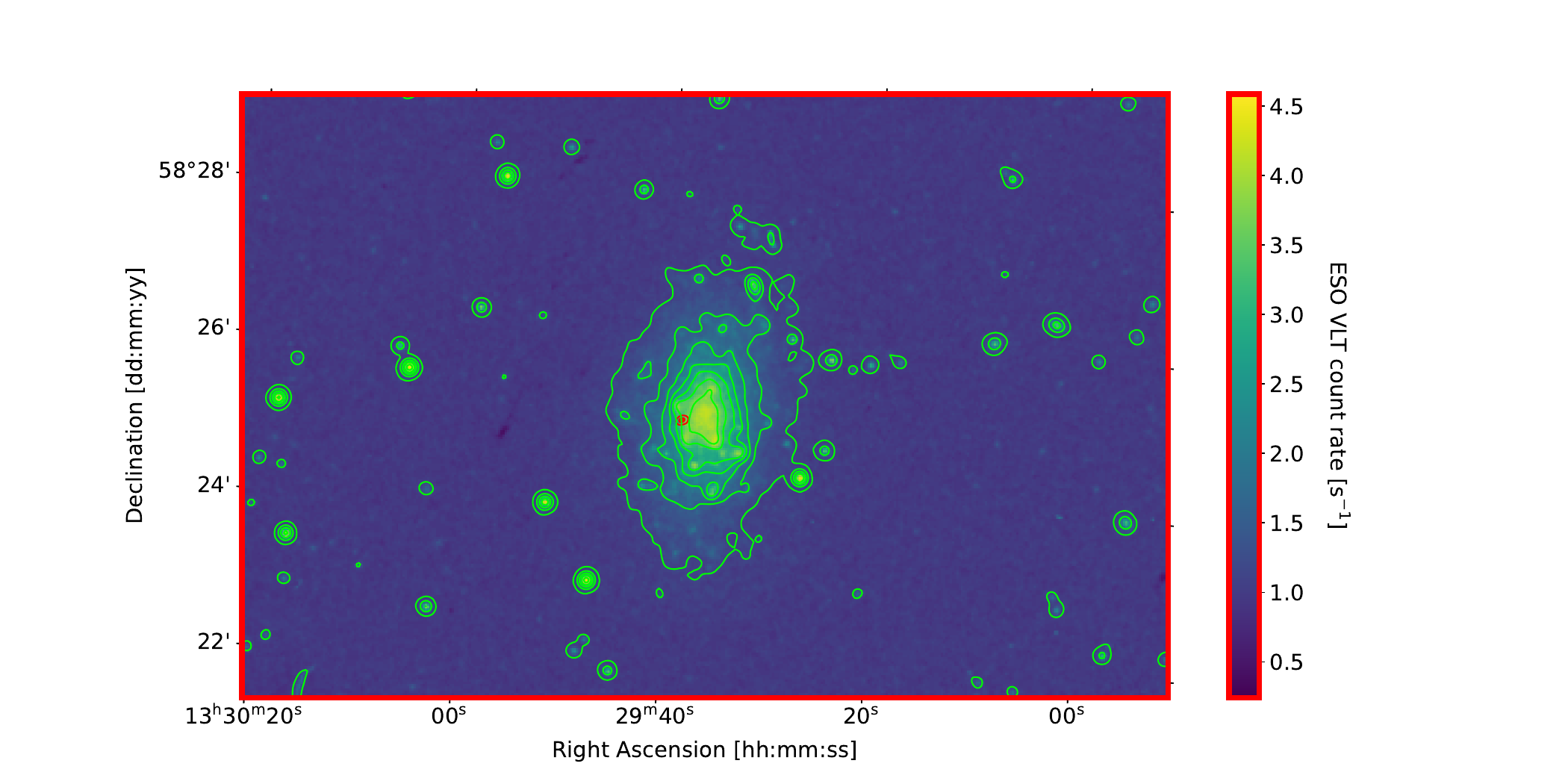}
\includegraphics[scale=0.24]{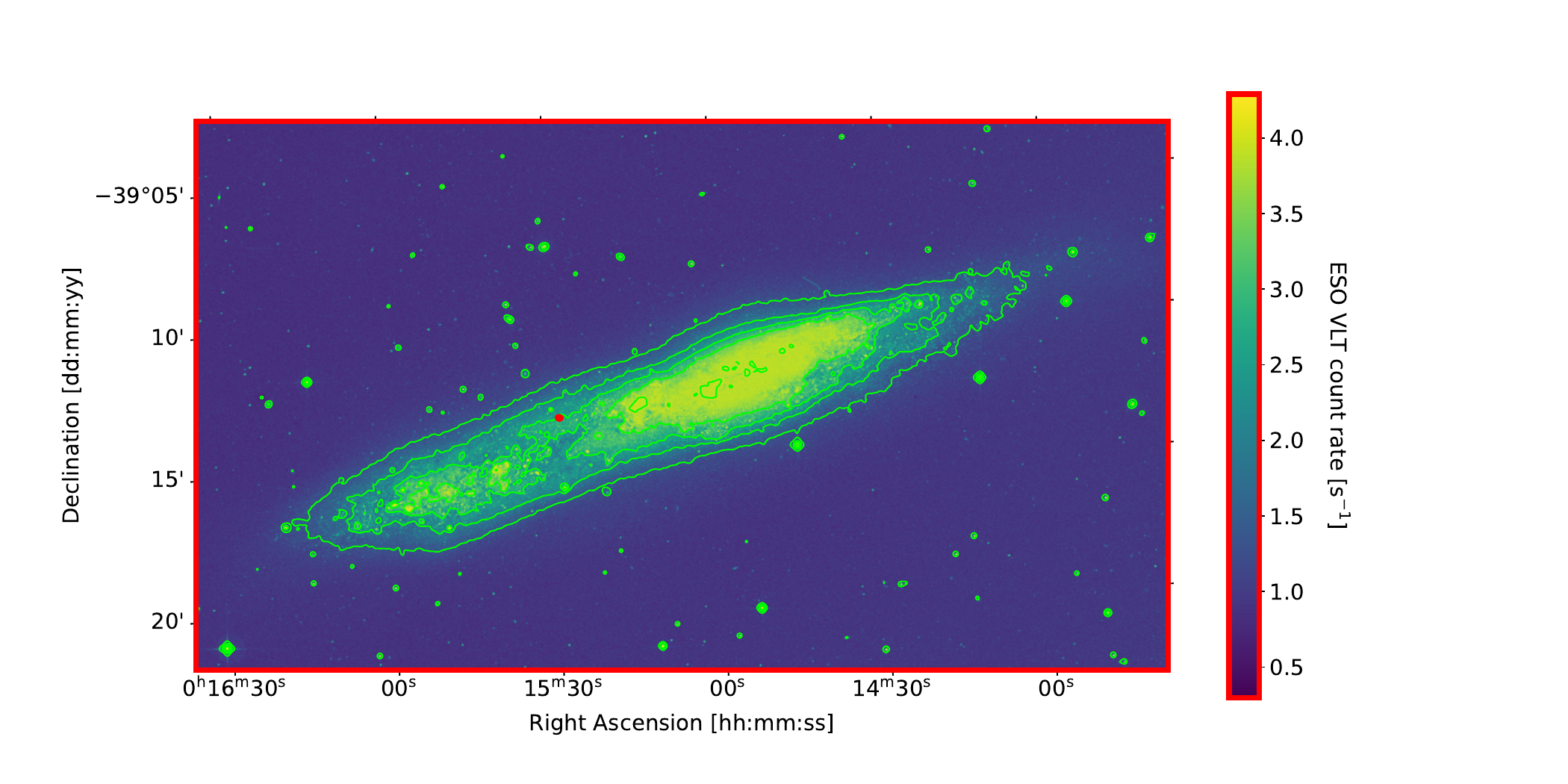}
\caption{{\it{ESO--VLT}} images of the host galaxies containing the two ULXs of this study, namely: NGC 5204 X--1 (top: $\alpha = 202.410833^{\circ}$, $\delta = +58.41833^{\circ}$) and NGC 55 ULX--1 ($\alpha = 3.870375^{\circ}$, $\delta = -39.221889^{\circ}$) with the sky co--ordinates shown in the ICRS--FK5 reference frame. The smaller red contours (from {\it{Chandra}} images) mark the precise sky locations of the ULXs residing in their host galaxies.}
\label{fig:host_galaxies}
\end{figure}

\subsection{NGC 5204 X--1}

NGC 5204 X--1 has a sky location ($\alpha = 202.410833^{\circ}$, $\delta = +58.41833^{\circ}$; \citealt{roberts_2000}) with X--ray luminosities ranging from 2--6 $\times 10^{39}$ erg s$^{-1}$ (\citealt{roberts_2005}), and is found in the outskirts of a Magellanic spiral galaxy with a secure optical counterpart that appears consistent with a B0Ib supergiant, as resolved by high angular resolution {\it{HST}} observations; \citealt{goad_2002}). This confirmed its nature as a young high mass X--ray binary (HMXB). {\it{NuSTAR}} observations (\citealt{Mukherjee_2015}) revealed a clear downturn in its broadband X--ray continuum at energies above 10 keV, identifying the source to be in an ultraluminous spectral state with two thermal components, associated with emission from a radiatively inefficient super--Eddington accretion flow and optically thick winds. The hardness--intensity diagram from long--term {\it{Swift}} monitoring suggests that it switches between the hard ultraluminous (HUL) state, soft--bright and soft--dim ultraluminous (SUL) states. Interestingly, the detection of a tentative 200 day super--orbital periodicity in its light--curve (\citealt{gurpide_2021}) suggests that it could be driven by mass transfer rate modulations from the presence of a third body owing to changes in orbital eccentricity and the Roche Lobe radius of the companion, aside from disc precession (\citealt{stefano2019}).

\subsection{NGC 55 ULX--1}

NGC 55 ULX--1 is a well studied, variable ULX (sky location: $\alpha = 3.870375^{\circ}$, $\delta = -39.221889^{\circ}$; \citealt{barra_2022}), with a peak X--ray luminosity of a few $\times 10^{39}$ erg s$^{-1}$, and displays spectral state transitions between the super--soft state and the soft--ultraluminous state. It also shows dips on timescales of a few 100 seconds in its long--term lightcurve, with the spectrum peaking primarily in the 2.0--4.5 keV band and showing a tail at higher energies (\citealt{stobbart_roberts_2004}; \citealt{barra_2022}). It is associated with a barred Magellanic Sculptor type galaxy, gravitationally bound to a similar type of galaxy, namely NGC 300 (\citealt{garcia1993}; \citealt{karachentsev2003}).

\begin{figure}
\centering
\includegraphics[scale=0.35]{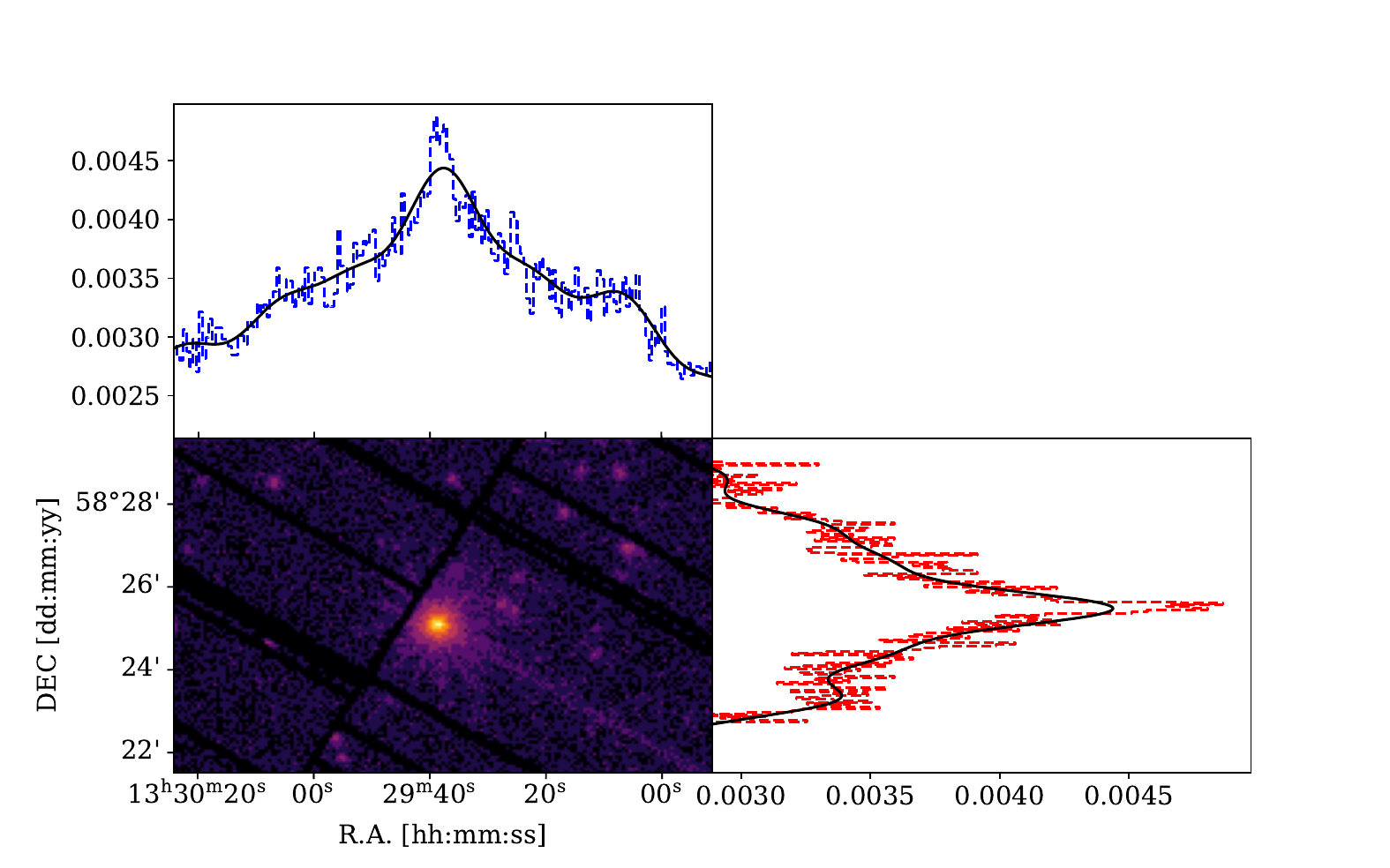}
\includegraphics[scale=0.35]{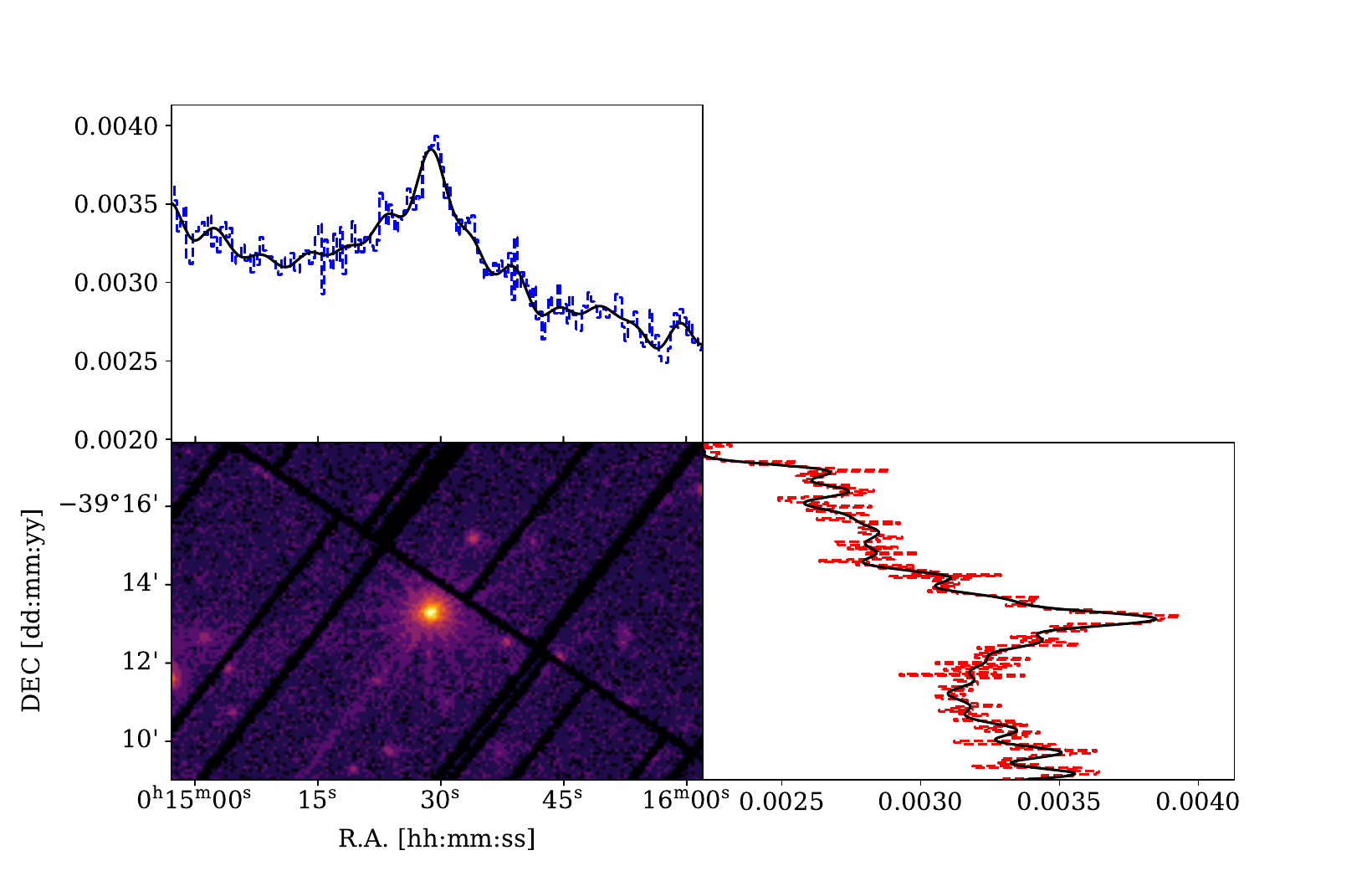}
\caption{{\it{XMM--Newton}} EPIC--pn images of the field surrounding the two ultraluminous X--ray sources: NGC 5204 X--1 ({\it{top}} panel) and NGC 55 ULX--1 ({\it{bottom}} panel), with the images showing the logarithm of the number of counts per pixel and having been smoothed with a Gaussian kernel.}
\label{Fig:im_ngc5204}
\end{figure}

\begin{figure*}
\centering
\includegraphics[scale=0.5]{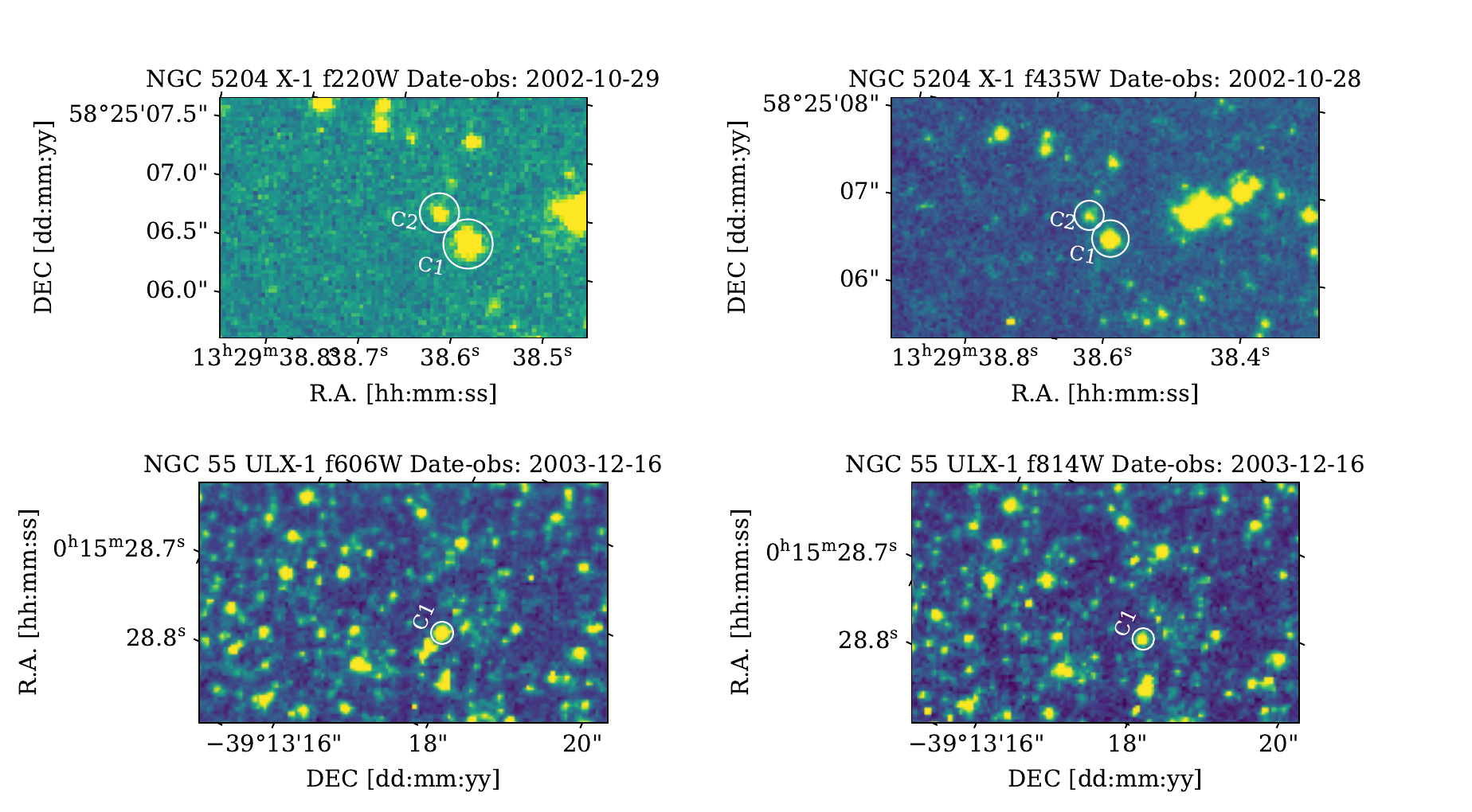}
\caption{Drizzled {\it{HST}} images of the optical counterparts corresponding to (i, {\it{top}}) NGC 5204 X--1 and (ii, {\it{bottom}}) NGC 55 ULX--1. The possible counterparts to the ULX are marked by the solid white circle in all four images. NGC 5204 X--1 has two counterparts, but C1 remains the likely candidate.}
\label{fig:im_ngc55}
\end{figure*}

\section{Data reduction}

\subsection{{\it{XMM-Newton}} observations}

The complete set of {\it{XMM-Newton}} observations analysed in this work are outlined in Table~\ref{Tab:obsids}. We used data from the EPIC--pn and EPIC--MOS detectors, which were processed with the Science Analysis Software (SAS v19.0.0). The event files were barycentered using the SAS task \href{https://www.cosmos.esa.int/web/xmm-newton/sas-thread-timing-gui}{{\tt{barycen}}}, with the sky locations of both ULXs obtained with high angular resolution {\it{Chandra}} observations. We use the sky--locations $\alpha$ = 202.410$^{\circ}$, $\delta$ = 58.418$^{\circ}$ for NGC 5204 X--1 and $\alpha$ = 3.87022$^{\circ}$, $\delta$ = -39.2219$^{\circ}$ for NGC 55 ULX--1. The events were spatially filtered using a circular extraction region of 30 arcsecond radius centered on the source location. We excluded events associated with bad pixels by implementing the standard flagging criteria ({\tt{PATTERN}} $\leq 4$ for EPIC-PN, {\tt{PATTERN}} $\leq 12$ for EPIC-MOS and {\tt{FLAG}} $==0$), and further removed periods of soft proton flares by selecting an optimal (10--12 keV) count--rate threshold below which the background level remains steady via {\tt{bkgoptrate}}, which uses a sigma--clipping algorithm to excise events above the said threshold. Using the cleaned event files, we carried out a comprehensive search for periodic pulsations, quasi--periodic oscillations (QPOs) and time delays as outlined in the sub--sections below. We took care to subtract the channel--of--interest from the reference band when computing time lags and covariance spectra. In order to extract time averaged spectra over the 0.3--12 keV energy range, we made use of the {\tt{evselect}} tool. The spectra were grouped into a minimum of 50 counts per energy bin after subtracting the background and were deconvolved using the response matrix files and auxiliary response files derived with the {\tt{arfgen}} and {\tt{rmfgen}} tasks. We adopted a distance of 5.24 Mpc (i.e. redshift $z=0.0011$) to NGC 5204 X--1 and 1.94 Mpc (i.e. $z=0.00042$) to NGC 55 ULX--1 \footnote{\url{https://ned.ipac.caltech.edu/cgi-bin/nDistance?name=NGC+5204}}\footnote{\url{https://ned.ipac.caltech.edu/cgi-bin/nDistance?name=NGC+55}}, adopting a standard $\Lambda$CDM cosmology with Hubble parameter $H_{0} = 67-74$ km s$^{-1}$ Mpc $^{-1}$ and use typical matter, radiation and dark energy densities: $\Omega_{r,0} \sim 0, \Omega_{m,0} = 0.315$ and $\Omega_{\Lambda,0} = 0.685$ (\citealt{desi_jul2025}).

\subsection{{\it{HST}} and Gemini--North observations}
\begin{table*}
\caption{{\it{HST}} observation log of the two ULXs considered in this study. We tabulate the aperture--corrected magnitudes and fluxes of the ULX counterparts in their respective wide--band filters by cross--matching with {\it{Chandra}} images.}
\begin{tabular}{rrrrrr}
\hline
\hline
\centering
ULX & Date & Exposure Time & Filter & Vega magnitude & Flux [erg cm$^{-2}$ s$^{-1}$ \AA$^{-1}$] \\
\hline
\hline
NGC 5204 X--1 & 29--10--2002 & 2720 & F220W & $21.5 \pm 0.9$ & $(1.0 \pm 0.9) \times 10^{-17}$ \\
NGC 5204 X--1 & 28--10--2002 & 1300 & F435W & $23.7 \pm 0.6$ & $(2 \pm 1) \times 10^{-18}$ \\
\hline
NGC 55 ULX--1 & 16--12--2003 & 676 & F606W & $23.3 \pm 0.3$ & ($1.4 \pm 0.3$) $\times 10^{-18}$ \\
NGC 55 ULX--1 & 16--12--2003 & 700 & F814W & $23.1 \pm 0.4$ & ($6 \pm 2$) $\times 10^{-19}$ \\
\hline
\hline
\end{tabular}
\label{Tab:obsids_hst}
\end{table*}
The field surrounding the counterpart containing the ULXs studied in this work was observed during several {\tt{HST}} visits with the WFC3, WFC2 and ACS detectors in two independent wide--band filters. We summarise the details of these observations in Table~\ref{obsids_hst}. We ran the {\tt{calwf3}} pipeline (version 3.3\footnote{https://hst-crds.stsci.edu/}) on the raw event files of all WFC3 observations using the latest calibration files obtained through the {\tt{STScI}} website \footnote{The equivalent pipeline {\tt{calacs}} was used to reduce the ACS data}. The pipeline produces event files that have been bias subtracted, flat fielded and corrected for charge transfer inefficiency (CTI). We ensured that all sub-exposures were aligned to within 0.1" accuracy using the {\tt{drizzlepac}} tool {\tt{TweakReg}}v1.4. We then implemented the PSF-fitting algorithm using {\tt{dolphot}} and {\tt{sextractor}} (\citealt{bertin2013}), in addition to simple aperture photometry (via {\tt{photutils}}) to derive instrumental magnitudes for the optical counterpart. This method identifies point sources above a preset detection threshold from a stack of flat--fielded and cosmic--ray cleaned images and makes use of the {\tt{TinyTim}} PSF models (\citealt{r33}; \citealt{r34}) to simultaneously fit the PSF of each source and the local background. This approach is preferred over aperture photometry in scenarios where the PSFs of two closely spaced point sources overlap, which is an indication of severe crowding (e.g. in longer wavelength optical wavebands). We carried out the necessary pre--processing steps prior to running the photometry routine, including the masking of bad pixels affected by cosmic rays and image combination. The instrumental magnitudes were initially measured using a circular aperture of three pixel radius, and converted to an equivalent 10 pixel (or 0.4") magnitude by applying aperture corrections. We selected relatively bright and well isolated stars in the field-of-view, with signal-to-noise ratios larger than that of the ULX counterpart. We then used the inverse sensitivity (i.e. {\tt{photflam}}) values to derive the optical fluxes in units of erg cm$^{-2}$ s$^{-1}$ \AA$^{-1}$. 
\newline
\newline
We reduced the Gemini--North (GMOS-N) observations of NGC 5204 X--1 (not available for NGC 55 ULX--1) by bias--subtracting, flat--fielding, wavelength and flux calibrating the raw exposure frames using {\tt{pyraf}} run on a virtual machine in the same manner as \cite{sathyaprakash2026}.  
\begin{table*}
\caption{Characteristics of the latest {\it{XMM-Newton}} observations of NGC 5204 X--1 and NGC 55 ULX--1. $^{\text{1}}$Fractional variability, $^{\text{2}}$HUL (Hard ultraluminous), $^{\text{3}}$SUL (Soft ultraluminous) and $^{\text{4}}$BD (Broadened disc)}
\begin{tabular}{rrrrrrrrr}
\hline
\hline
\centering
Source & Instrument & ObsID & Date & Exposure & Count-rate [s$^{-1}$] & $F_{\text{var}}$$^{\text{1}}$ & Hardness--ratio (HR) & Class$^{\text{2} \text{3} \text{4}}$ \\
       &            &       & [dd-mm-yy] & [ks] & [0.3--10 keV] & [0.3--10 keV] & [F$_{\text{hard}}$/F$_{\text{soft}}$] \\
\hline
\hline
NGC 5204 X--1 & EPIC-pn & 0142770101 & 06-01-2003 & 15.3 & 0.55 $\pm$ 0.01 & 0.09 $\pm$ 0.02 & 0.1 $\pm 0.001$ & HUL \\
NGC 5204 X--1 & EPIC-pn & 0142770301 & 25-04-2003 & 16.6 & 0.77 $\pm$ 0.01 & 0.14 $\pm$ 0.06 & 31.6 $\pm 0.04$ & HUL \\
NGC 5204 X--1 & EPIC-pn & 0150650301 & 01-05-2003 & 9.1 & 0.99 $\pm$ 0.01 & 0.12 $\pm$ 0.03 & 23.99 $\pm 0.03$ & SUL \\
NGC 5204 X--1 & EPIC-pn & 0405690101 & 16-11-2006 & 27.3 & 1.32 $\pm$ 0.03 & 0.14 $\pm$ 0.02 & 512.9 $\pm 0.6$ & SUL \\
NGC 5204 X--1 & EPIC-pn & 0405690201 & 19-11-2006 & 43.5 & 0.96 $\pm$ 0.005 & 0.12 $\pm$ 0.02 & 18.62 $\pm 0.02$ & SUL \\
NGC 5204 X--1 & EPIC-pn & 0405690501 & 25-11-2006 & 41.1 & 0.72 $\pm$ 0.004 & 0.10 $\pm$ 0.04 & 6.76 $\pm 0.02$ & SUL \\
NGC 5204 X--1 & EPIC-pn & 0693851401 & 21-04-2013 & 15.1 & 0.54 $\pm$ 0.01 & 0.09 $\pm$ 0.02 & 20.89 $\pm 0.05$ & HUL \\
NGC 5204 X--1 & EPIC-pn & 0693850701 & 29-04-2013 & 15.1 & 0.58 $\pm$ 0.01 & 0.09 $\pm$ 0.02 & 10.72 $\pm 0.03$ & HUL \\
NGC 5204 X--1 & EPIC-pn & 0741960101 & 27-06-2014 & 21.4 & 0.53 $\pm$ 0.01 & 0.08 $\pm$ 0.02 & 4.27 $\pm 0.01$ & HUL \\
NGC 5204 X--1 & EPIC-pn & 0921360101 & 18-05-2023 & 115.0 & 0.65 $\pm$ 0.002 & 0.10 $\pm$ 0.03 & 3470 $\pm 10$ & SUL \\
NGC 5204 X--1 & EPIC-pn & 0921360201 & 10-11-2023 & 112.0 & 0.92 $\pm$ 0.003 & 0.13 $\pm$ 0.04 & 2400 $\pm 4$ & SUL \\
\hline
NGC 55 ULX--1 & EPIC-pn & 0028740101 & 24-05-2010 & 125.0 & 0.92 $\pm$ 0.02 & 0.09 $\pm$ 0.01 & 17 $\pm 3$ & BD \\
NGC 55 ULX--1 & EPIC-pn & 0028740201 & 27-11-2019 & 4.8 & 1.68 $\pm$ 0.03 & 0.11 $\pm$ 0.02 & 10 $\pm 1$ & BD \\
NGC 55 ULX--1 & EPIC-pn & 0655050101 & 27-12-2019 & 4.8 & 1.69 $\pm$ 0.03 & 0.11 $\pm$ 0.02 & 1 $\pm 1$ & BD \\
NGC 55 ULX--1 & EPIC-pn & 0824570101 & 24-05-2020 & 129.0 & 0.92 $\pm$ 0.01 & 0.16 $\pm$ 0.004 & 25 $\pm 1$ & BD \\
NGC 55 ULX--1 & EPIC-pn & 0864810101 & 12-12-2021 & 127.0 & 1.14 $\pm$ 0.01 & 0.08 $\pm$ 0.01 & 54 $\pm 1$ & BD \\
NGC 55 ULX--1 & EPIC-pn & 0883960101 & 27-11-2019 & 4.08 & 1.02 $\pm 0.01$ & 0.23 $\pm 0.005$ & 5 $\pm 1$ & SUL \\
NGC 55 ULX--1 & EPIC-pn & 0852610101 & 27-12-2019 & 4.06 & 0.99 $\pm 0.01$ & 0.25 $\pm 0.004$ & 6 $\pm 3$ & SUL \\ 
NGC 55 ULX--1 & EPIC-pn & 0852610201 & 11-05-2020 & 4.9 & 0.38 $\pm 0.004$ & 0.22 $\pm 0.01$ & 3 $\pm 3$ & SUL \\
NGC 55 ULX--1 & EPIC-pn & 0852610301 & 19-05-2020 & 4.1 & 0.126 $\pm 0.001$ & 0.09 $\pm 0.07$ & 8 $\pm 4$ & - \\
NGC 55 ULX--1 & EPIC-pn & 0852610401 & 19-05-2020 & 4.1 & 0.126 $\pm 0.001$ & 0.09 $\pm 0.07$ & 8 $\pm 4$ & - \\
\hline
\hline
\end{tabular}
\label{Tab:obsids_xmm}
\end{table*}

\section{Timing results}
\begin{figure*}
\centering
\includegraphics[scale=0.3]{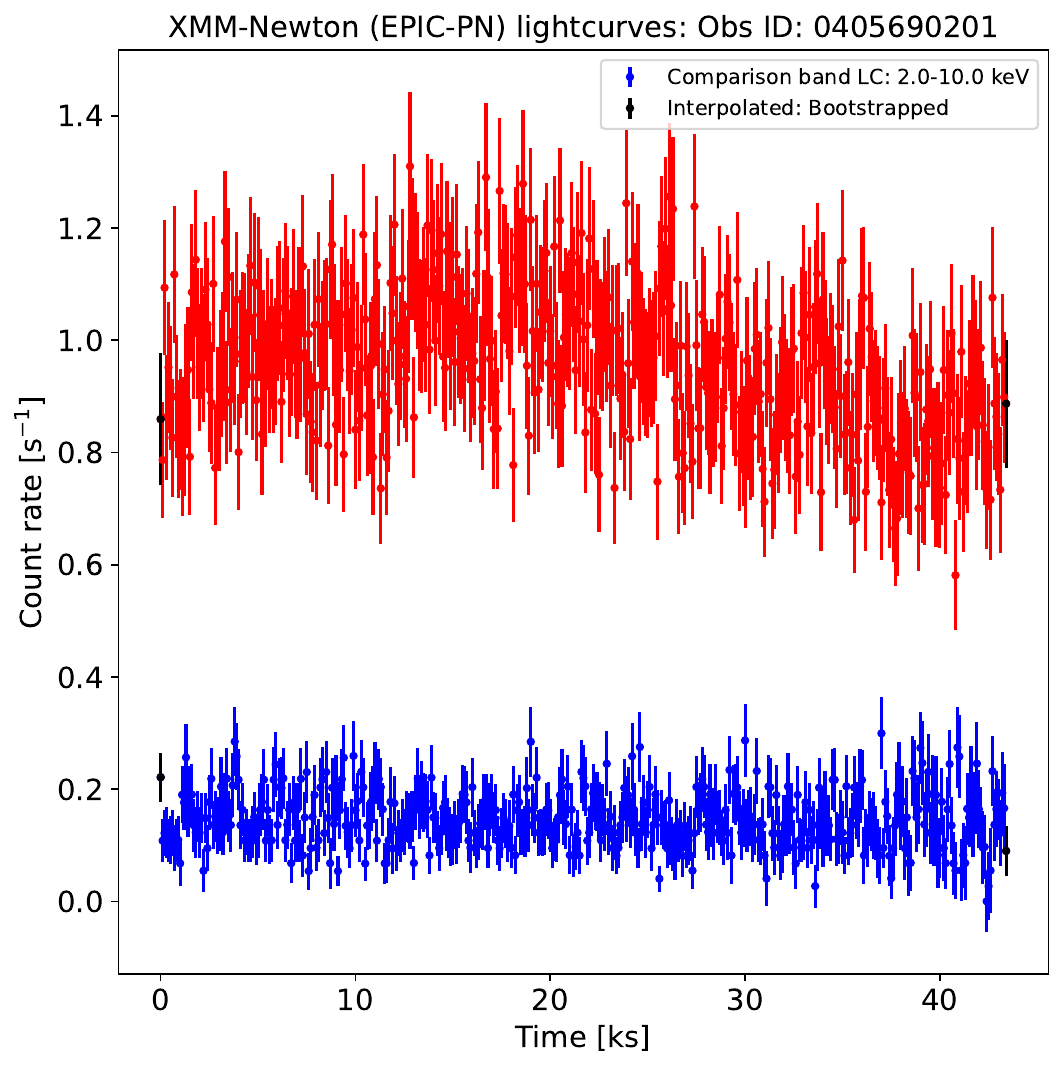}
\includegraphics[scale=0.3]{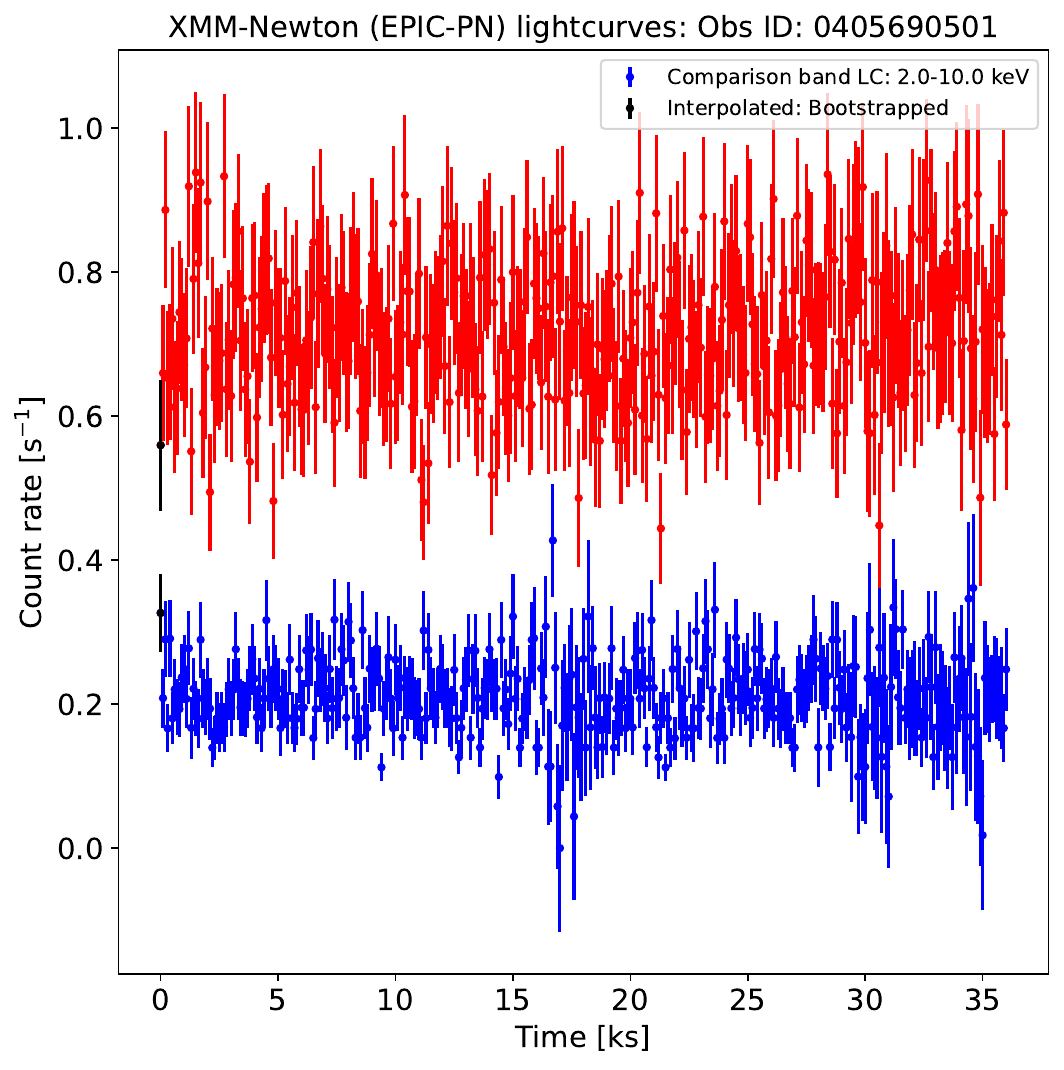}
\includegraphics[scale=0.3]{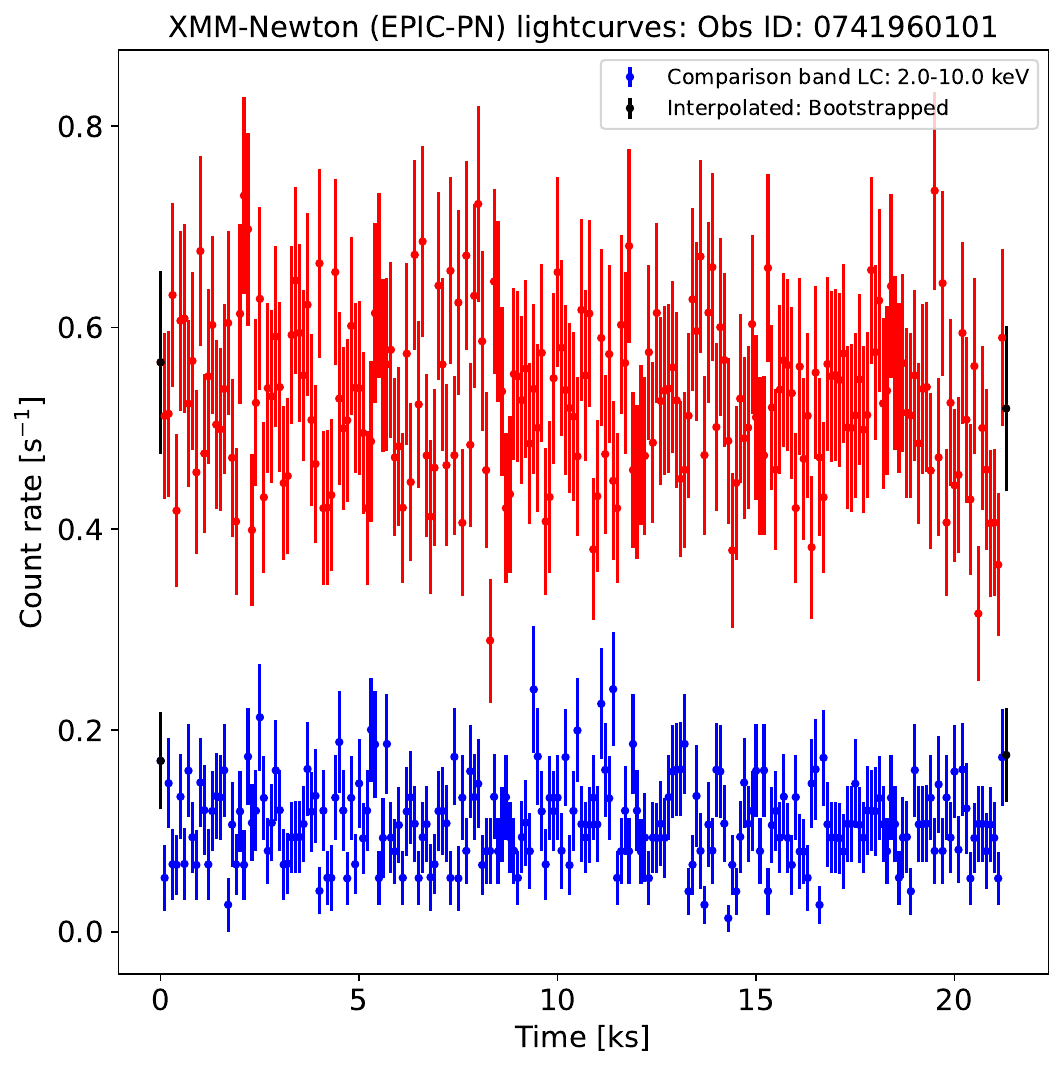}
\includegraphics[scale=0.3]{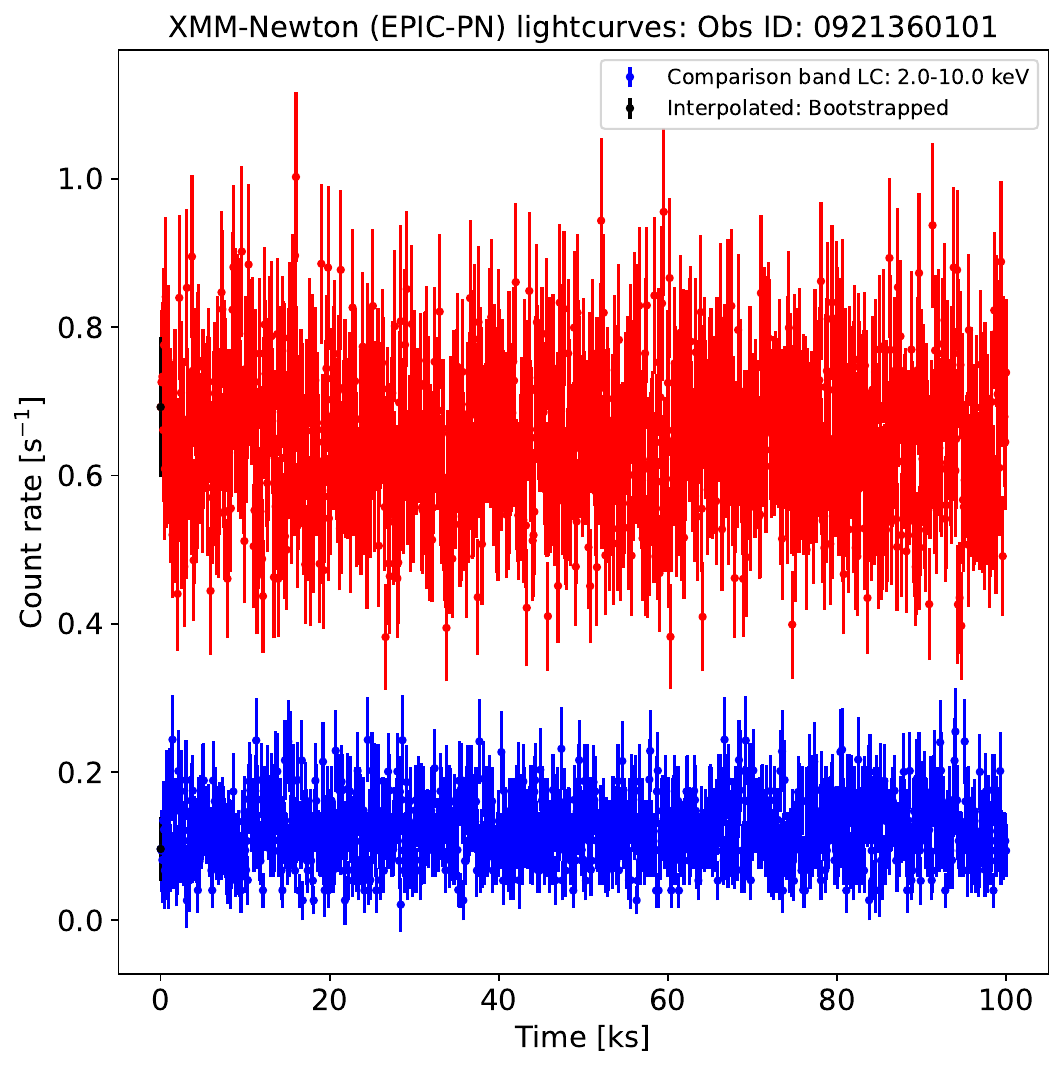}
\includegraphics[scale=0.3]{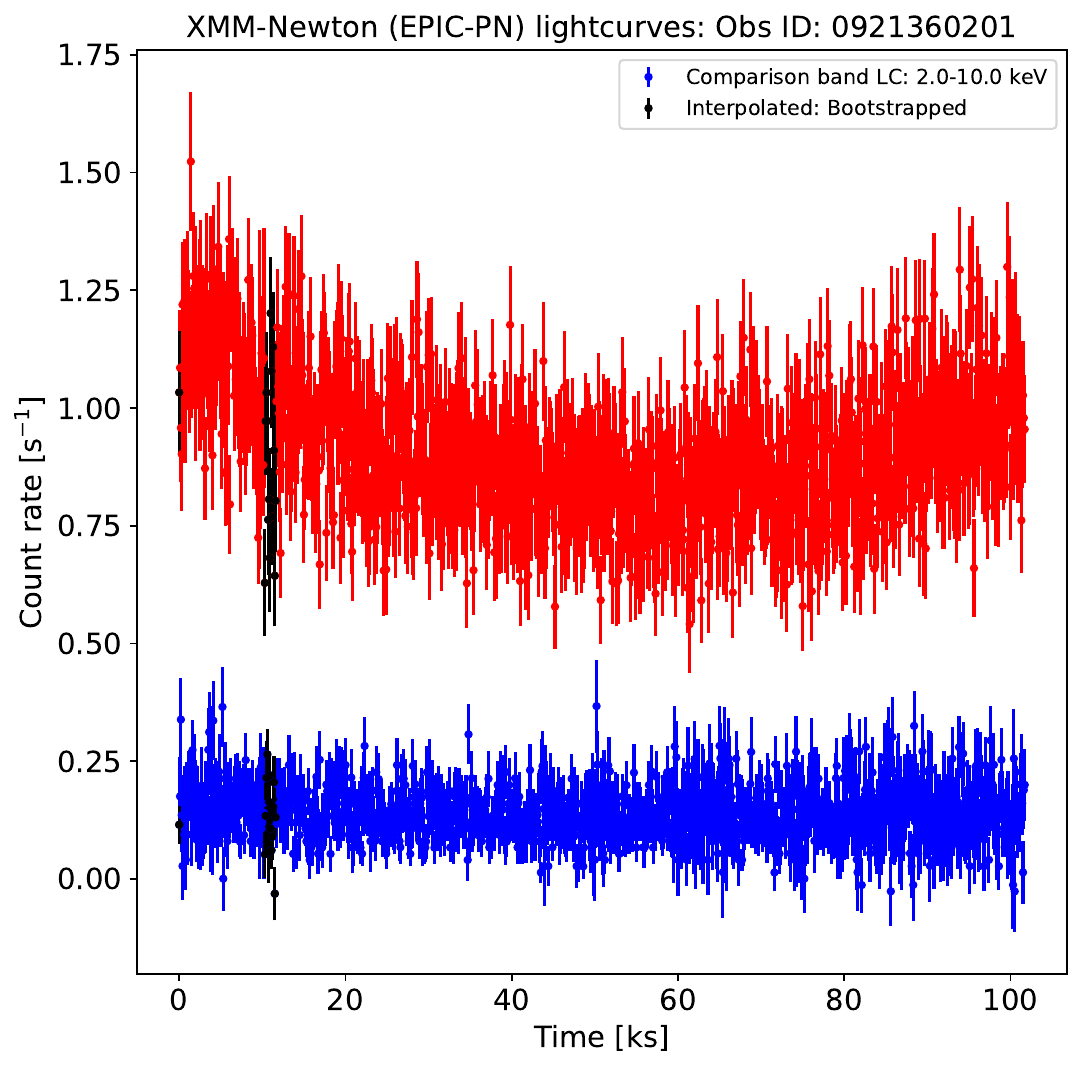}
\caption{EPIC--pn lightcurves of {\it{XMM-Newton}} observations of NGC 5204 X--1. In all cases, the reference band light-curve (0.3--12 keV) is shown by the solid red line, the comparison band lightcurves by the solid blue line while the interpolated segments of the light--curve are shown by the black markers. In the final observation, we see low--frequency variability but in all other cases there is considerable high--frequency noise, as typical for HUL sources.}
\label{Fig:XMM_lc_ngc5204}
\end{figure*}

\begin{figure*}
\centering
\includegraphics[scale=0.3]{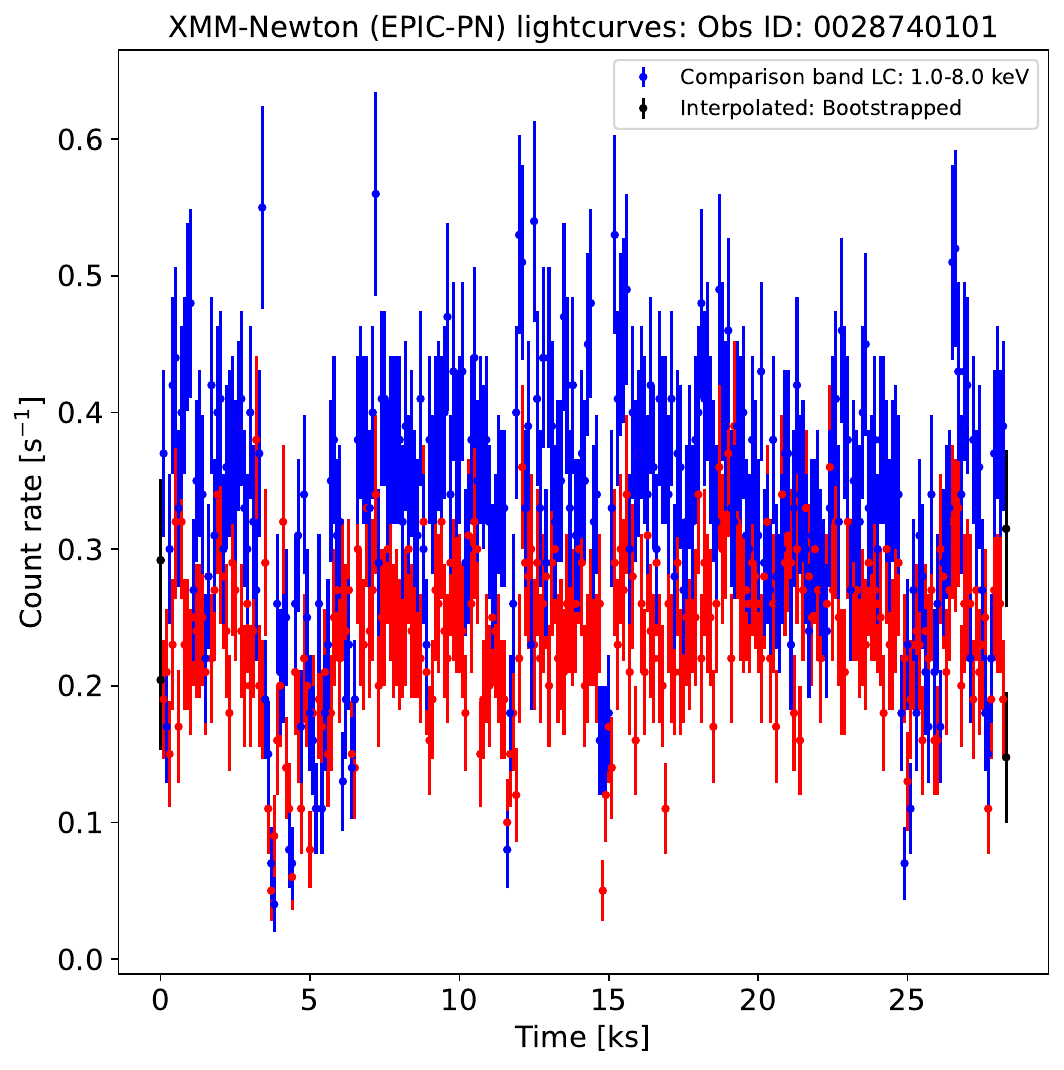}
\includegraphics[scale=0.3]{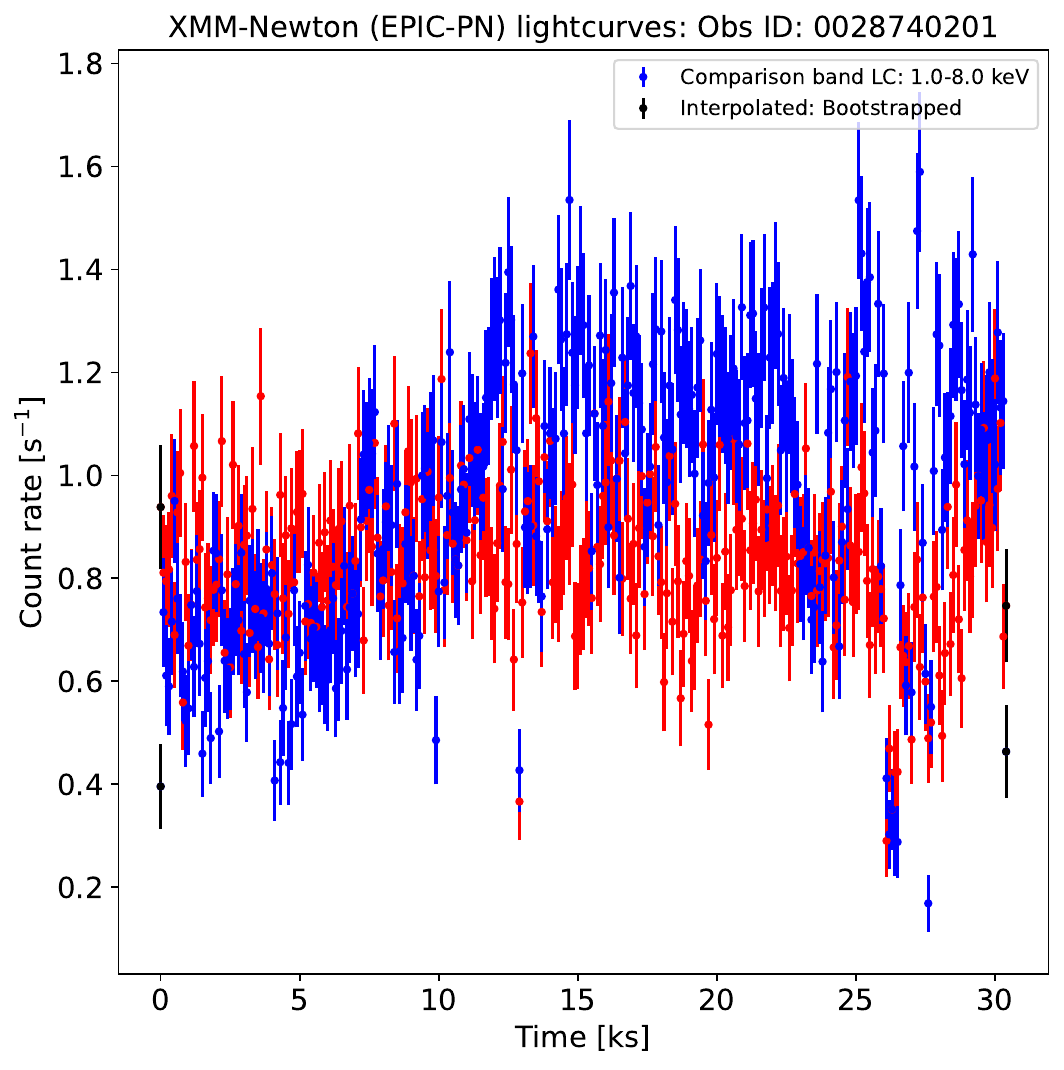}
\includegraphics[scale=0.3]{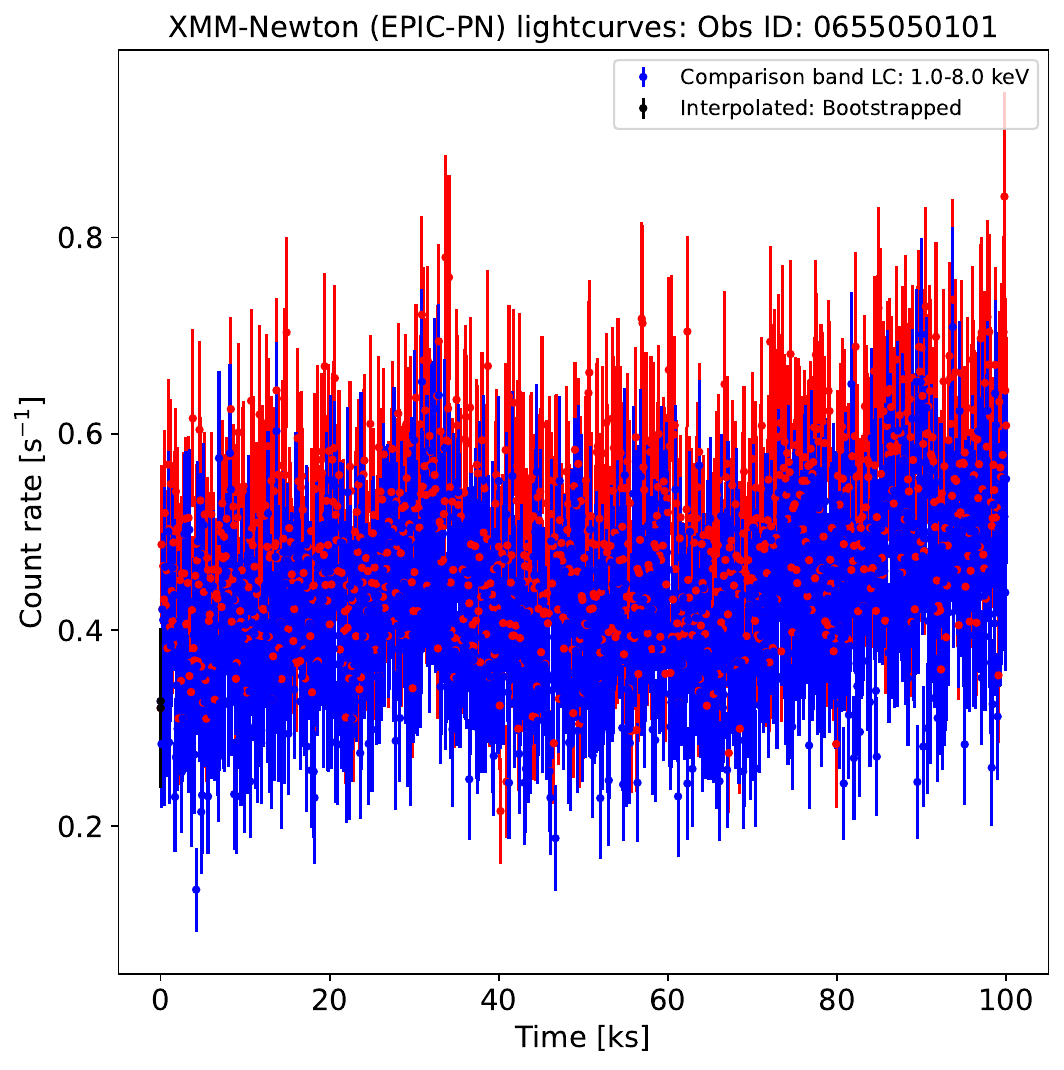}
\includegraphics[scale=0.3]{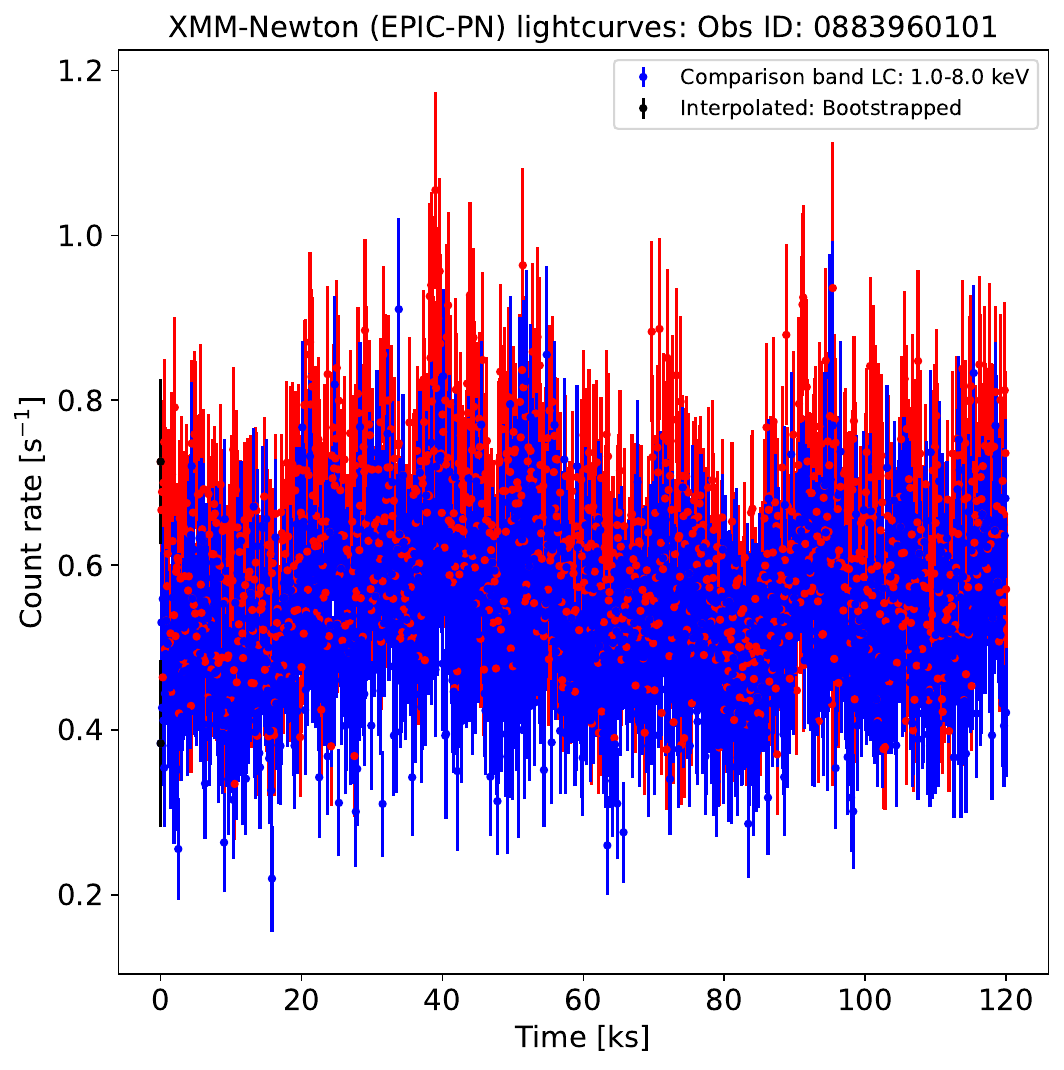}
\includegraphics[scale=0.3]{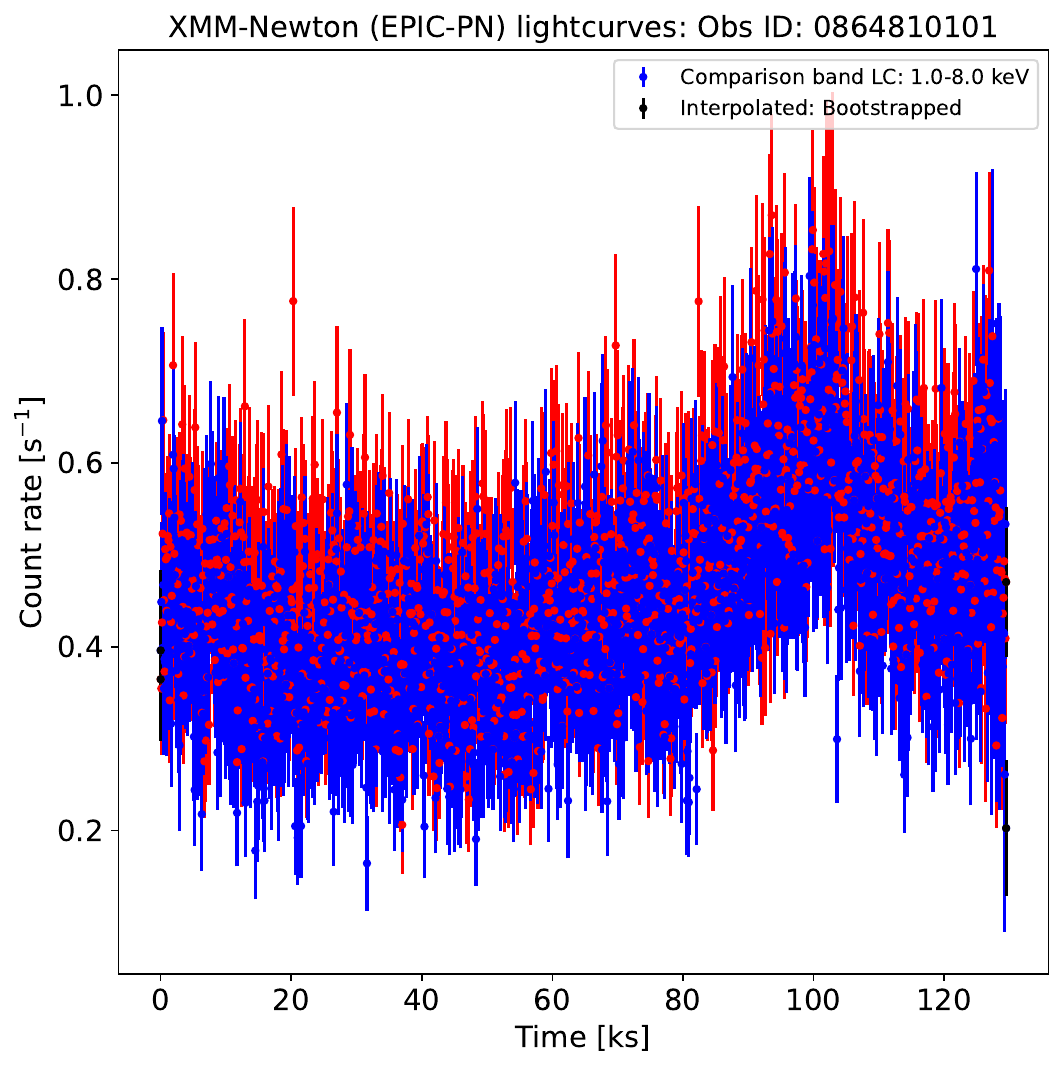}
\caption{EPIC-pn lightcurves of {\it{XMM-Newton}} observations of NGC 55 ULX--1, that show possible dips in some observations, revealing the inherently turbulent nature of the accretion disc.}
\label{Fig:XMM_lc_ngc55}
\end{figure*}

\subsection{Constraints on the lag--energy profiles in the X--ray observations}

In order to search for frequency and energy dependent time lags in the latest {\it{XMM-Newton}} (EPIC-pn and EPIC-MOS) datasets, we extracted background subtracted light-curves in fifteen energy bands spaced on a logarithmic--scale, and a reference light-curve in the 0.3--2 keV energy band, all with 100 second time bins. We describe the procedure to extract the lag--energy spectrum below. We first compute the frequency and energy dependent time lags $\tau(f,E)$, formally defined as the argument of the cross power spectral density and subsequently average $\tau(f,E)$ over a frequency channel: $10^{-4}$ -- $5 \times 10^{-4}$ Hz. Then, for each light-curve $x(t)$ we compute the Fast Fourier Transform $X(f)$ using the {\tt{FFT}} routine in {\tt{numpy}}, followed by the power spectral densities (PSDs) that are then averaged over $N$ segments and binned geometrically in frequency space to ensure that the uncertanties are normally distributed. We then derive the cross power spectral density (CPSD) $\langle C_{xy}(f) \rangle$ between the reference band $Y(f)$ and the comparison band $X(f)$. For both the PSD and the CPSD, we use the fractional rms--squared normalisation, such that integrating the PSD between two frequency limits yields the fractional variability $F_{\text{var}}$ (see \citealt{heil_2012}). The CPSD is a measure of how well variability in a particular energy band is correlated with that in the reference band and can thus be used to filter out uncorrelated noise, enabling a probe of coherent variability processes from different regions of the accretion flow. The frequency dependent time--lag can then be defined mathematically as the argument of the complex CPSD:

\begin{multline}
	\langle C_{xy} (f) \rangle = \langle X^{*}(f) Y(f) \rangle \\
	\langle C_{xy} (f) \rangle = A_{\text{x}} A_{\text{y}} \exp(i \phi(f)) \\
	\tau (f) = \frac{\phi(f)}{2\pi f} = \frac{1}{2\pi f} \arctan \bigg [ \frac{\text{Im} \langle C_{xy} (f) \rangle}{\text{Re} \langle C_{xy} (f) \rangle} \bigg ] \\
\end{multline}

\begin{figure*}
\centering
\includegraphics[scale=0.25]{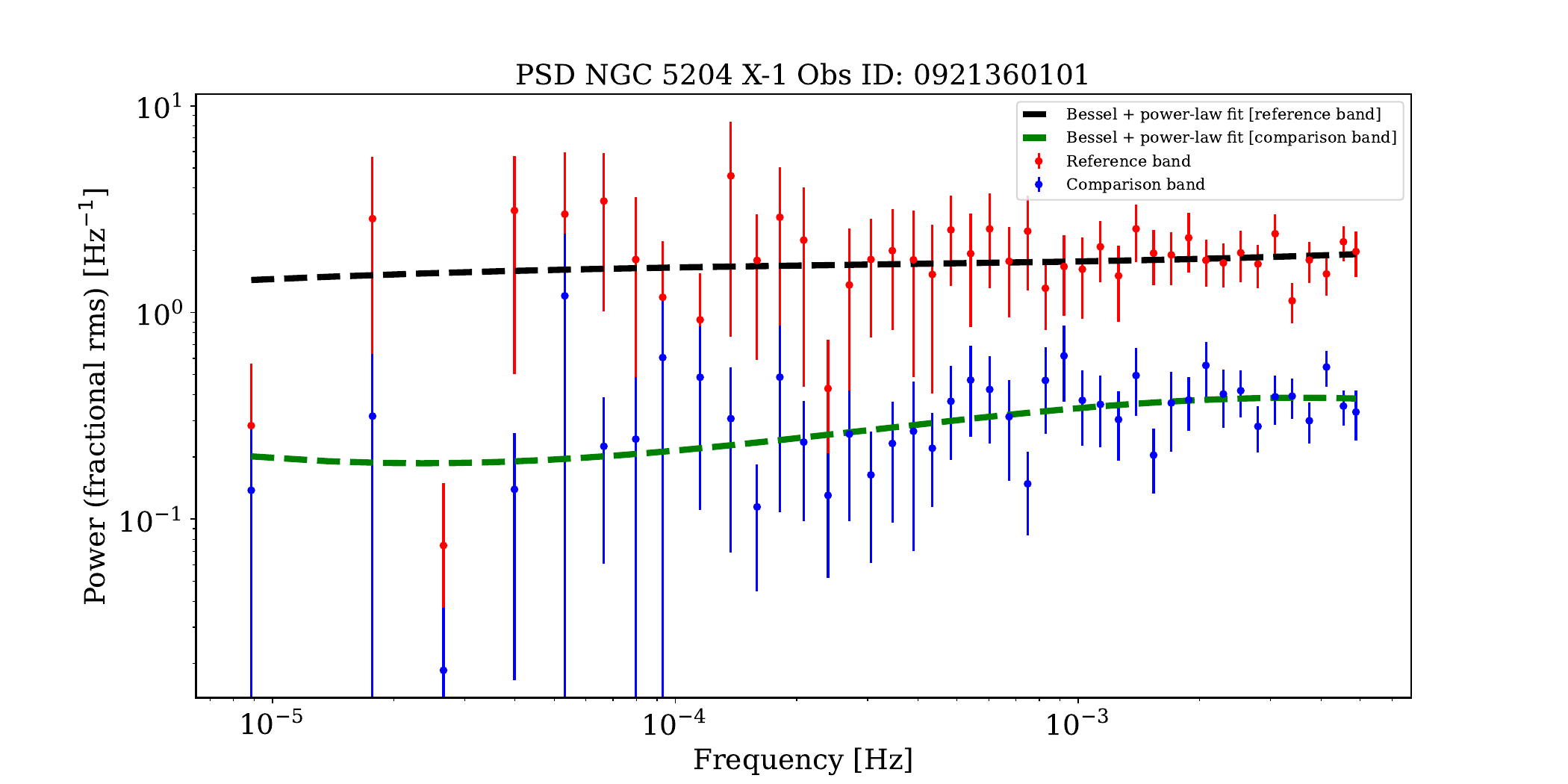}
\includegraphics[scale=0.25]{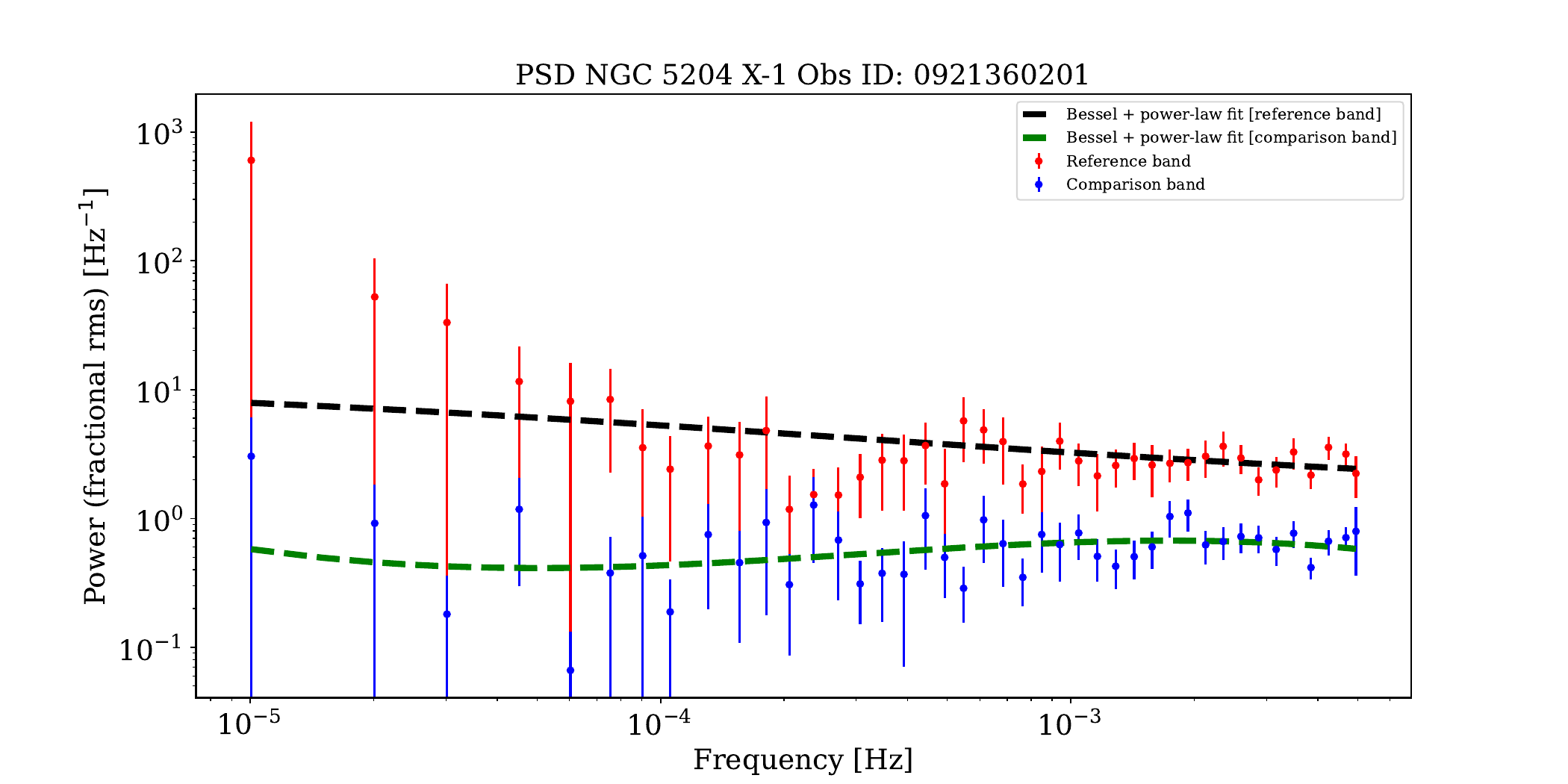}
\includegraphics[scale=0.25]{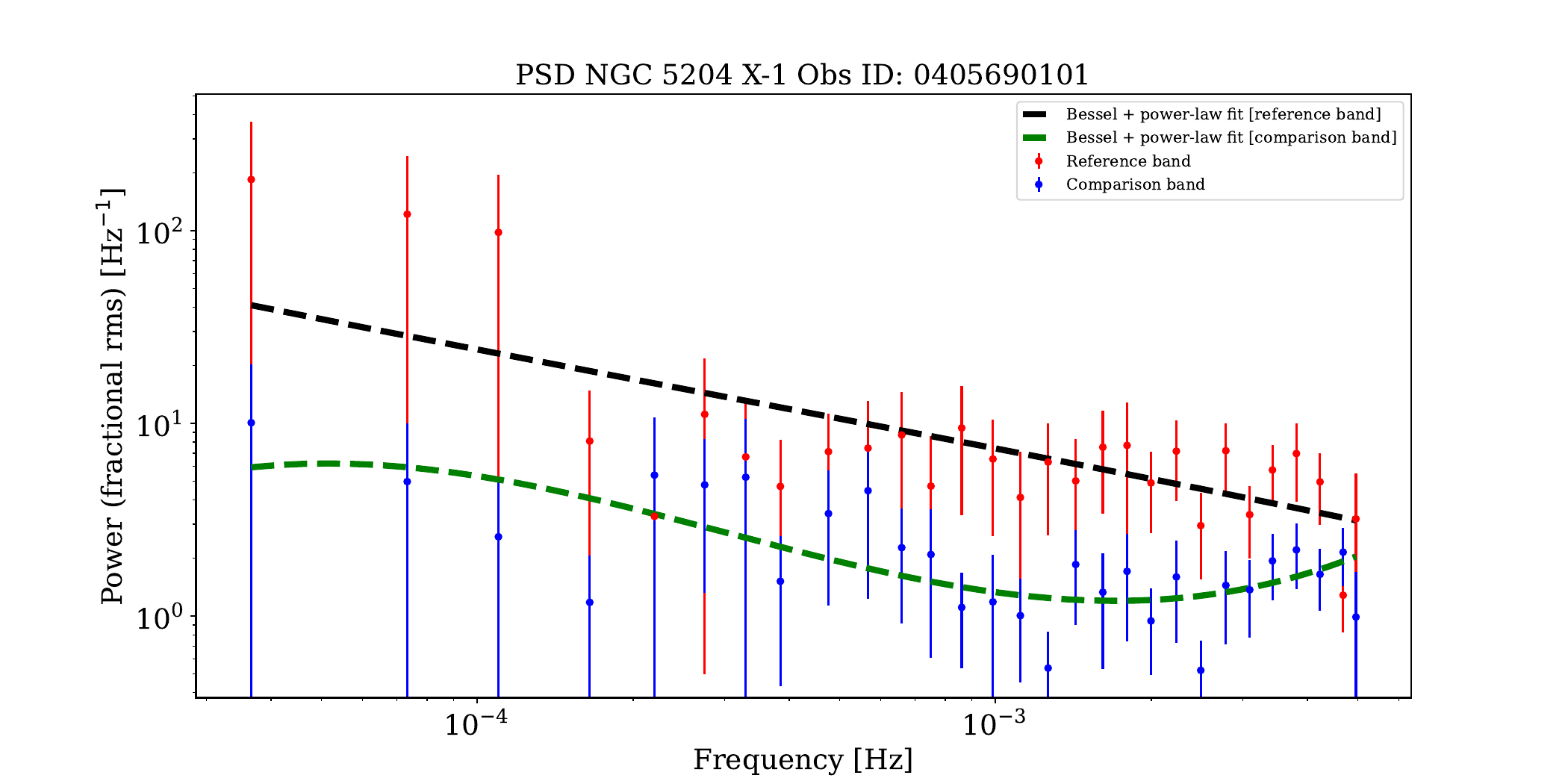}
\includegraphics[scale=0.25]{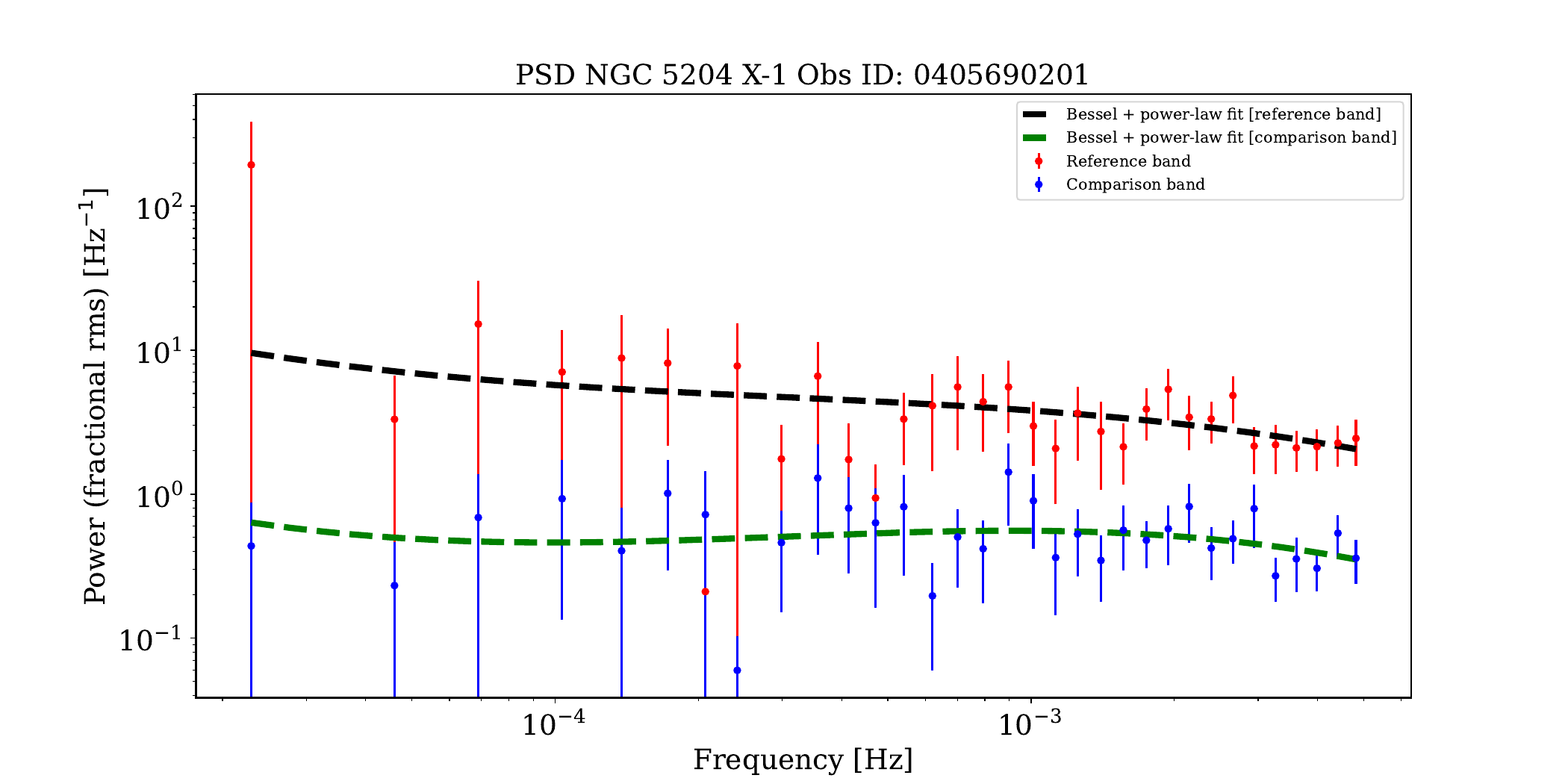}
\includegraphics[scale=0.25]{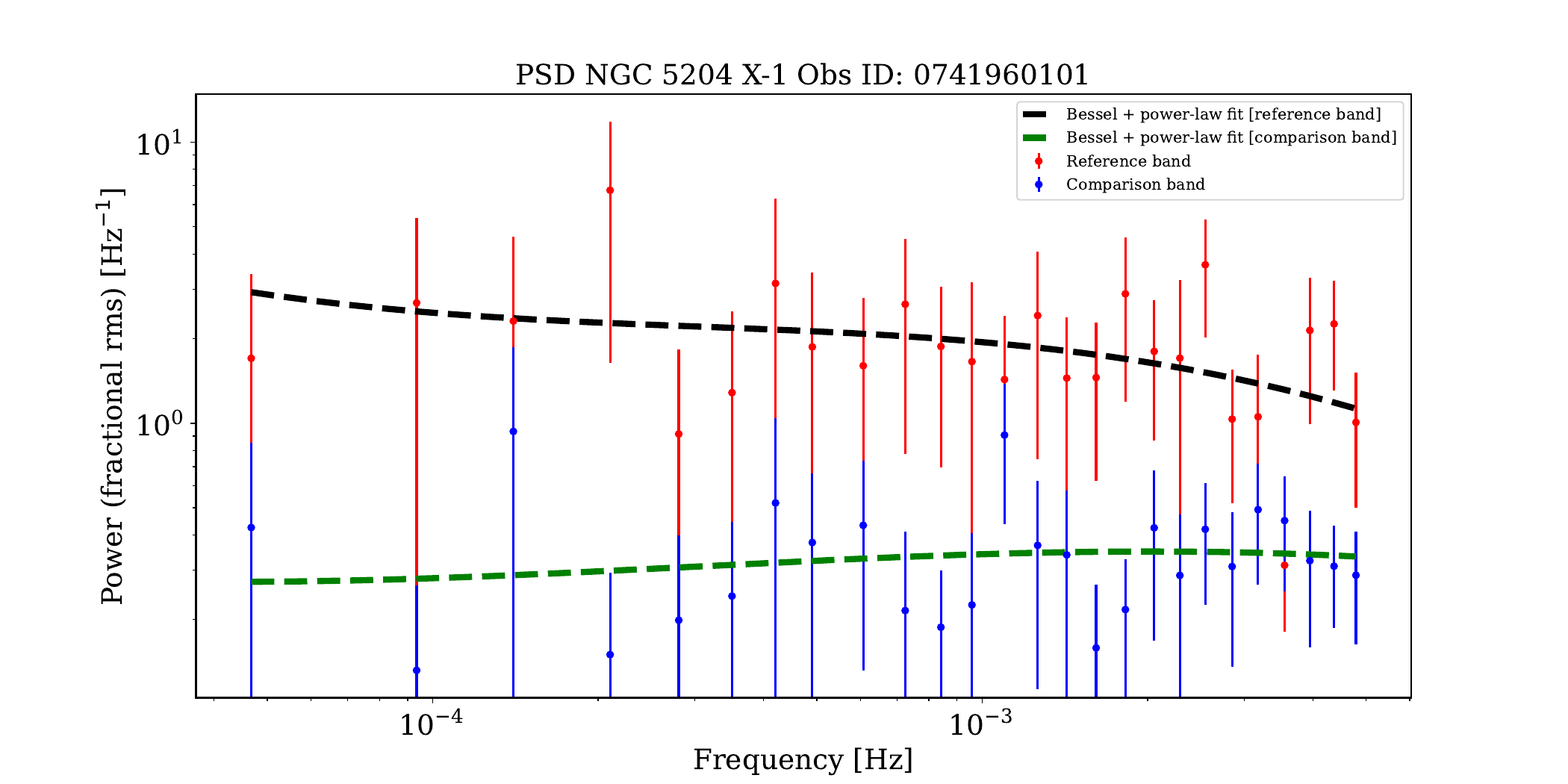}
\caption{We show the power-spectral densities (PSDs) for the highest signal--to--noise observations of NGC 5204 X--1, binned geometrically, to place limits on the presence of low--frequency (Type C) quasi--periodic oscillations, which have previously been found in other ULXs at similar accretion states (NGC 5408 X--1: \citealt{deMarco2013}; TDEs). Here, the PSD continua are phenomenologically well modelled with a simple power--law plus Bessel function of the first kind. We find no quasi--periodicities in these spectra at any epoch for a range of different geometric binning factors.}
\label{fig:XMM_psds_ngc5204}
\end{figure*}

\begin{figure*}
\centering
\includegraphics[scale=0.25]{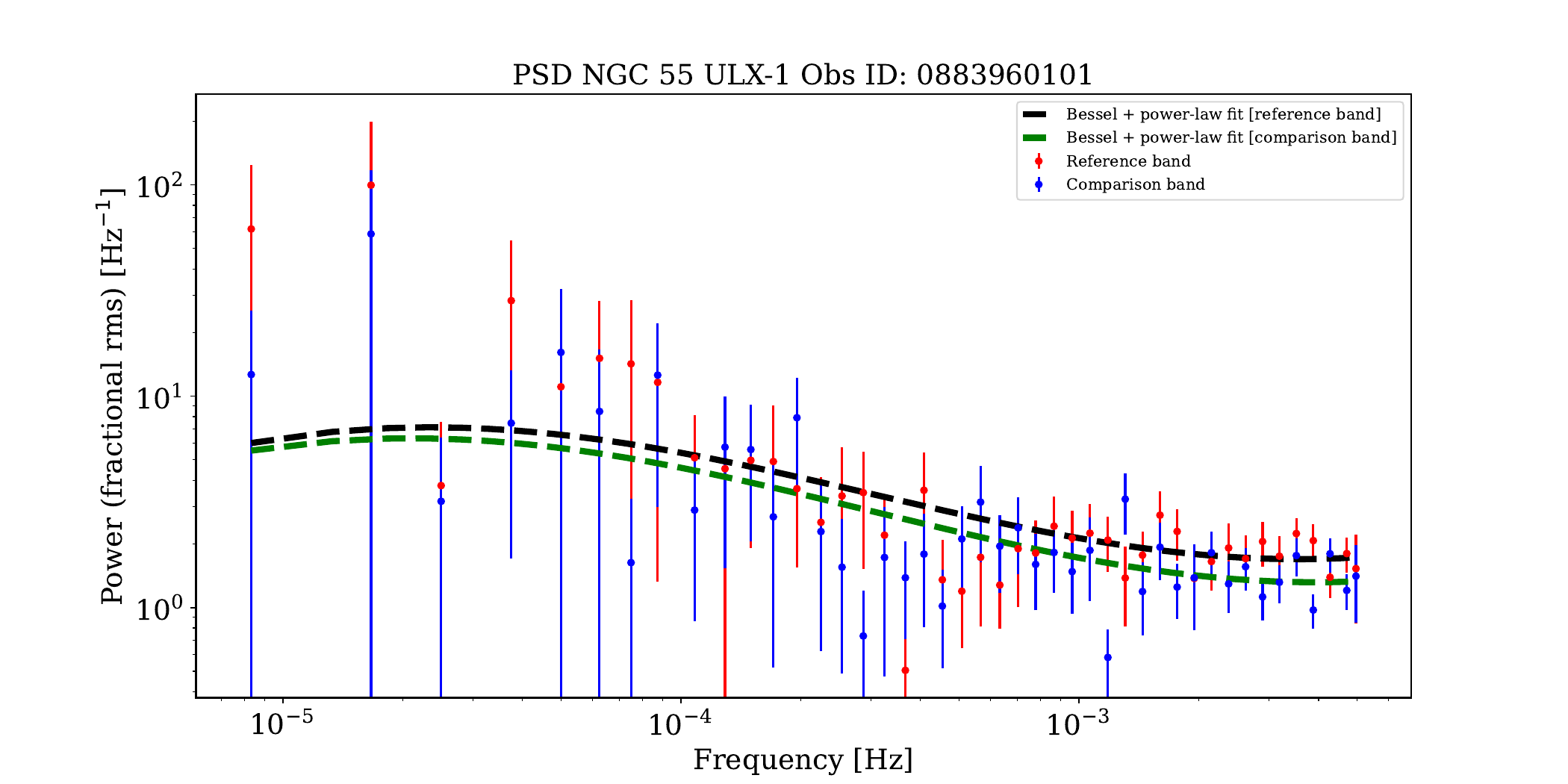}
\includegraphics[scale=0.25]{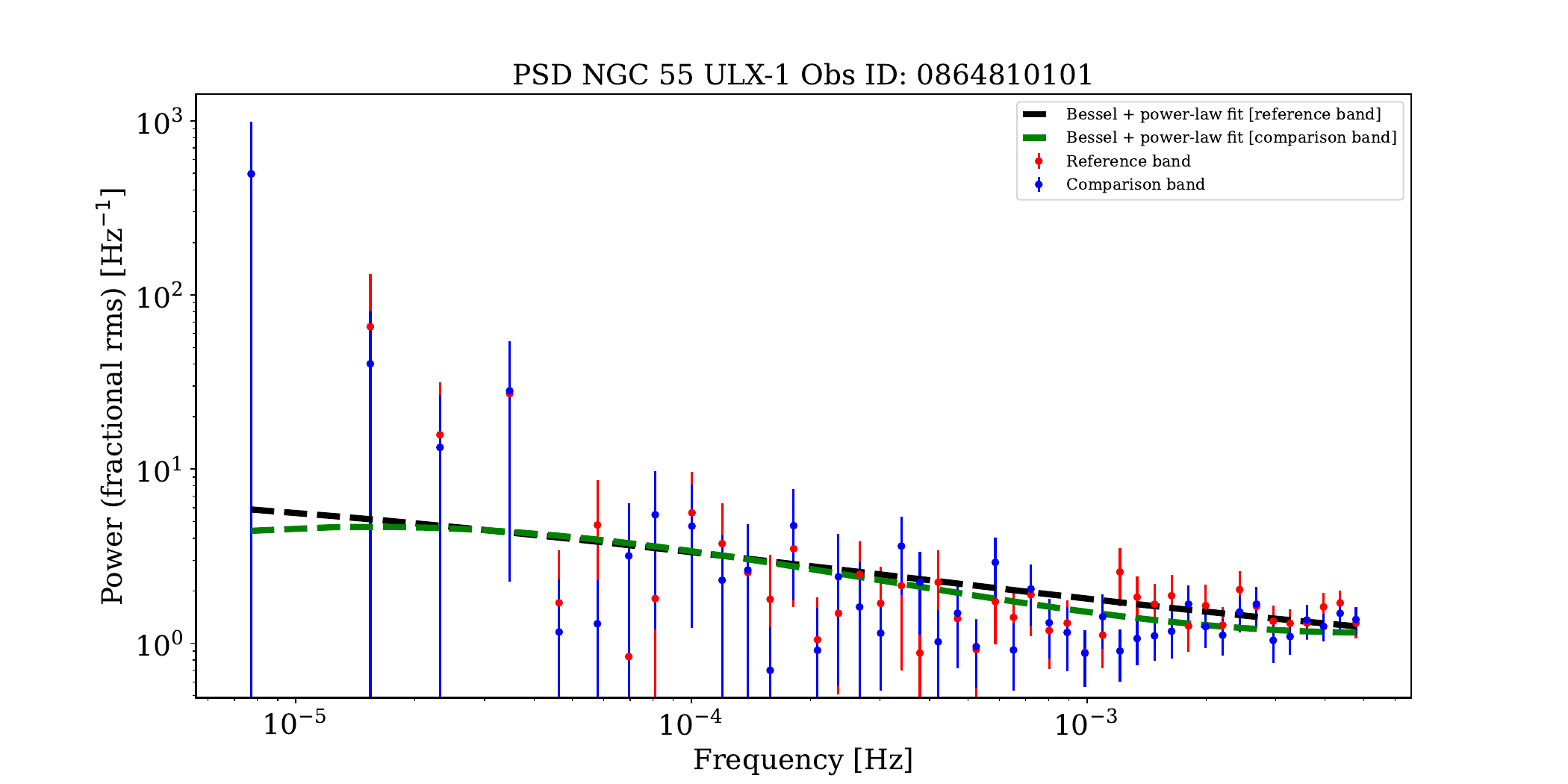}
\includegraphics[scale=0.25]{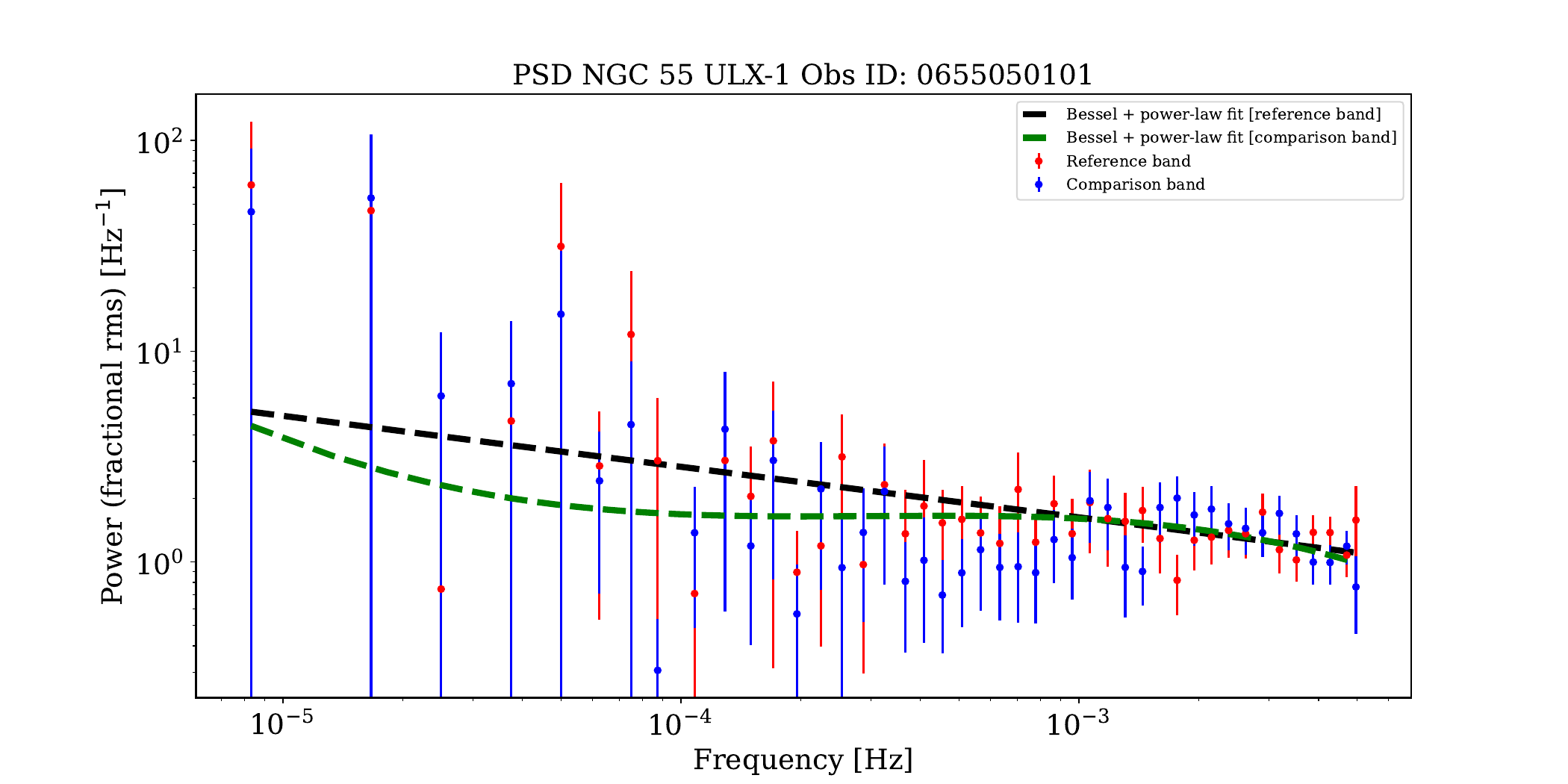}
\includegraphics[scale=0.25]{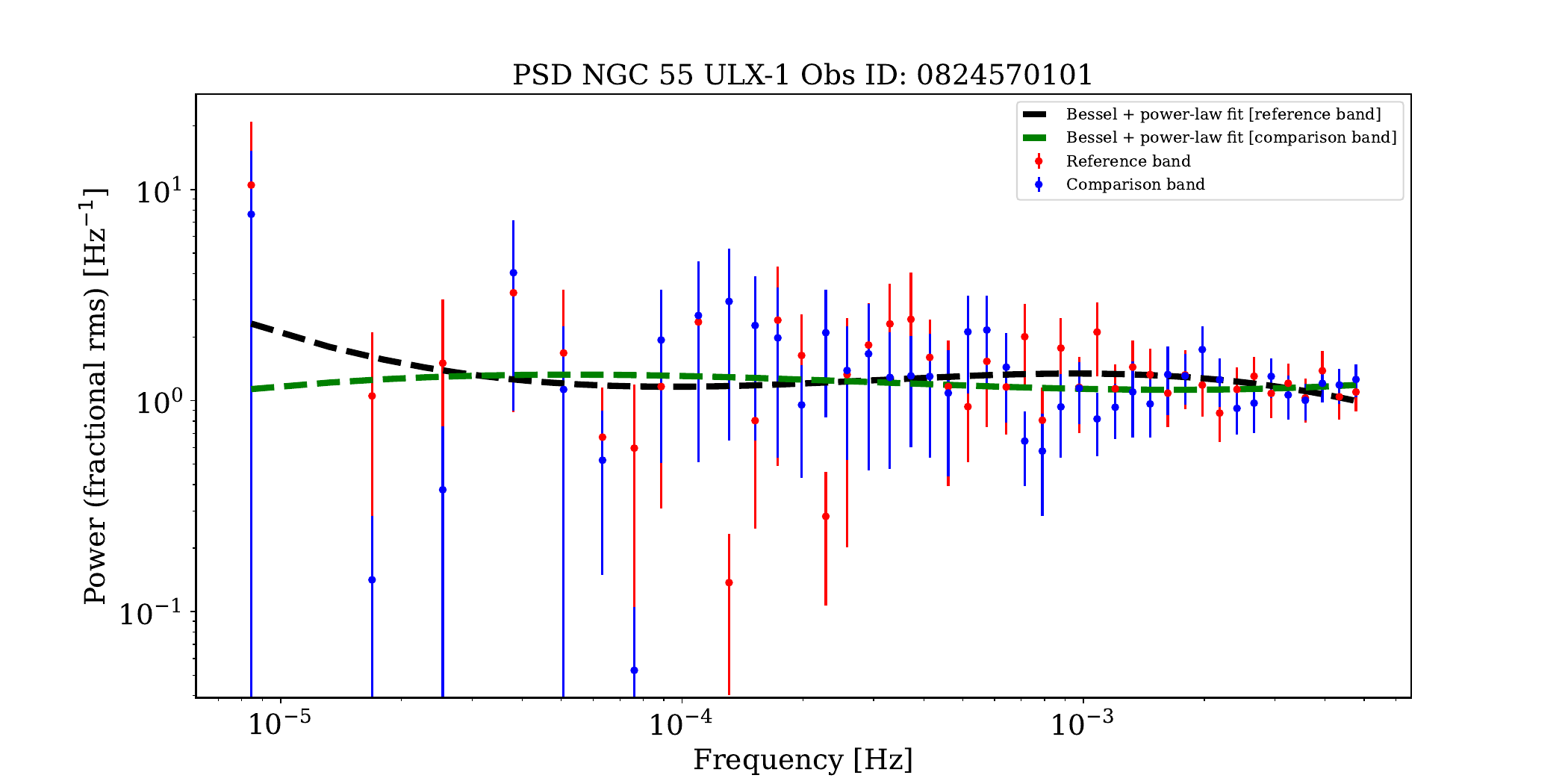}
\caption{Same as Figure~\ref{fig:XMM_psds_ngc5204} but for NGC 55 ULX--1.}
\label{fig:XMM_psds_ngc55}
\end{figure*}

\begin{figure*}
\centering
\includegraphics[scale=0.35]{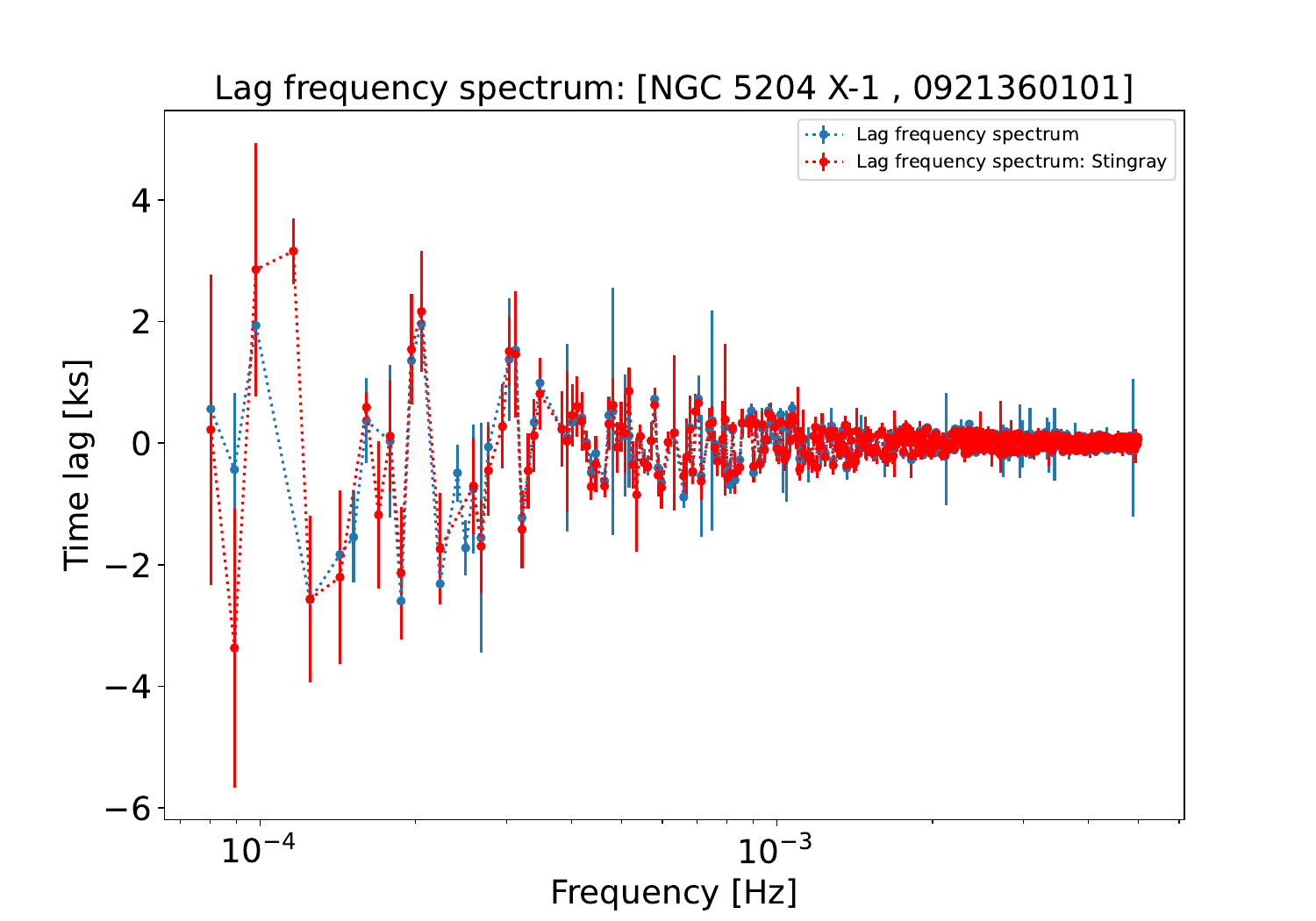}
\includegraphics[scale=0.35]{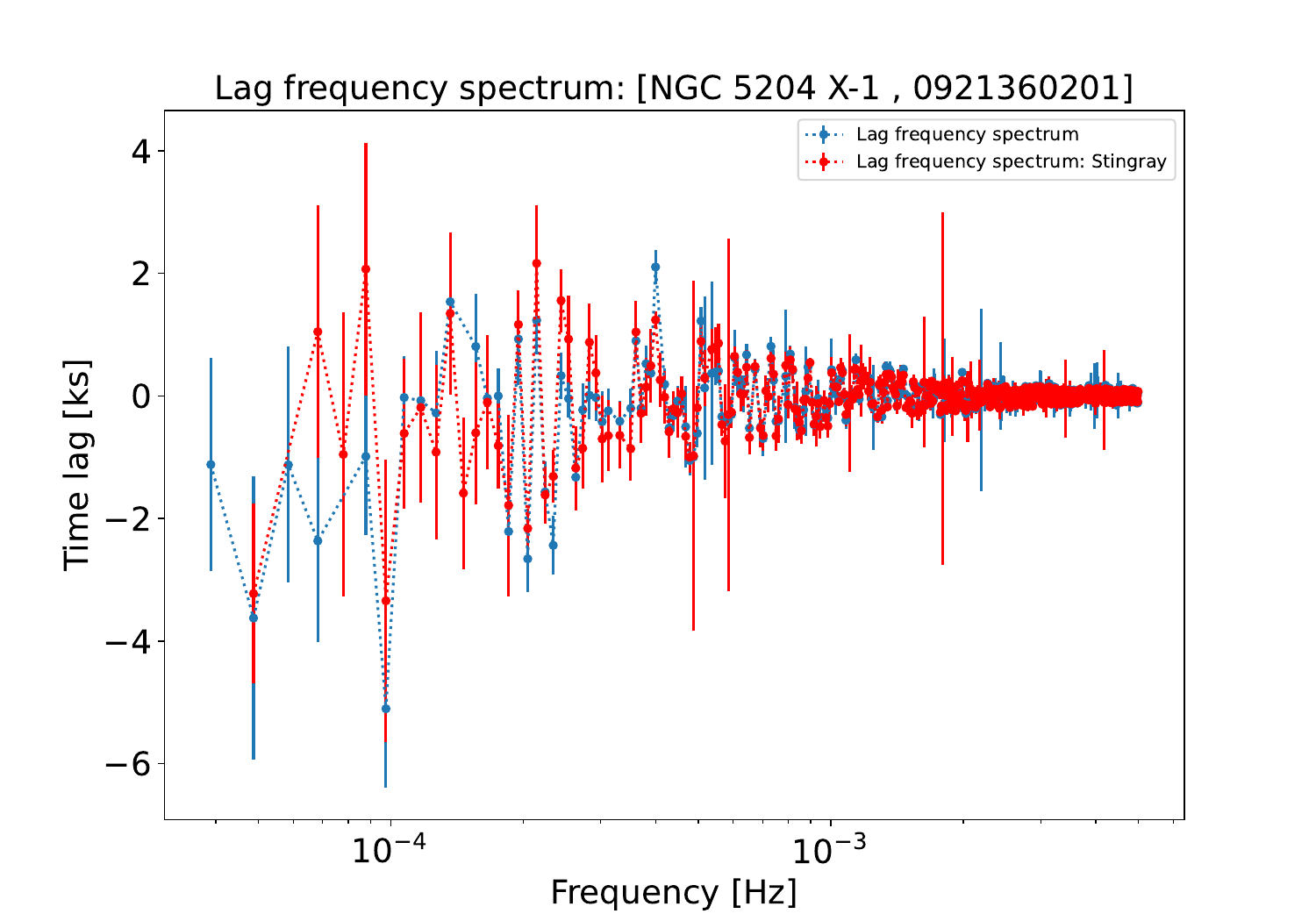}
\includegraphics[scale=0.35]{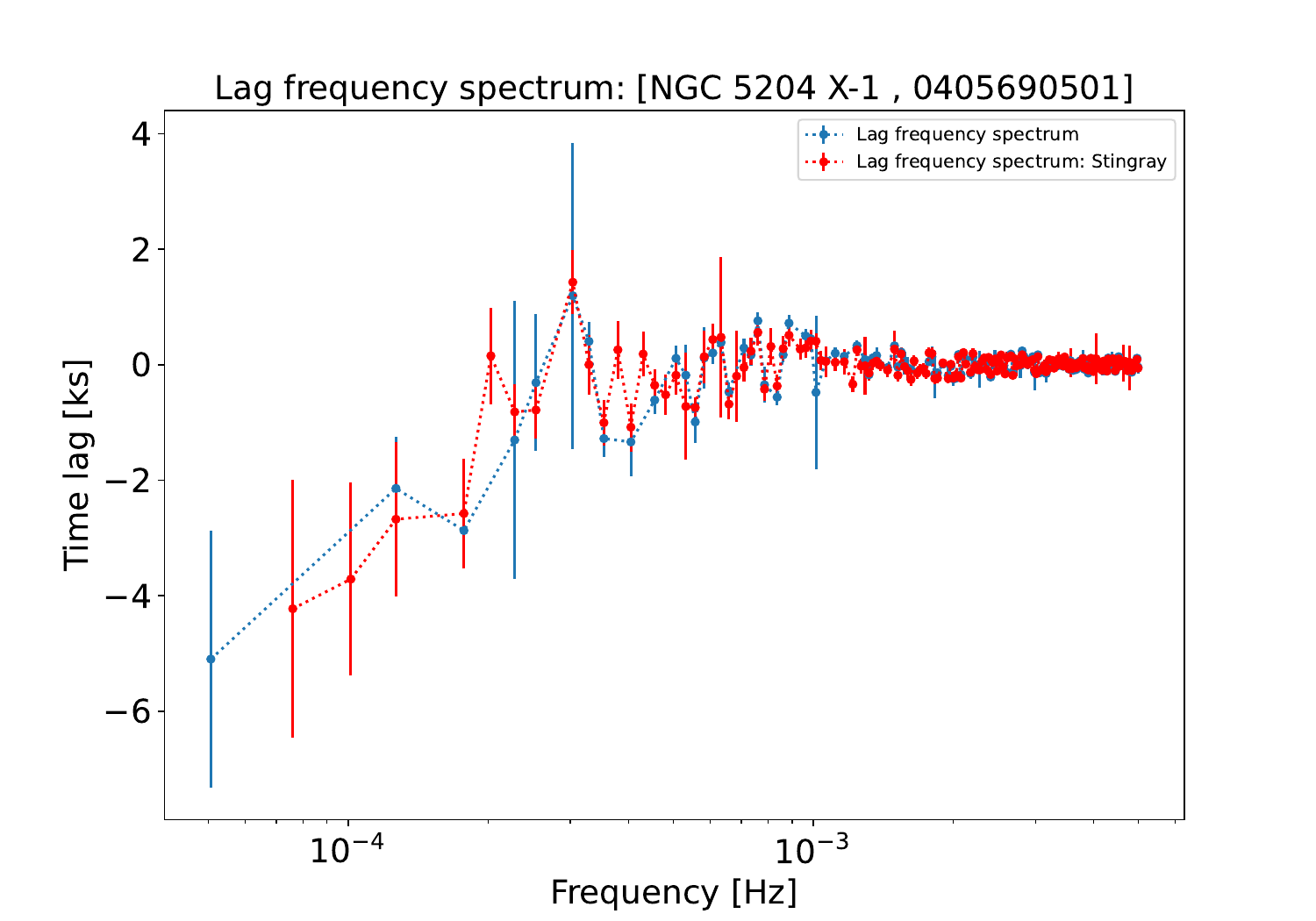}
\includegraphics[scale=0.35]{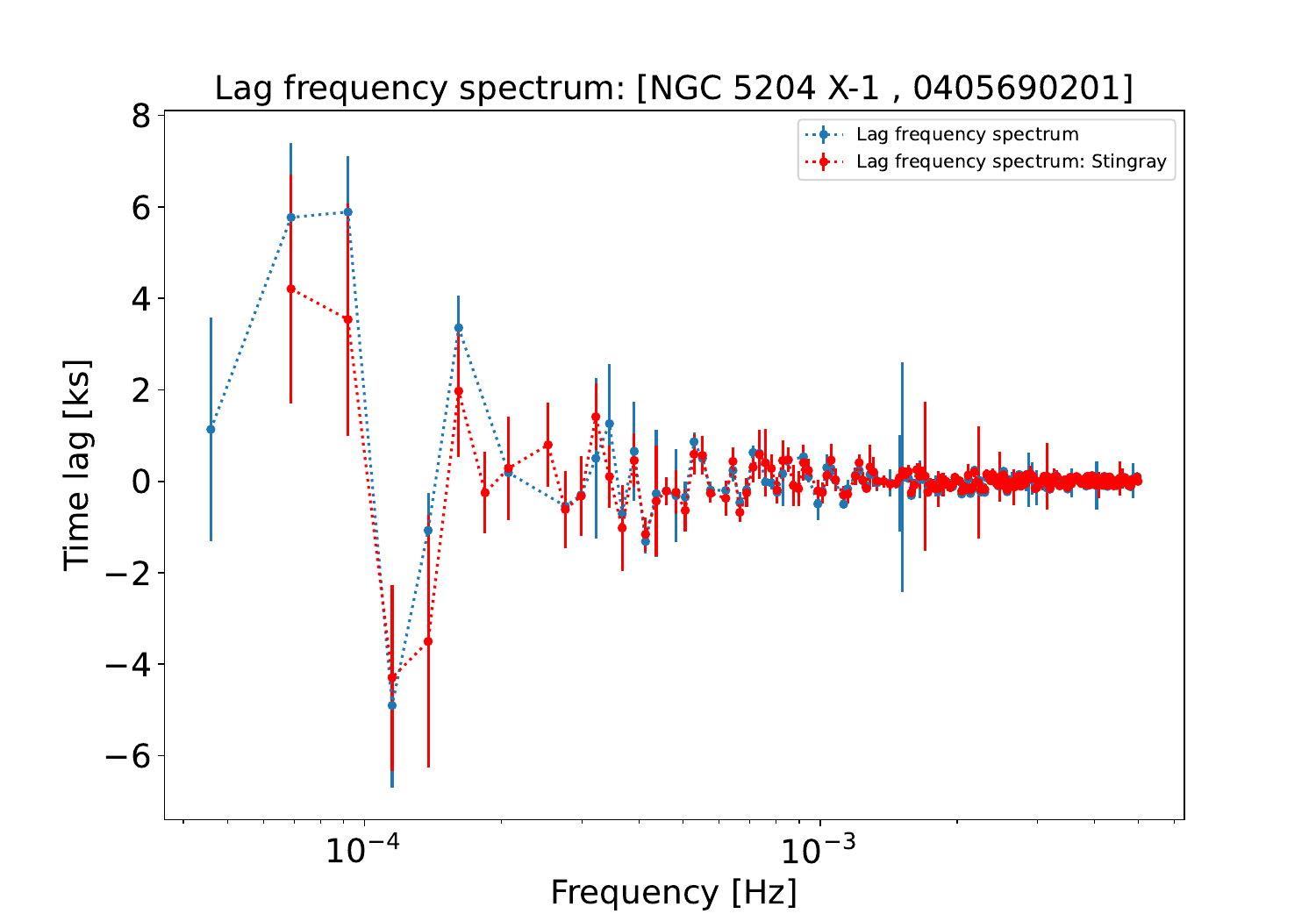}
\includegraphics[scale=0.35]{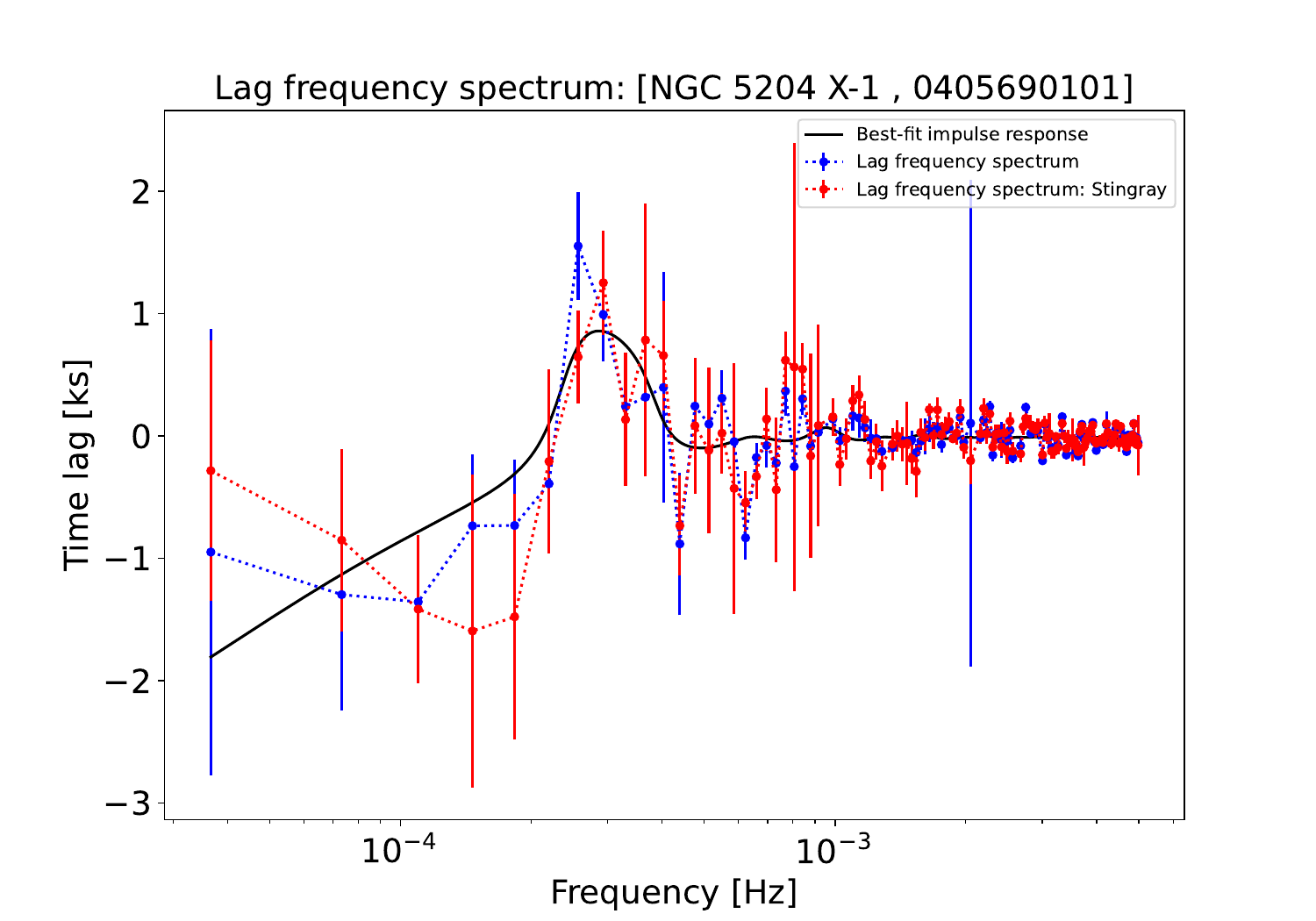}
\caption{A selection of lag--frequency spectra of NGC 5204 X--1 computed between the soft (0.3--2.0) keV reference band and the hard (2.0--10 keV) comparison band light--curves using {\it{XMM--Newton}} observations, as outlined in section 3. The aim was to search for propagation and reverberation lags in order to derive constraints on the size--scale and structure of the accretion flow surrounding the compact object. In most observations, we find no obvious correlations in the time--delay profiles, although in the particular case of Obs ID: 0405690101, we model the spectra using the transfer functions highlighted in the text since the lags here are moderately significant. We use the thick disc static accretion geometry proposed for ULXs, which contains an advection domainted spherical envelope near $R_{\text{sph}}$ that smooths the hard X--rays from close to the compact object. Here, the figures are shown only for the longest exposure observations with the highest count--rates.}
\label{fig:XMM_lag_freq_ngc5204}
\end{figure*}

\begin{figure*}
\centering
\includegraphics[scale=0.35]{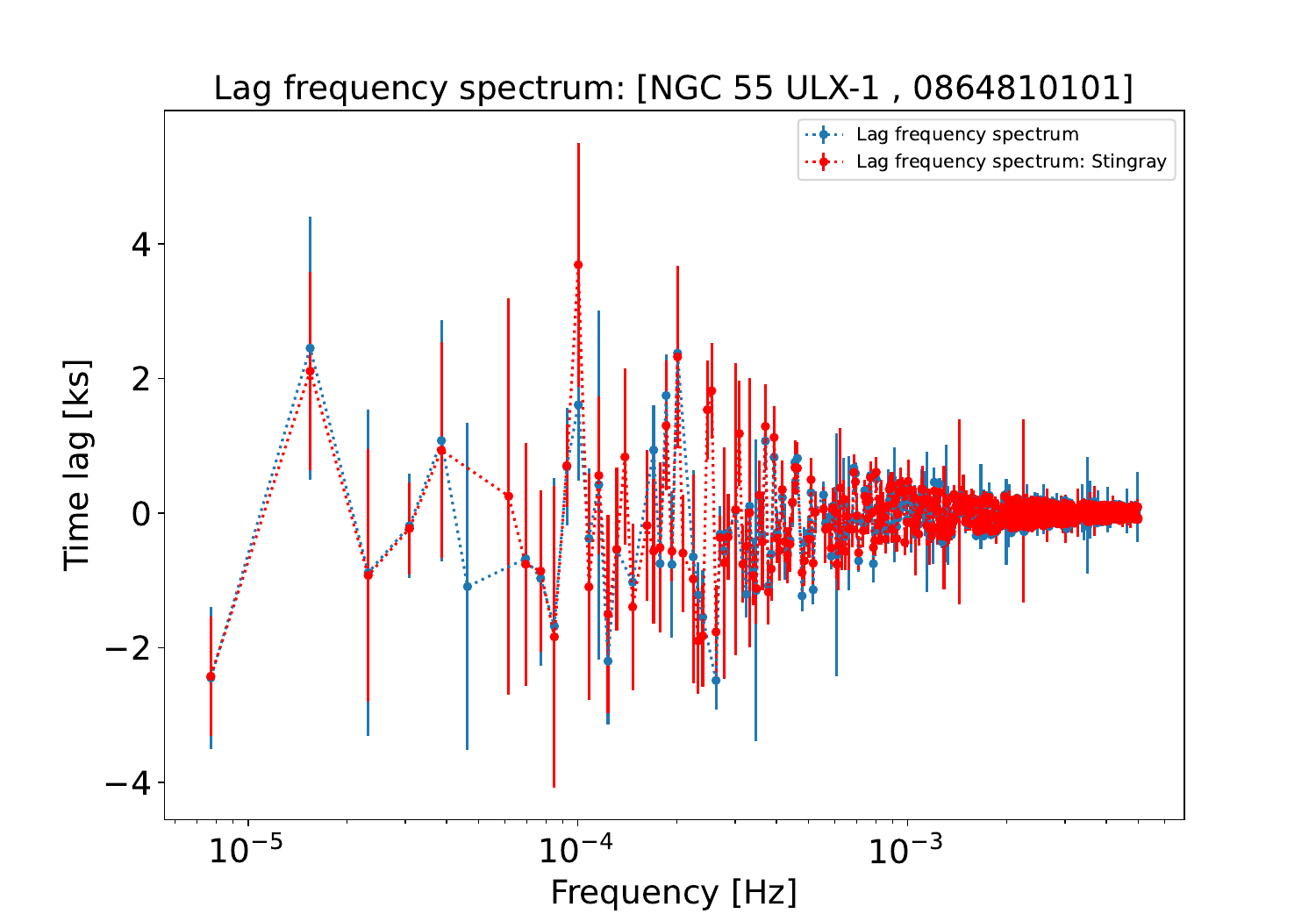}
\includegraphics[scale=0.35]{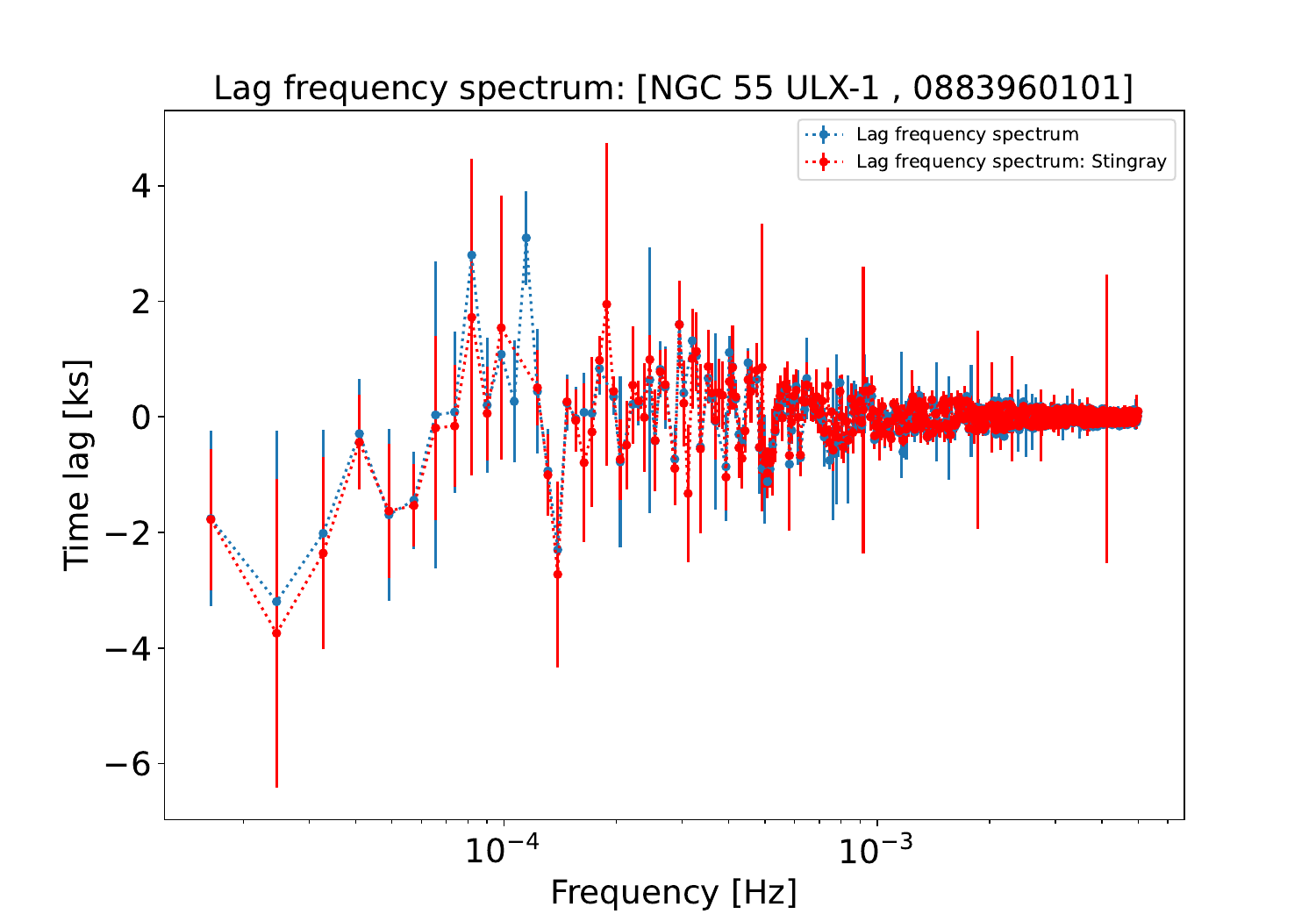}
\includegraphics[scale=0.35]{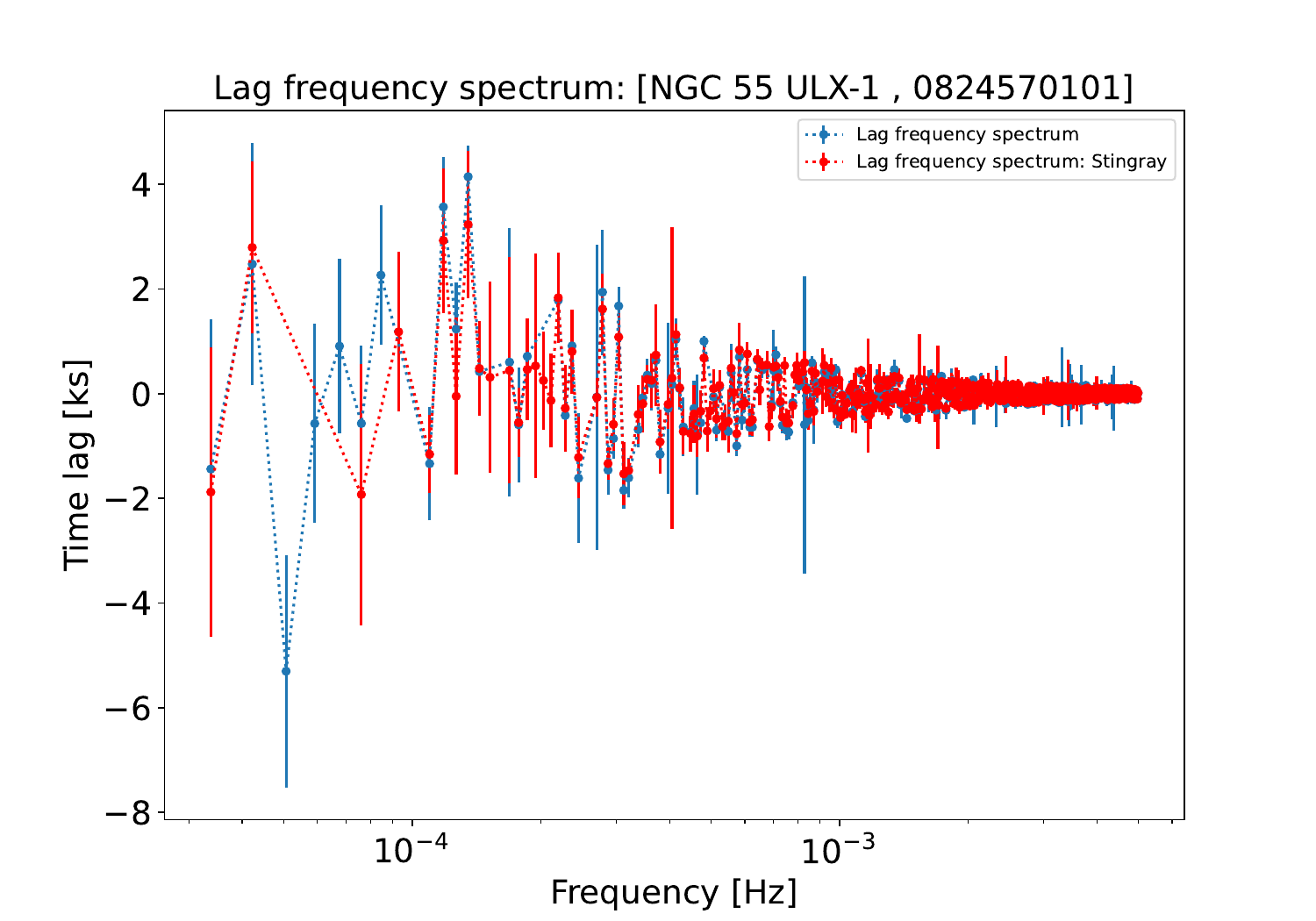}
\includegraphics[scale=0.35]{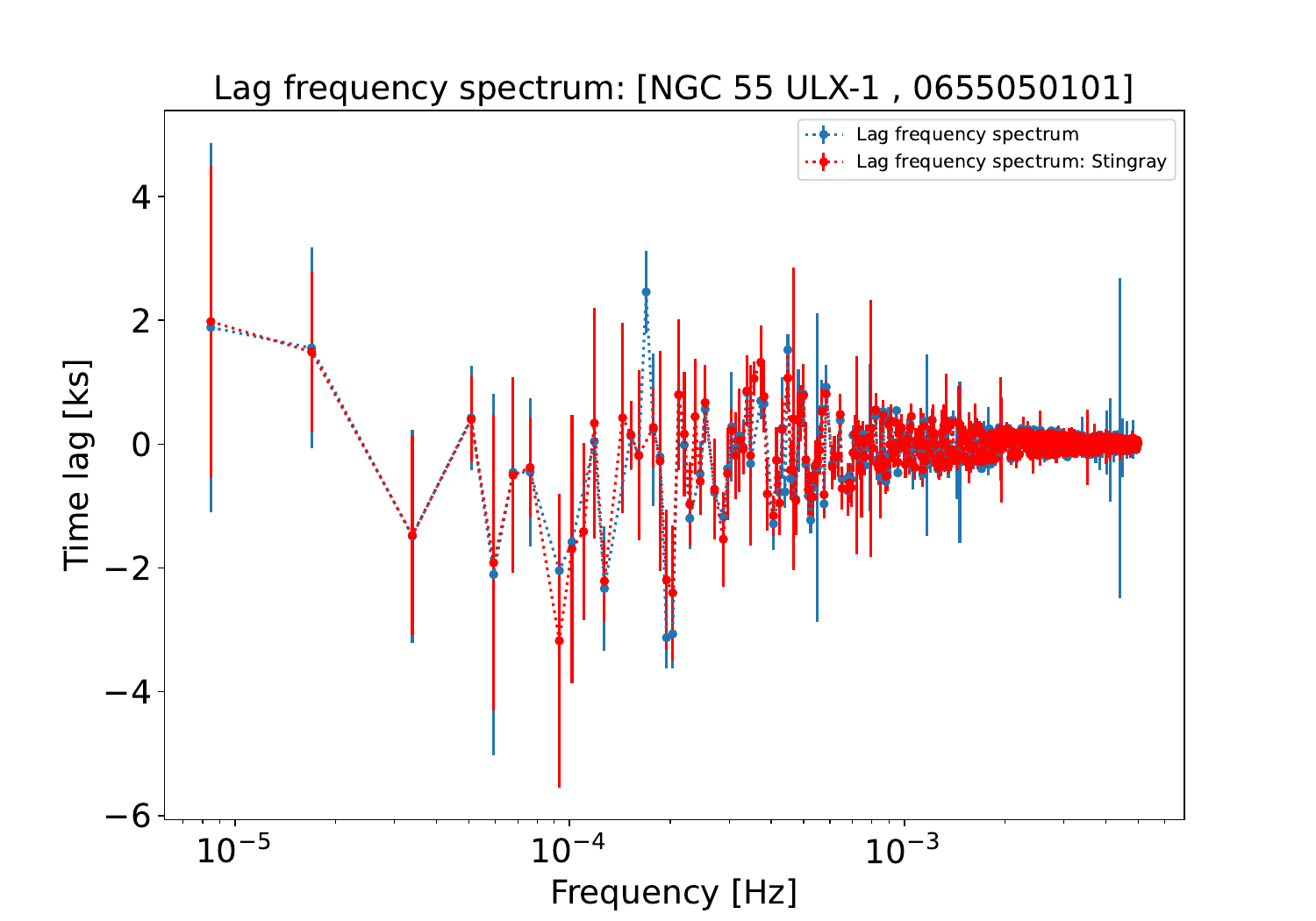}
\caption{Same as Figure~\ref{fig:XMM_lag_energy_ngc5204}, but for NGC 55 ULX--1, for which the reference band was 0.3--1.0 keV and the comparison band was 1--8.0 keV.}
\label{fig:XMM_lag_freq_ngc55}
\end{figure*}

\begin{figure*}
\centering
\includegraphics[scale=0.35]{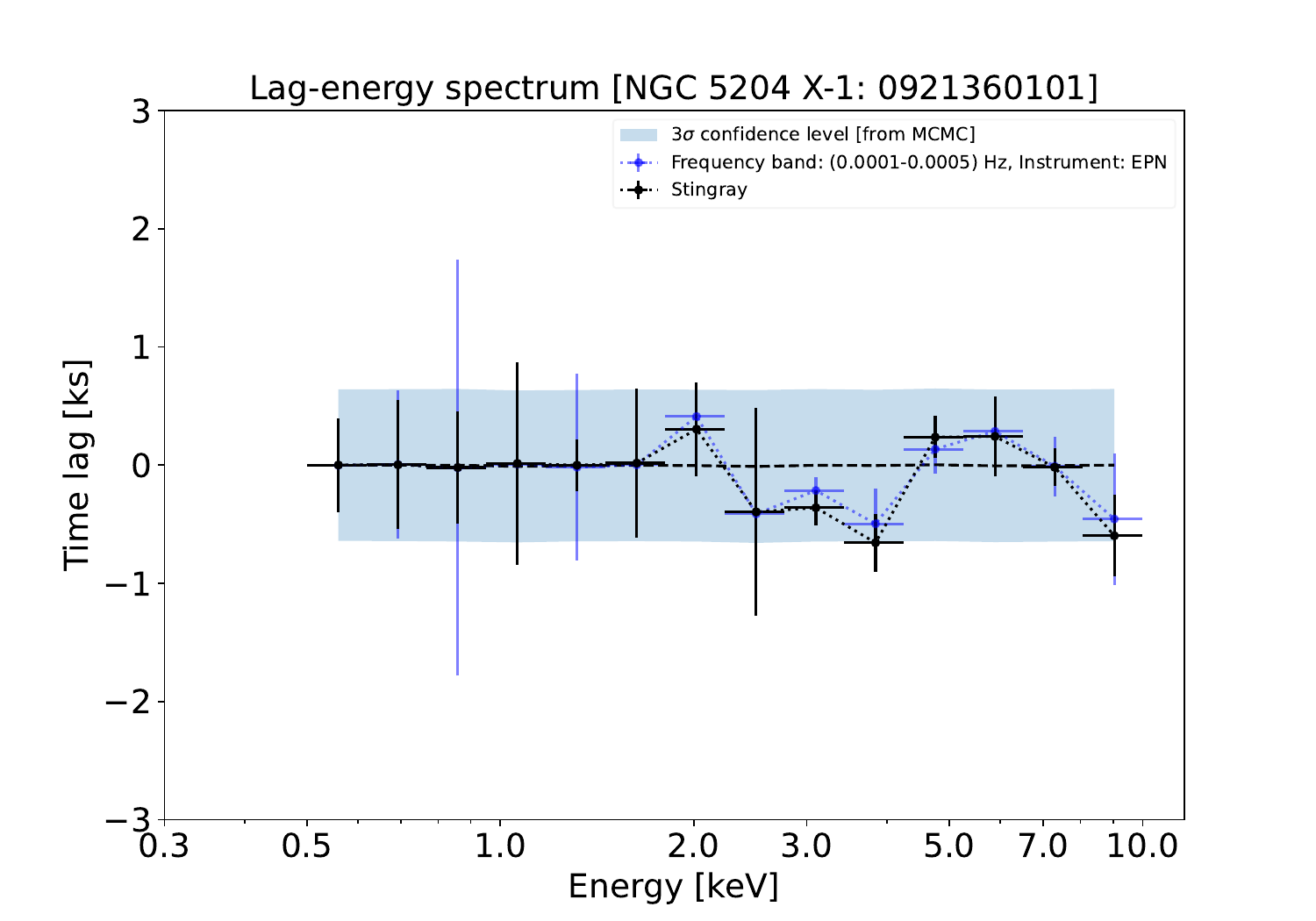}
\includegraphics[scale=0.35]{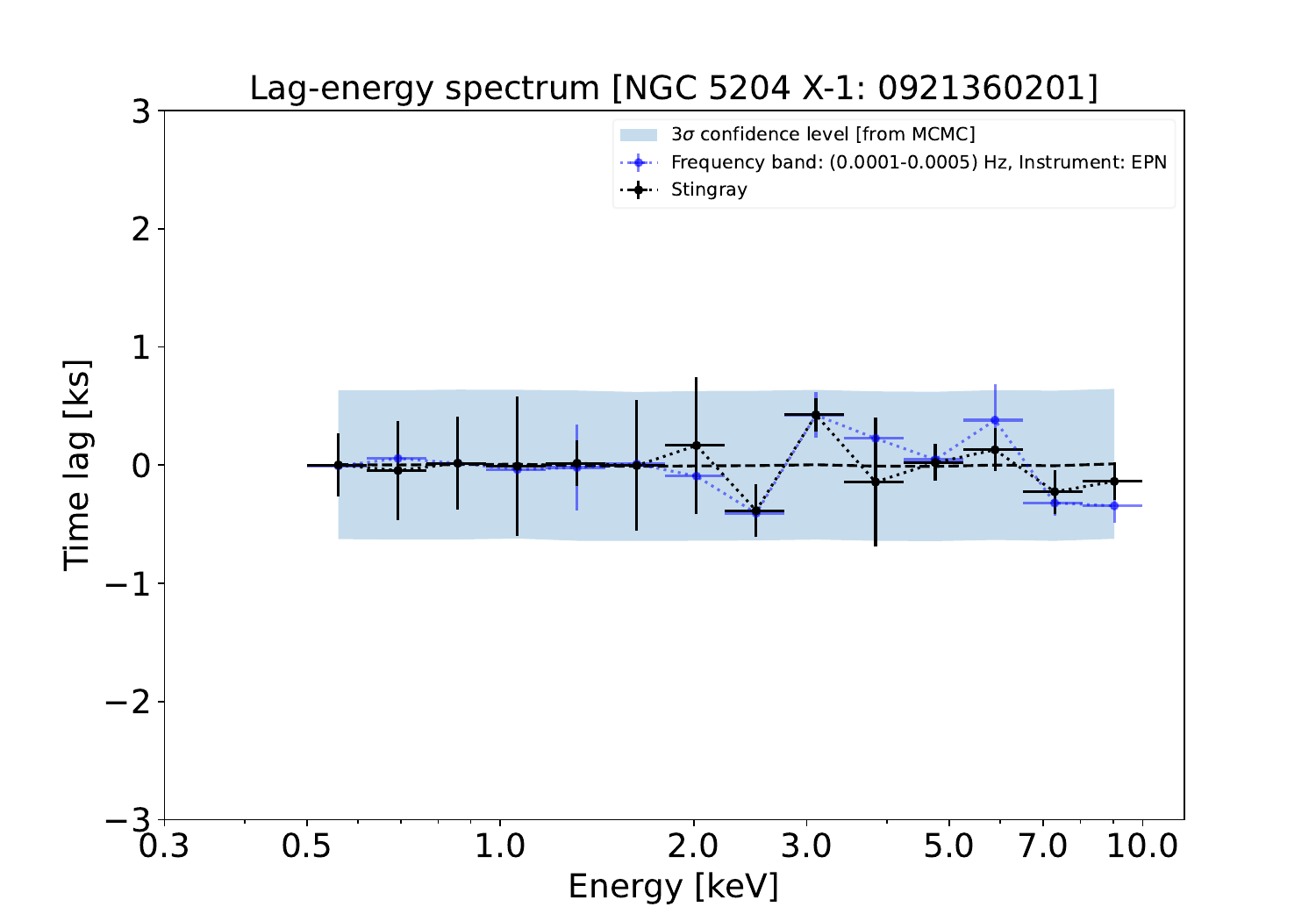}
\includegraphics[scale=0.35]{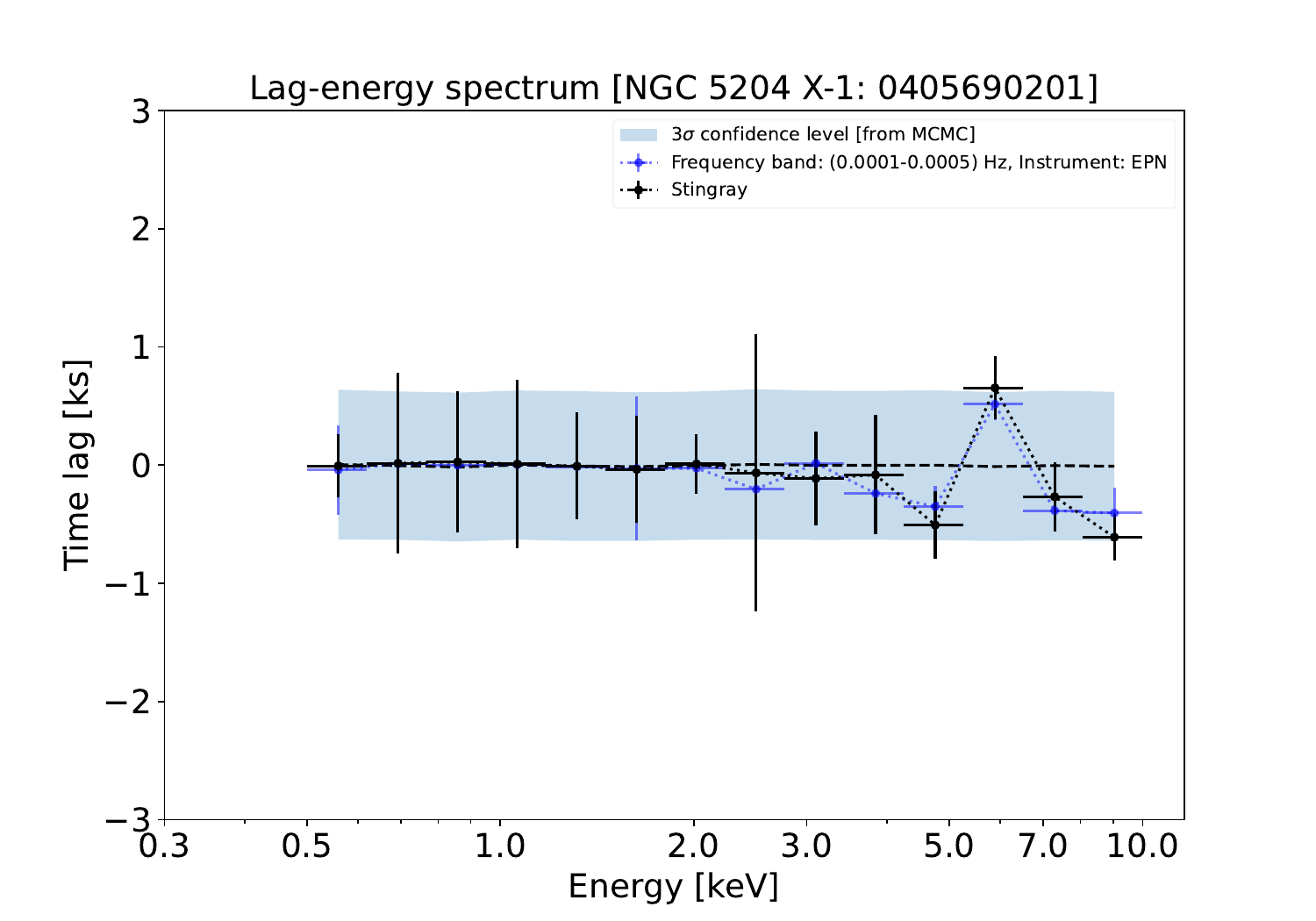}
\includegraphics[scale=0.35]{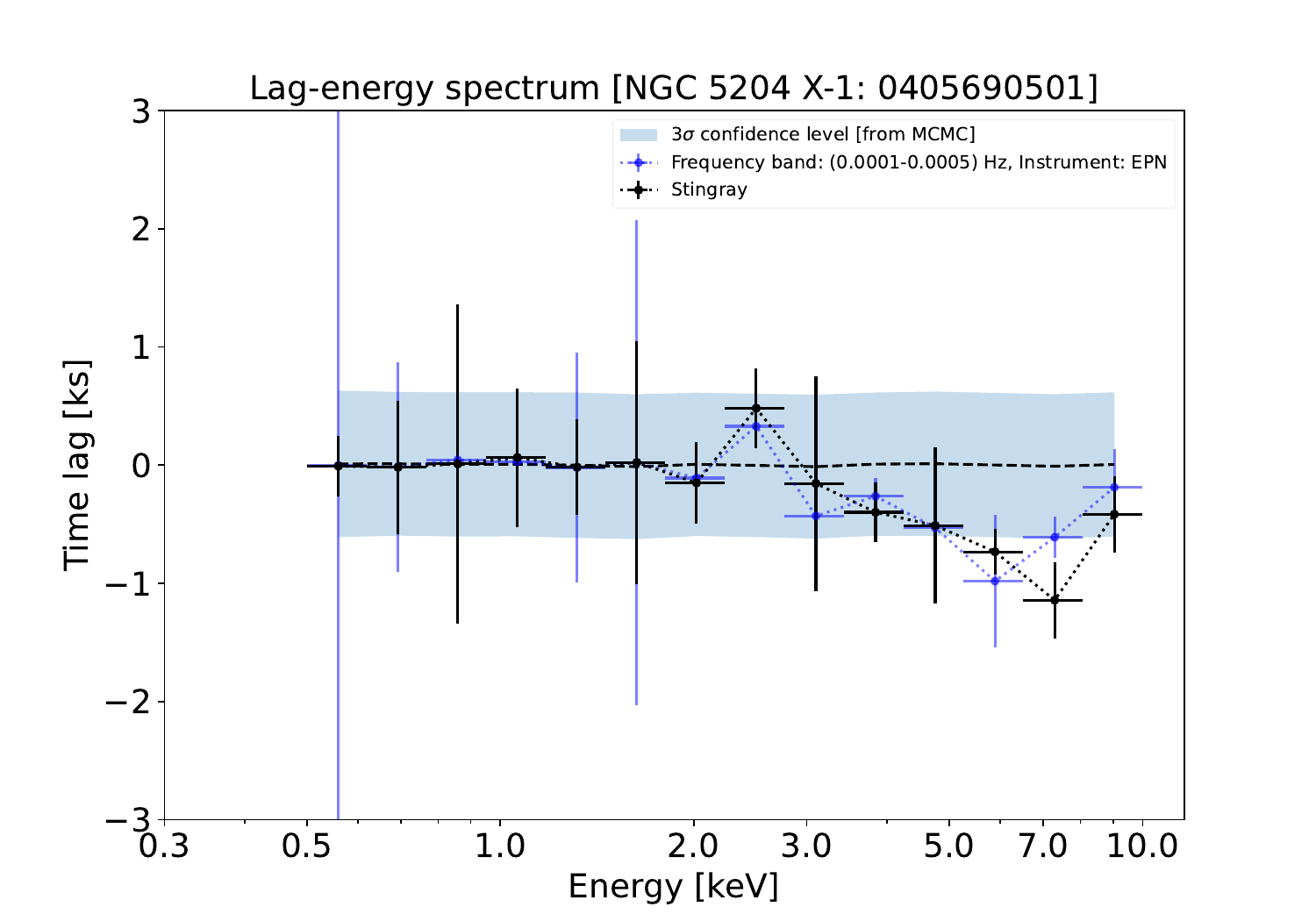}
\caption{A selection of Lag--energy spectra of NGC 5204 X--1, with the blue shaded region showing the 1$\sigma$ confidence level from MCMC simulations using the method of \citealt{timmer_koenig_1995}.}
\label{fig:XMM_lag_energy_ngc5204}
\end{figure*}

\begin{figure*}
\centering
\includegraphics[scale=0.35]{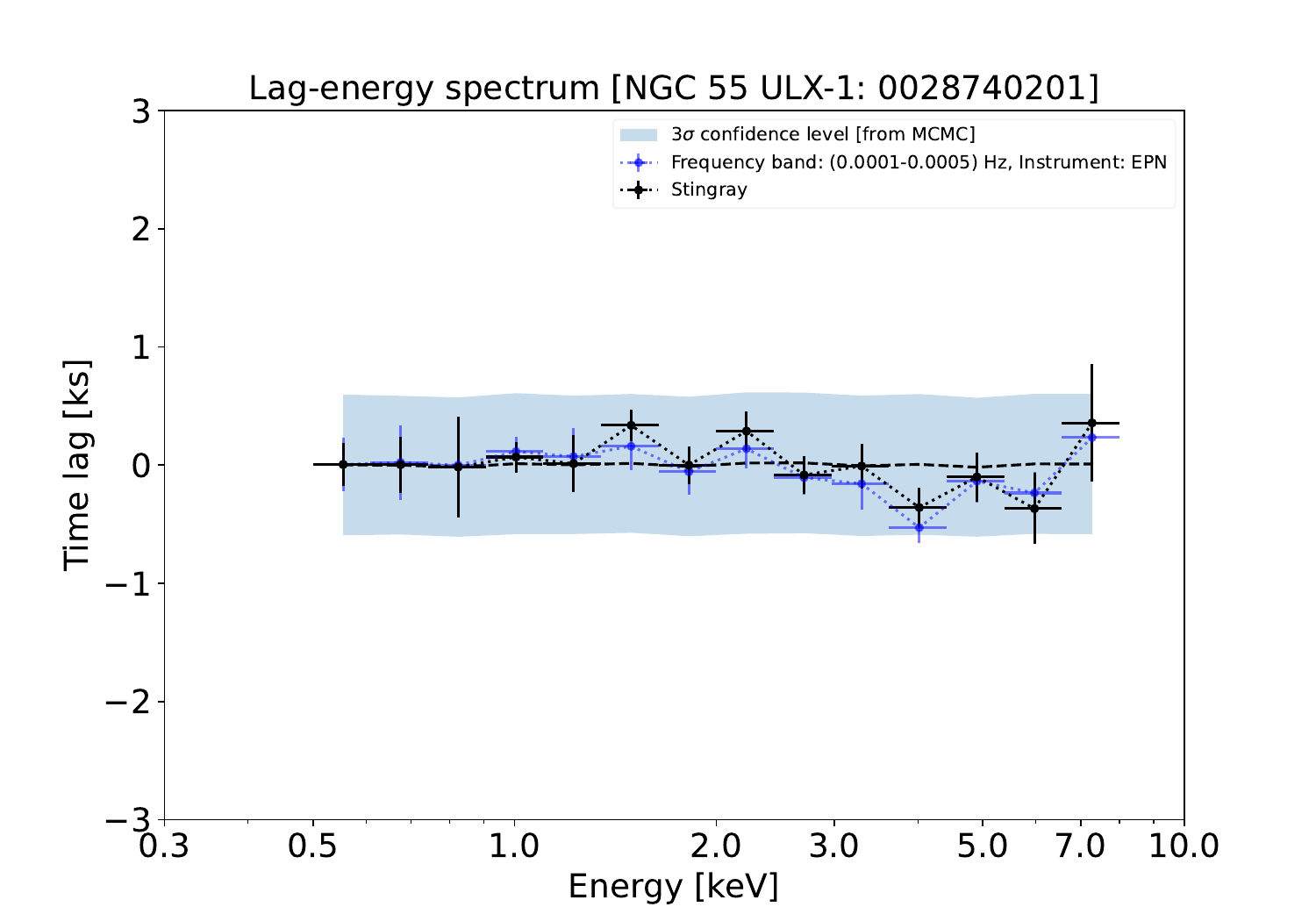} 
\includegraphics[scale=0.35]{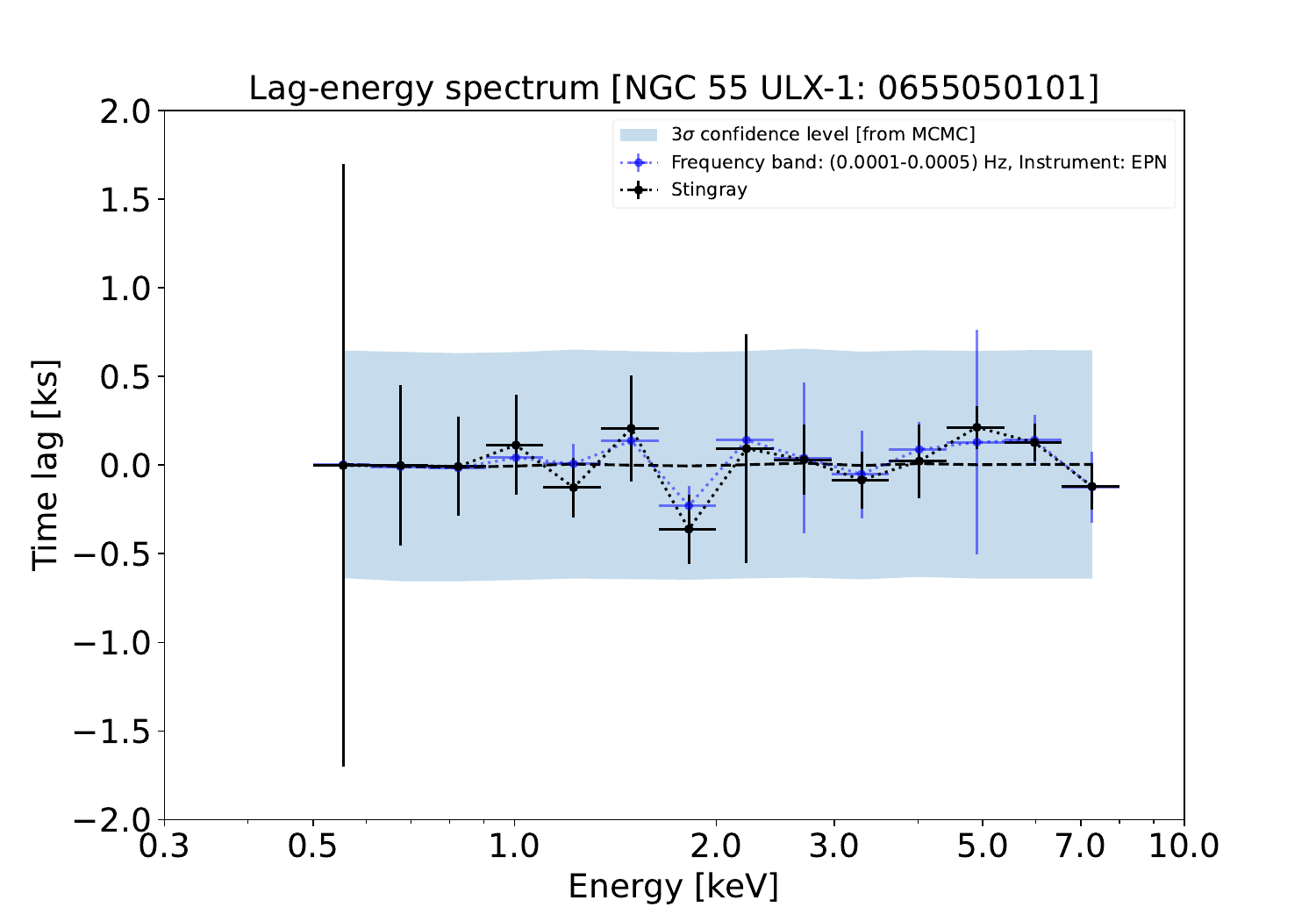}
\includegraphics[scale=0.35]{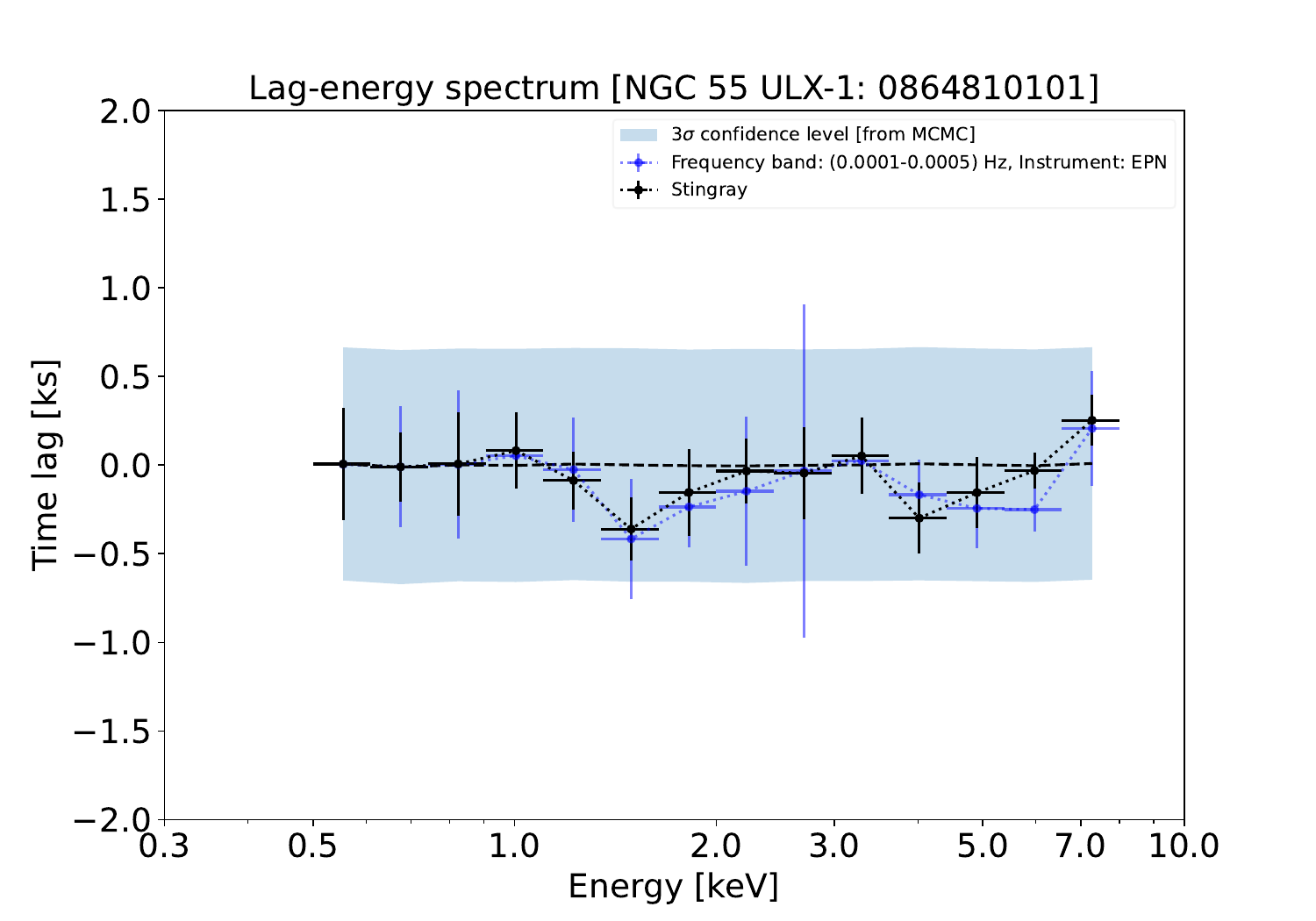}
\includegraphics[scale=0.35]{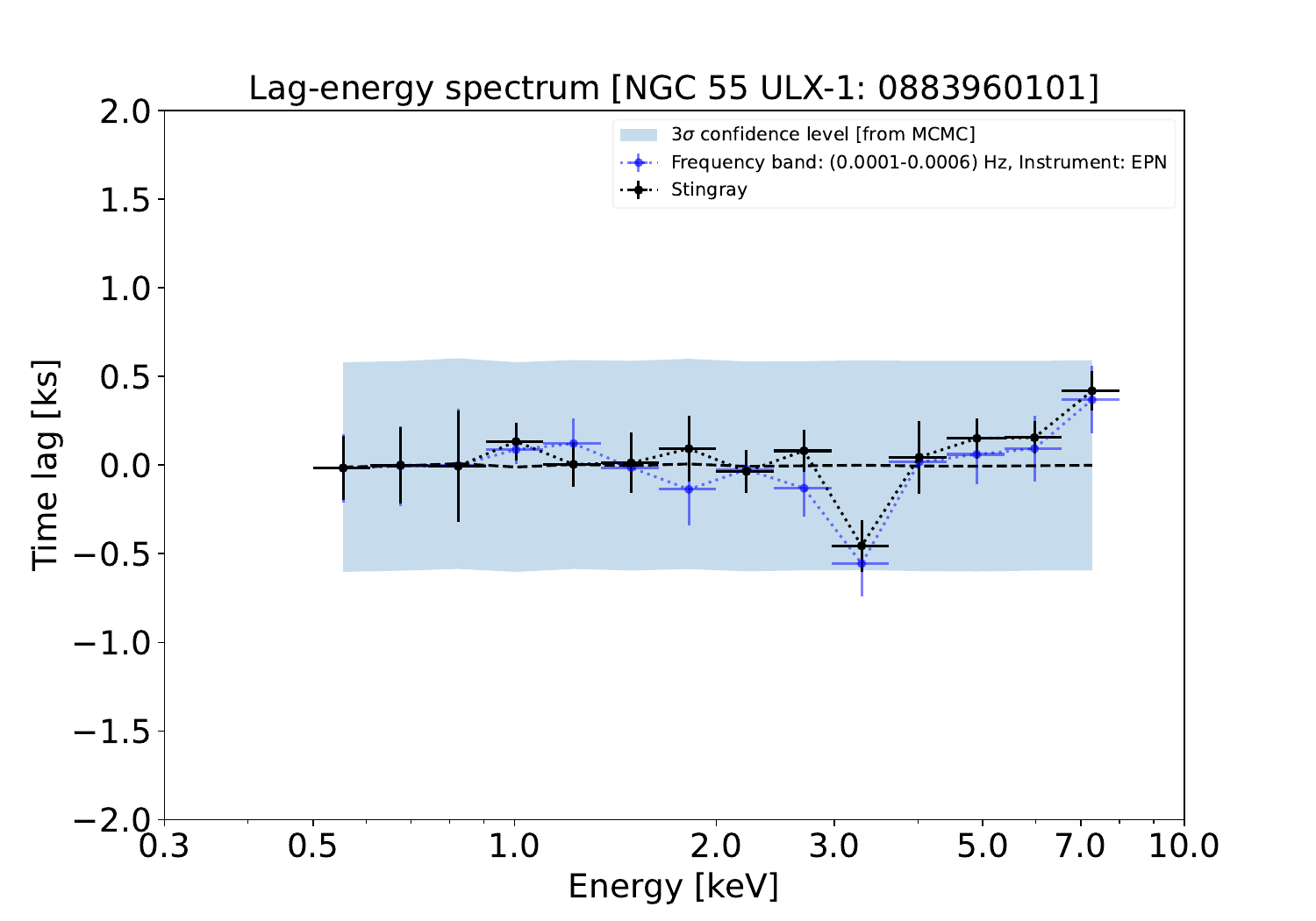}
\caption{Lag-energy spectra for the {\it{XMM-Newton}} observations of NGC 55 ULX--1 with the same method used to compute confidence levels as in Figure~\ref{fig:XMM_lag_energy_ngc5204}.}
\label{fig:XMM_lag_energy_ngc55}
\end{figure*}

where $A_{\text{x}}$ and $A_{\text{y}}$ are the amplitudes of the averaged CPSD for the comparison band and the reference band, $\phi(f)$ is the phase between the real and imaginary components of the CPSD and the time lag. The uncertainty in the time lag $\delta \tau$ can be intuitively thought of as being the scatter in the cross--spectrum vector, which is related to the raw coherence $\gamma$ between the reference band and comparison band light--curves and the number of light--curve segments $N$ over which the PSD is averaged: 
\newline
\newline
\begin{equation}
\delta \tau = \sqrt{\frac{1 - \gamma^{2}}{2\gamma^{2}N}}
\end{equation}
\newline
\newline
where
\begin{equation}
\gamma^{2} = \frac {\mid \langle {C_{xy}} (f) \rangle \mid^{2} - n_{\text{bias}}^{2}}{ (\langle P_{X} (f) \rangle - P_{X,\text{noise}}) (\langle P_{Y} (f) \rangle - P_{Y,\text{noise}}) }
\end{equation}
\newline
\newline
where $n^{2}_{\text{bias}}$ refers to any power in the CPSD that arises due to uncorrelated noise in the reference--band and comparison--band PSDs. Please refer to this \href{https://github.com/raj1294/XcovSpec}{link} for access to the original code/workflow (\citealt{XcovSpec}) that was used to produce the lag--energy and lag--frequency spectra. In order to improve the signal--to--noise of our lag measurements and correct for phase--wraps offset by units of $\pi$ in the phase--lag spectrum, we made use of all three {\it{XMM-Newton}} detectors (i.e. EPIC--pn, EPIC--MOS1 and EPIC--MOS2). Moreover, we included datasets that have an exposure threshold $>50$ ks and a fractional variability threshold $>0.1$. We show the reference band and comparison--band lightcurves of all observations analysed in this work in Figures~\ref{Fig:XMM_lc_ngc5204} and~\ref{Fig:XMM_lc_ngc55}. We also show the lag--energy spectra in Figures~\ref{fig:XMM_lag_energy_ngc5204} and~\ref{fig:XMM_lag_energy_ngc55}. In the case of NGC 5204 X--1 there is some evidence for a two--component lag--frequency spectrum in Obs ID: 0405690101 showing maximal amplitude $\sim 1$ ks below 2 keV at a significance above $3\sigma$ in the frequency channel: 0.5--1 $\times 10^{-4}$ Hz, which then switches sign, corresponding to soft (reverberation) lags in Obs ID: 0921360101 with upto $\sim 2$ ks amplitude in the same frequency channel. On the other hand for NGC 55 ULX--1, we find no significant time lags above a 3$\sigma$ threshold in any of the observations. We discuss the physical implications of these results in section 4.
\newline
\newline
We assessed the statistical significance of the time lags by performing Monte Carlo simulations using the method of \citet{timmer_koenig_1995}. In particular, we fit the averaged PSD for each energy band $\langle P_{X} (f) \rangle$ with a constant plus power--law model, ensuring that the fits were statistically acceptable. For each frequency bin $f_{i}$, we drew two samples of normally distributed random numbers with mean $\mu = 0$ and variance $\sigma^{2} = 0.5\mid \langle P_{X} (f_i) \rangle \mid$, assigned to represent the real and imaginary parts of the FFT corresponding to the averaged PSD. This FFT was then inverse Fourier transformed to generate stochastic light-curves for each energy band, including the reference band (i.e. 0.3--12 keV). Since the light--curves generated in this manner are randomised in both amplitude and phase, one should not expect a fake light--curve in a given energy band to be correlated with respect to a fake reference band light--curve: which would thus correspond to a delay of zero at all frequencies, distributed with some standard deviation $\sigma$. Hence, by computing the time lags of these artificial light-curves a large number of times, one obtained confidence limits on the lag--energy spectrum as shown in Figures~\ref{fig:XMM_lag_energy_ngc5204} and~\ref{fig:XMM_lag_energy_ngc55}.  

\subsection{{\it{Swift}} XRT and UVOT light--curves}

In order to further scrutinize the accretion modes of NGC 5204 X--1 and NGC 55 ULX--1, we use the hardness--intensity diagram of the two sources obtained over several years via {\it{Swift}}--XRT monitoring, also including the {\it{XMM-Newton}} observations. We extracted the light-curves of {\it{Swift}}--XRT observations in two different energy bands (0.3--2.0 keV and 2.0--12.0 keV) by using the online light-curve builder maintained by the \href{https://www.swift.ac.uk/user_objects/}{UKSSDC (UK Swift Science Data Centre)}, adopting the standard grade selection that flags bad pixels: i.e. grades $0$--$12$ in photon counting mode and grades 0--$2$ in windowed timing mode. We did not use inter--observation binning of the light-curves but rather the entire observation. The long--term light-curve alongside the hardness--intensity diagram is shown in Figure~\ref{fig:superorbital_period} with the {\it{XMM-Newton}} hardness--intensity plotted alongside the {\it{Swift}} observations. The {\it{XMM-Newton}} count--rates were converted into {\it{Swift}}--XRT count--rates using the \href{https://heasarc.gsfc.nasa.gov/cgi-bin/Tools/w3pimms/w3pimms.pl}{{\tt{WebPimms}}} tool, after carefully modelling the {\it{XMM-Newton}} spectrum using {\tt{xspec}} (see below).   
\newline
\newline
We also extracted {\it{Swift}}--UVOT lightcurves to search for time delays between X--ray and UV (U, UB1, UB2 and UM2 band) wavelengths. We measured instrumental fluxes of the {\it{Swift}} UVOT data local to the source location by running the task {\tt{uvotdetect}}, where the sky location was obtained through {\it{HST}} and {\it{Chandra}} imaging after applying astrometric corrections. The images were excised to exclude a majority of stars exterior to the target, which enables a more precise estimation of the local background. Once source detections and fluxes were acquired for more than 50 archival {\it{Swift}}--UVOT images, we concluded that the angular resolution and detection sensitivity of UVOT (by simple comparison {\it{HST}} at similar wavelengths) means that the aperture is contaminated by several bright sources and hence the emission from the ULX counterpart is difficult to resolve. We thus decided not to use {\it{UVOT}} observations in order to compute simultaneous X--ray and UV lag--frequency spectra. 

\subsection{Search for quasi--periodic oscillations and pulsations}

\begin{figure*}
\centering
\includegraphics[scale=0.4]{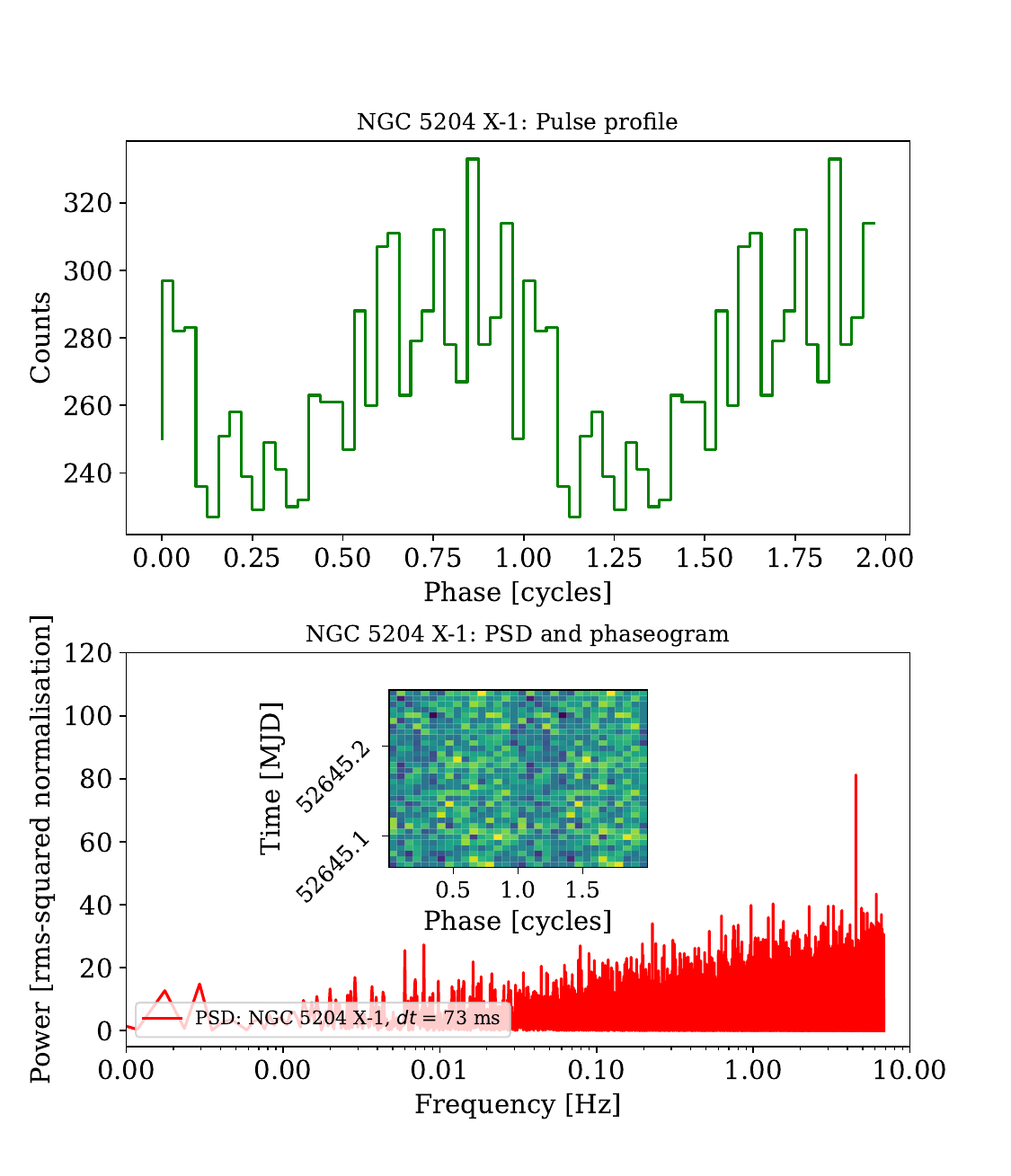}
\includegraphics[scale=0.4]{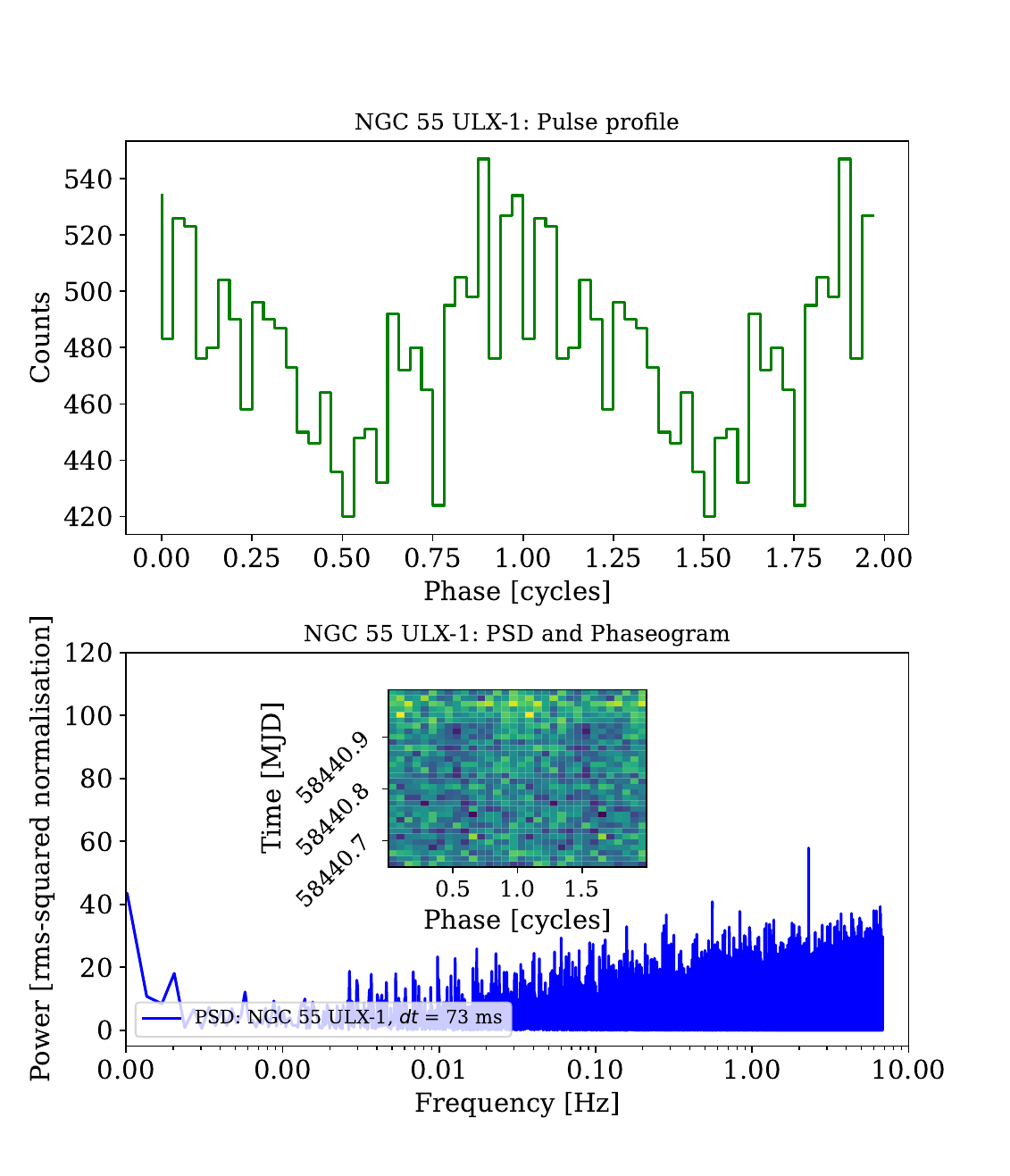}
\includegraphics[scale=0.4]{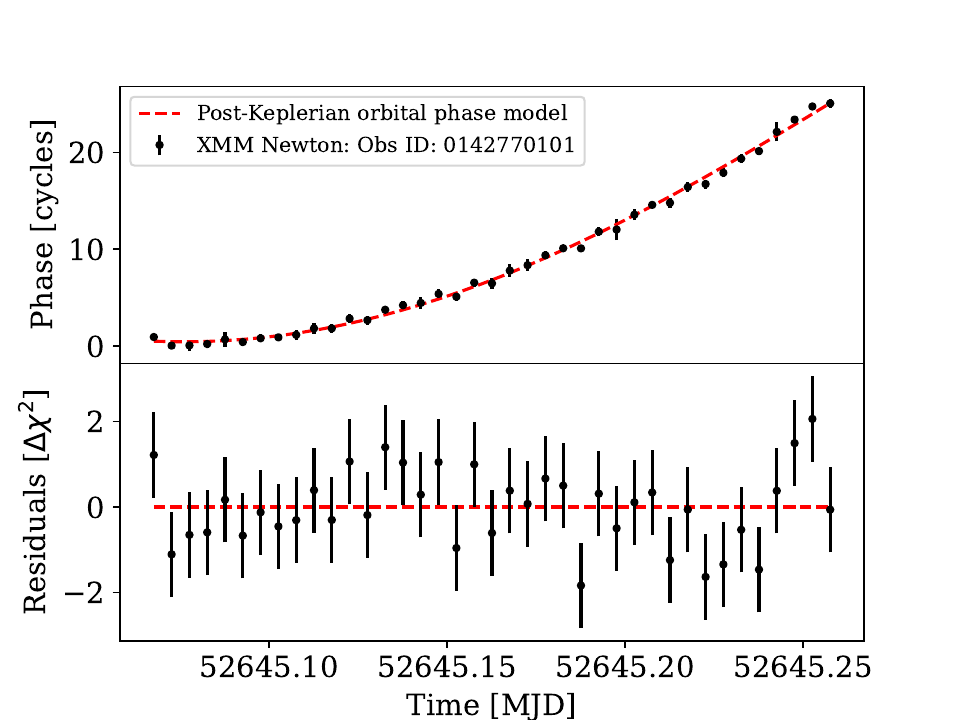}
\includegraphics[scale=0.4]{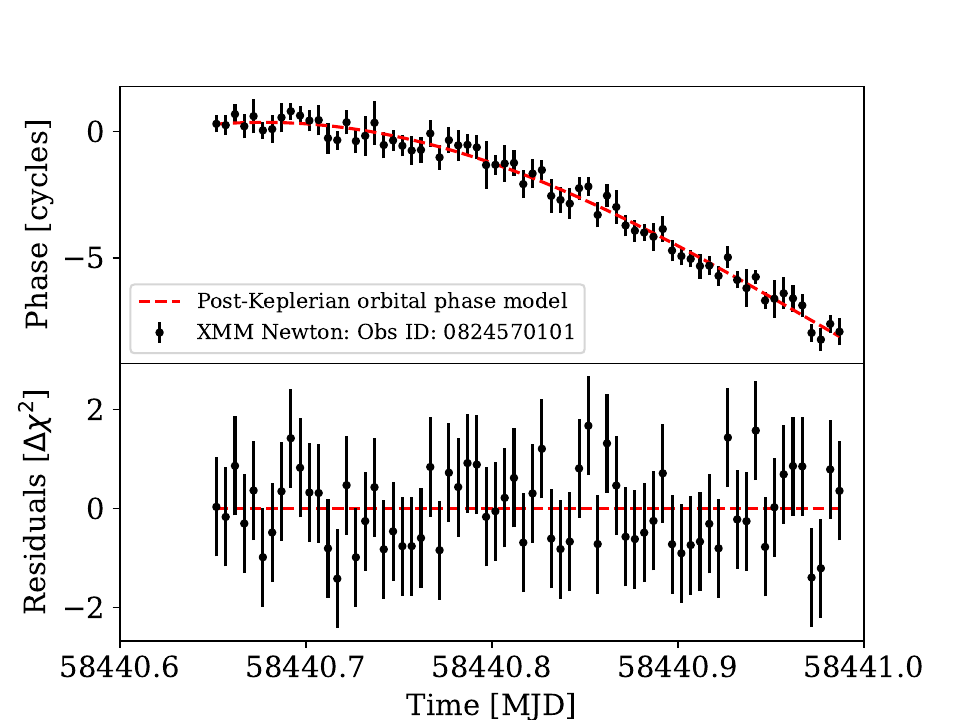}
	\caption{{\it{Left}}: Phaseogram describing the time--phase evolution of the candidate periodicity for the ULX NGC 5204 X--1, with the top panel showing the integrated pulse--profile. {\it{Right}}: Power spectral densities (PSDs) of the light--curves corresponding to two {\it{XMM-Newton}} observations of the ULX NGC 5204 X--1 (i.e. Obs IDs: 0142770101 and 0921360101), after correcting for the phase modulation of the time--series due to a spin--period derivative. {\it{Right}}: Same, but for the {\it{XMM--Newton}} observation of the ULX NGC 55 ULX--1 (Obs ID: 0824570101), also showing pulsations at a moderate significance level.}
\label{fig:pulsations_ngc5204_ngc55}
\end{figure*}

\begin{figure*}
\centering
\includegraphics[scale=0.35]{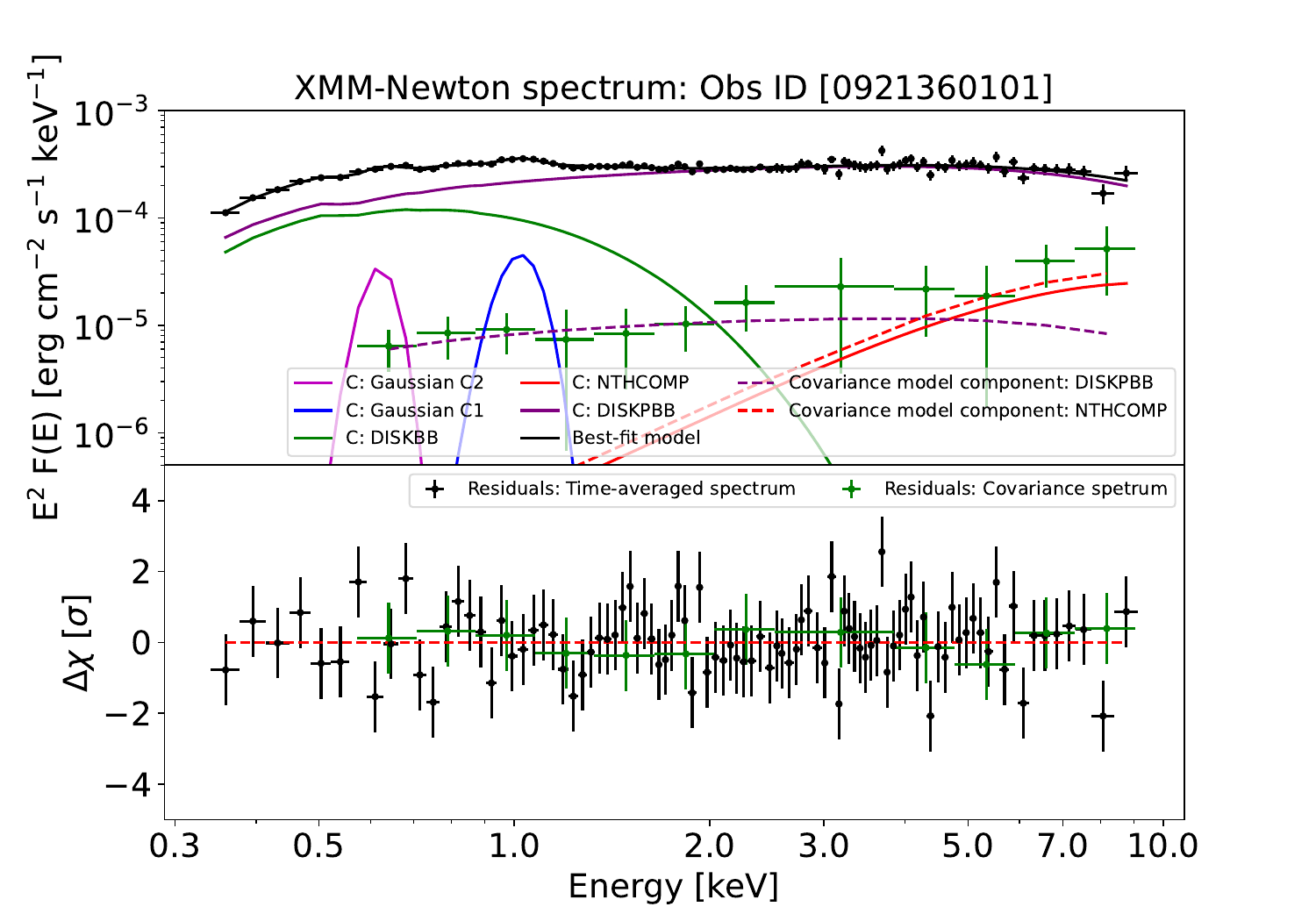}
\includegraphics[scale=0.35]{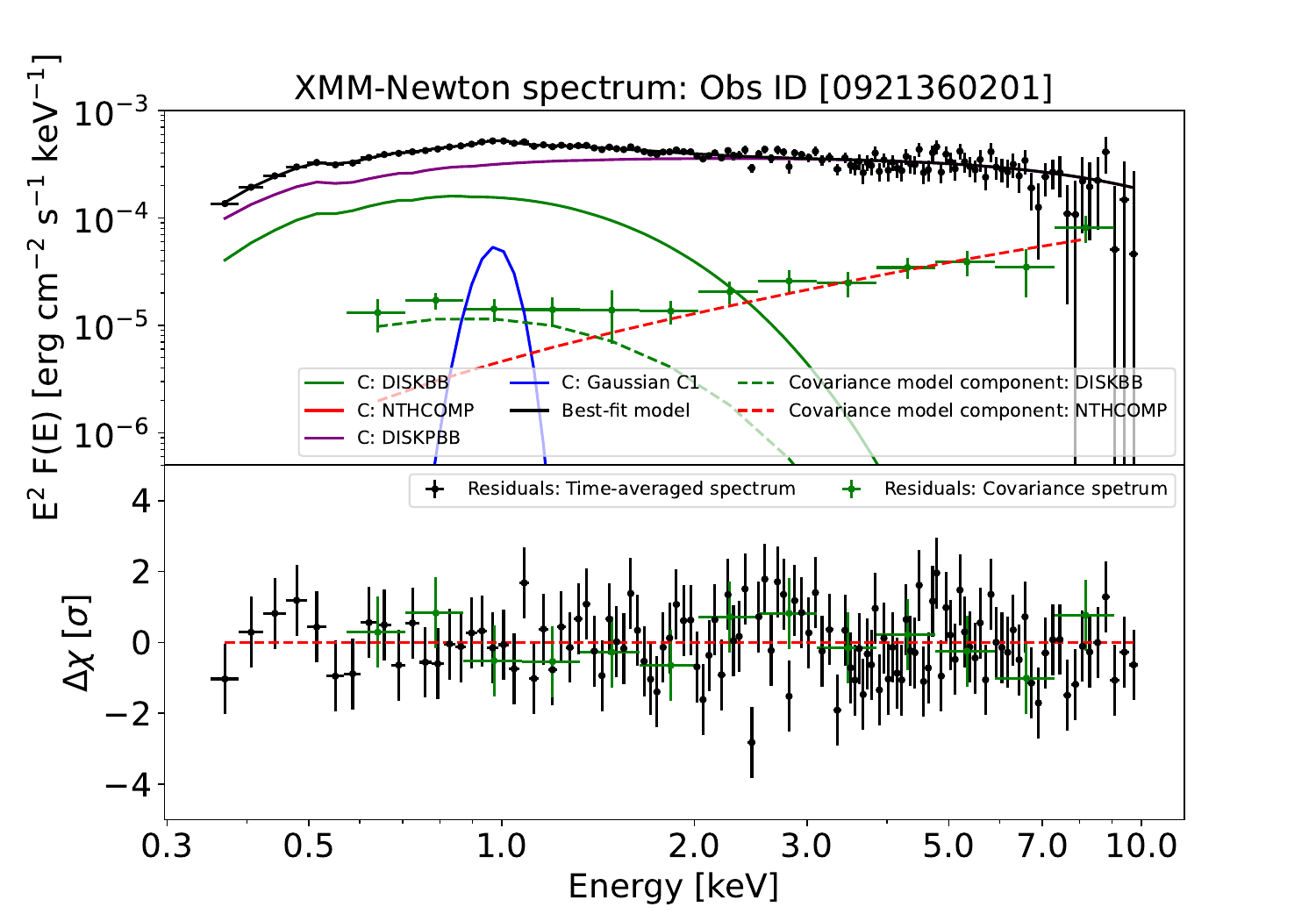}
\includegraphics[scale=0.35]{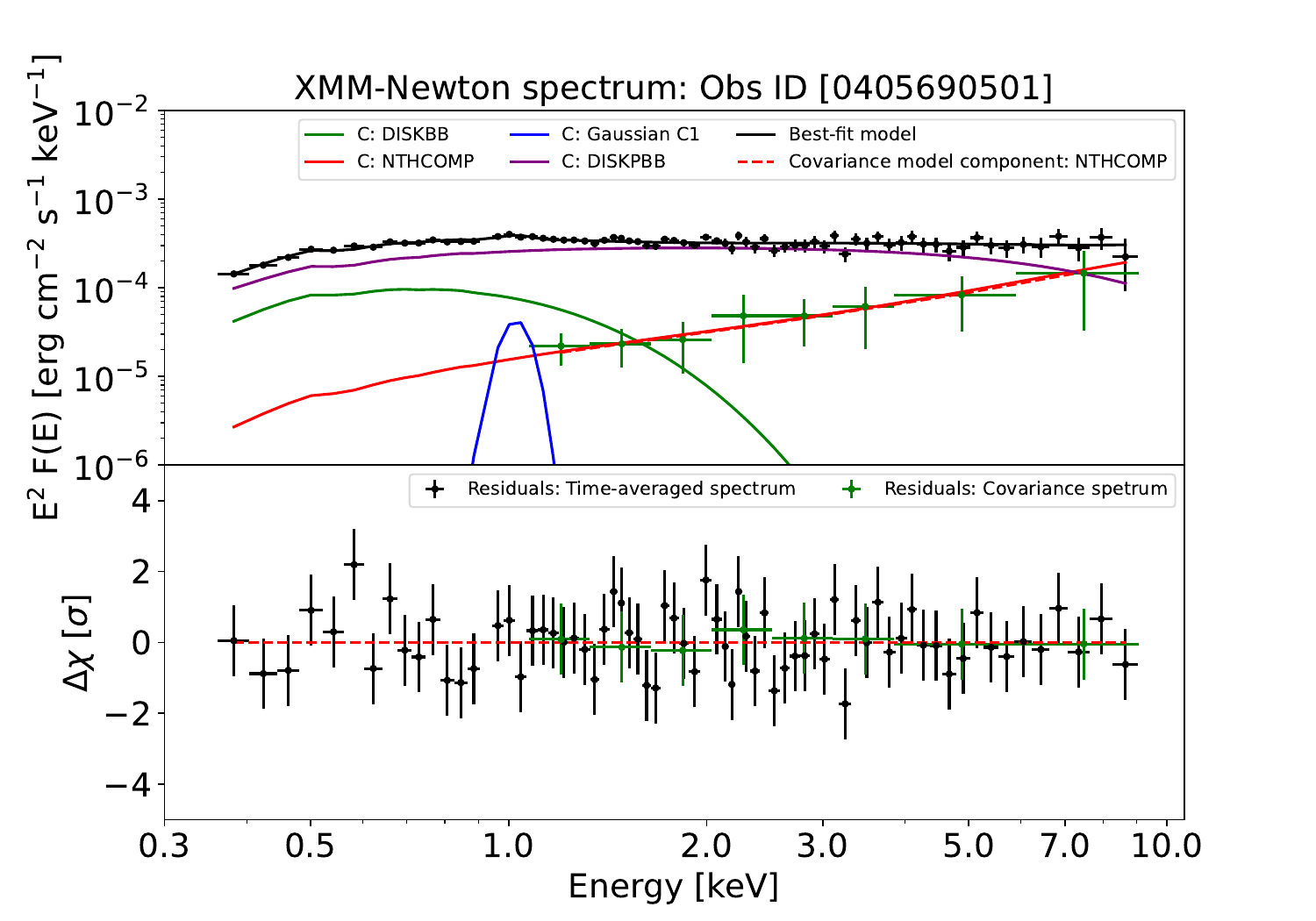}
\includegraphics[scale=0.35]{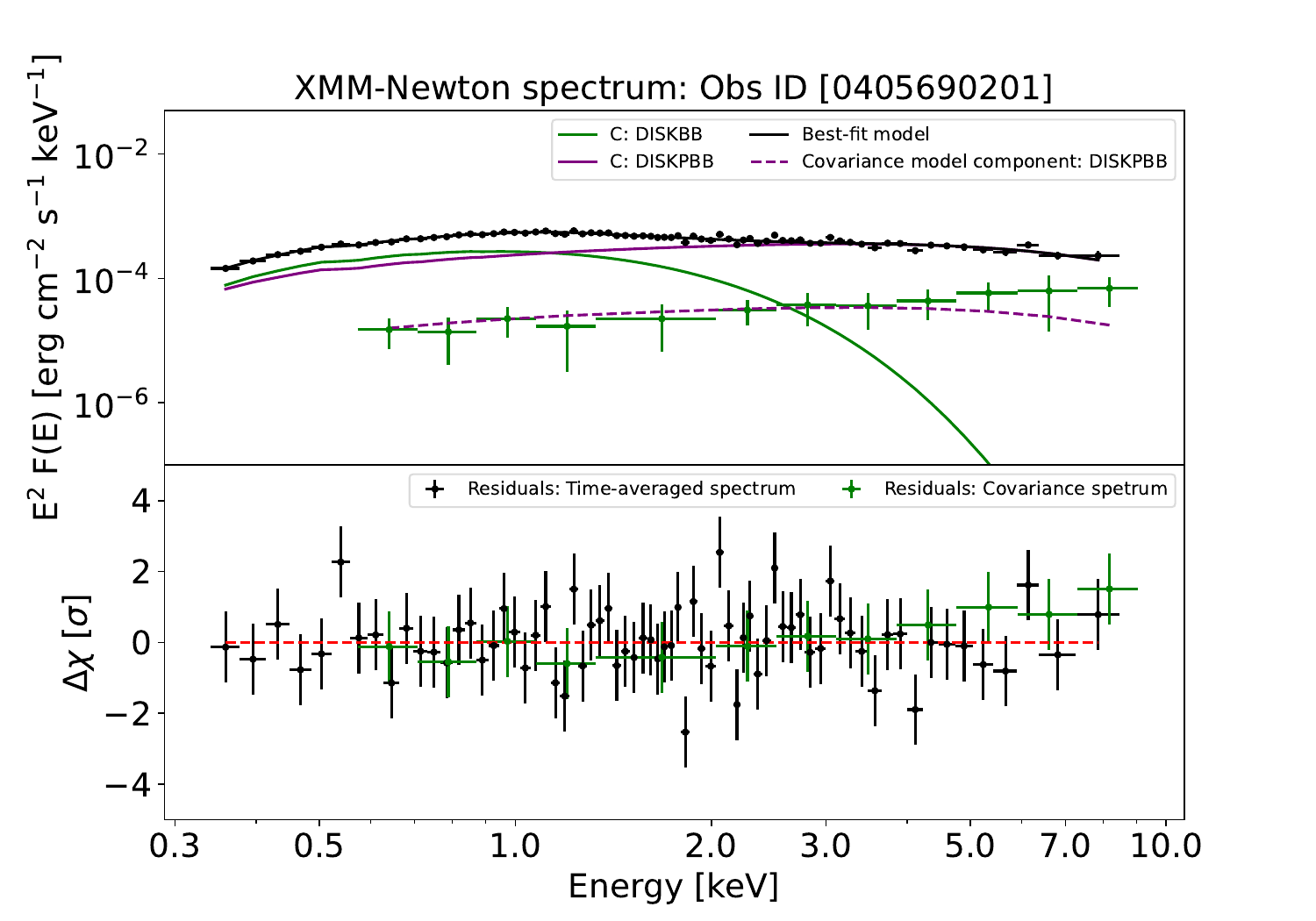}
\caption{{\it{Top panel}}: A selection of de--convolved EPIC--pn time averaged spectra (solid lines) and covariance spectra (dashed lines) of NGC 5204 X--1, fit simultaneously with multi--component thin disc (green), slim ($p$--free) thick disc (purple) and thermal comptonisation (red) models (see legend) using {\tt{xspec}}. We use these spectra in order to study the correlation between the peak disc flux and temperature ($F$--$T_{\text{eff}}$). {\it{Bottom panel}}: Residuals $\Delta \chi$ indicating that the models have an acceptable goodness--of--fit to the data.}
\label{Fig:covariance_ngc5204}
\end{figure*}

\begin{figure*}
\centering
\includegraphics[scale=0.35]{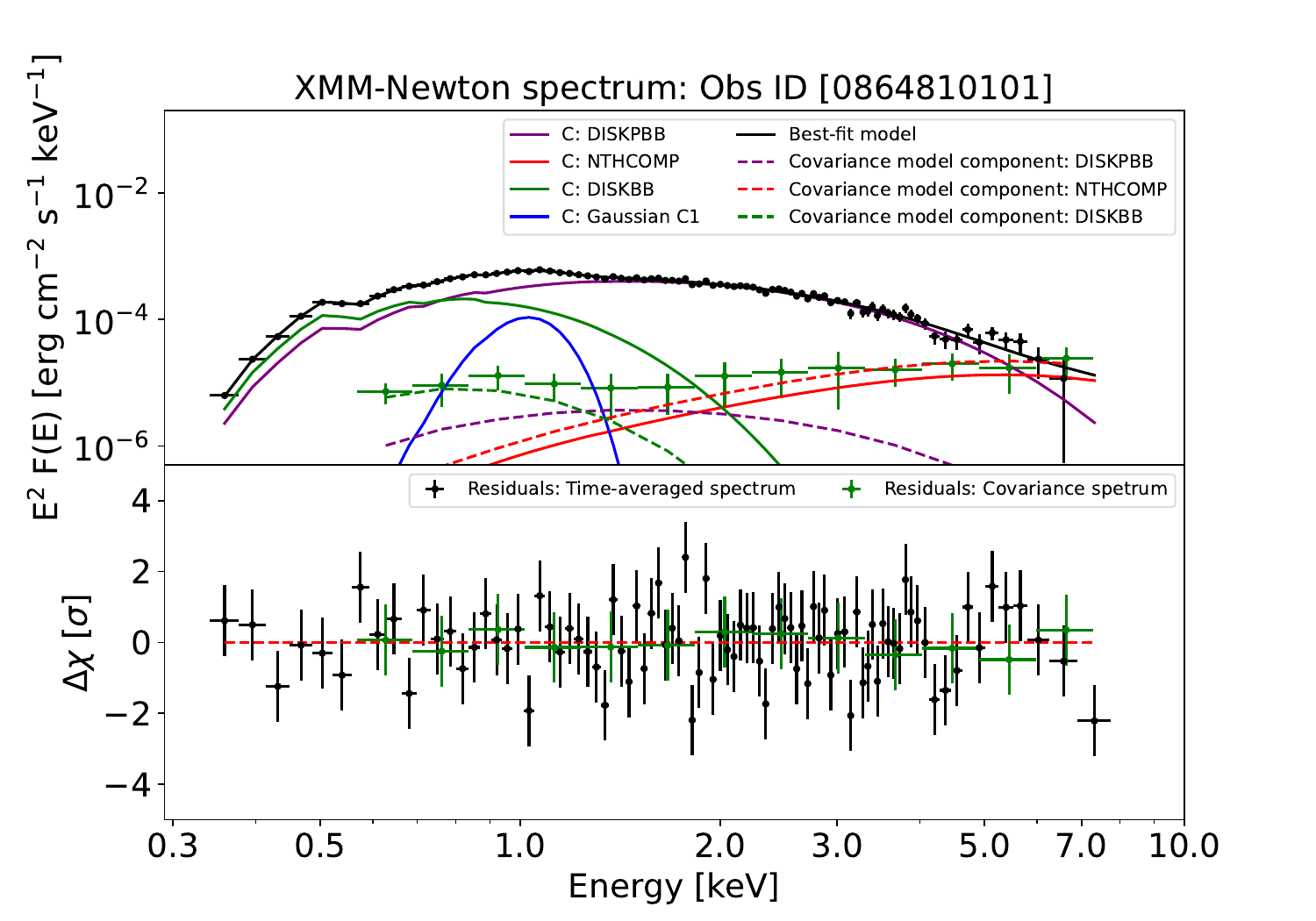}
\includegraphics[scale=0.35]{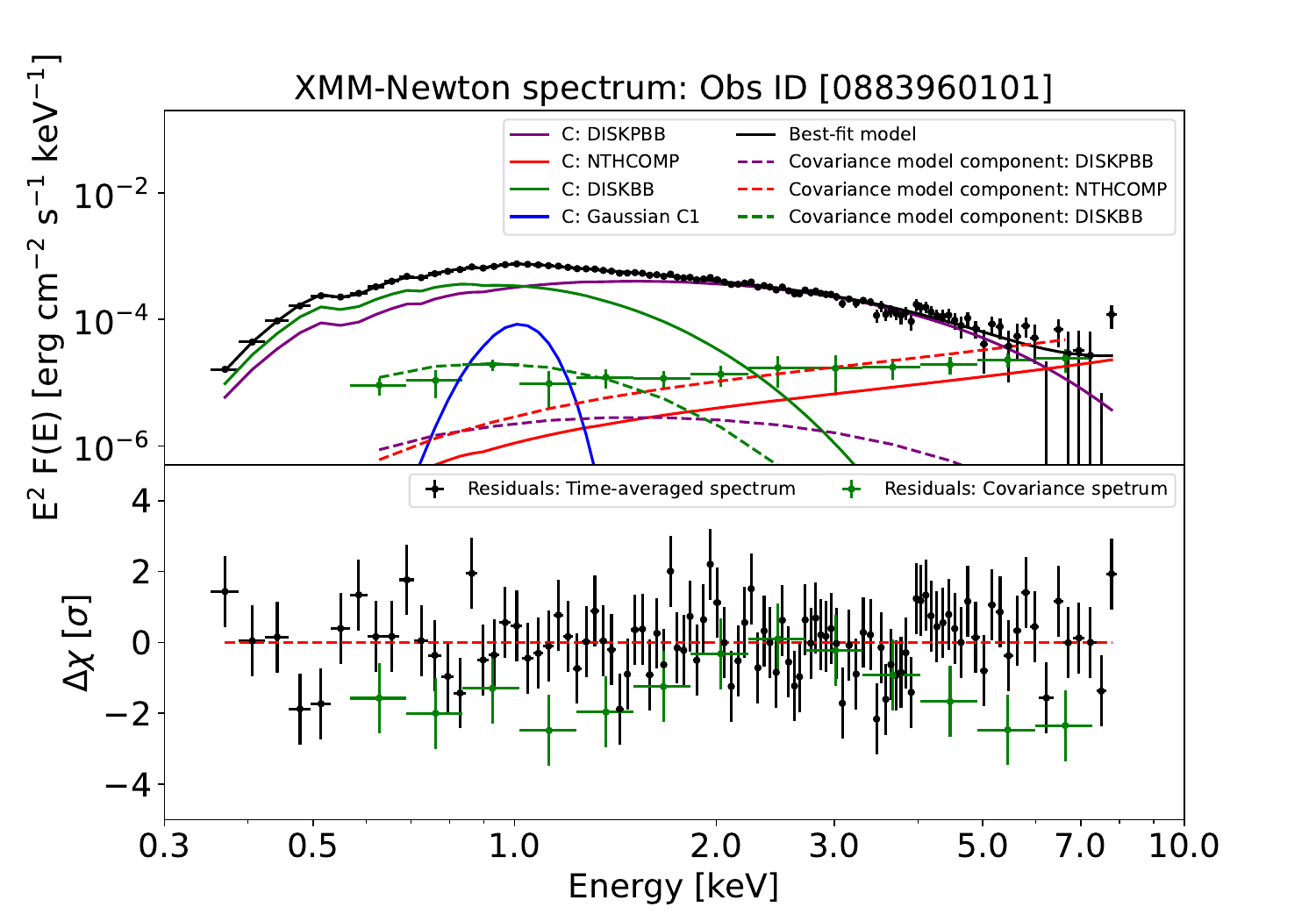}
\includegraphics[scale=0.35]{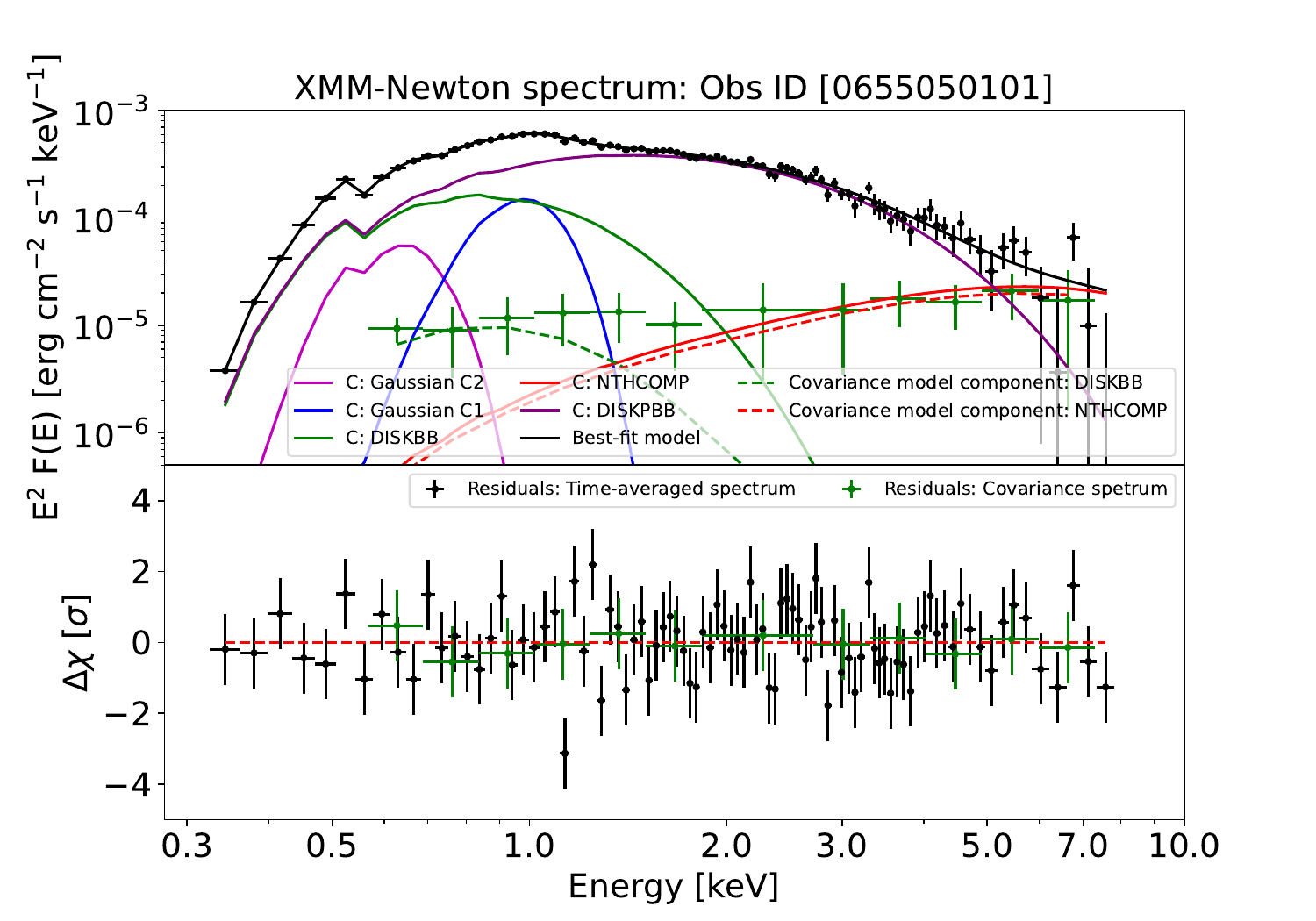}
\caption{{\it{Top panel}}: The same as Figure~\ref{Fig:covariance_ngc5204} but for NGC 55 ULX--1.}
\label{Fig:covariance_ngc55}
\end{figure*}

\begin{figure*}
\centering
\includegraphics[scale=0.4]{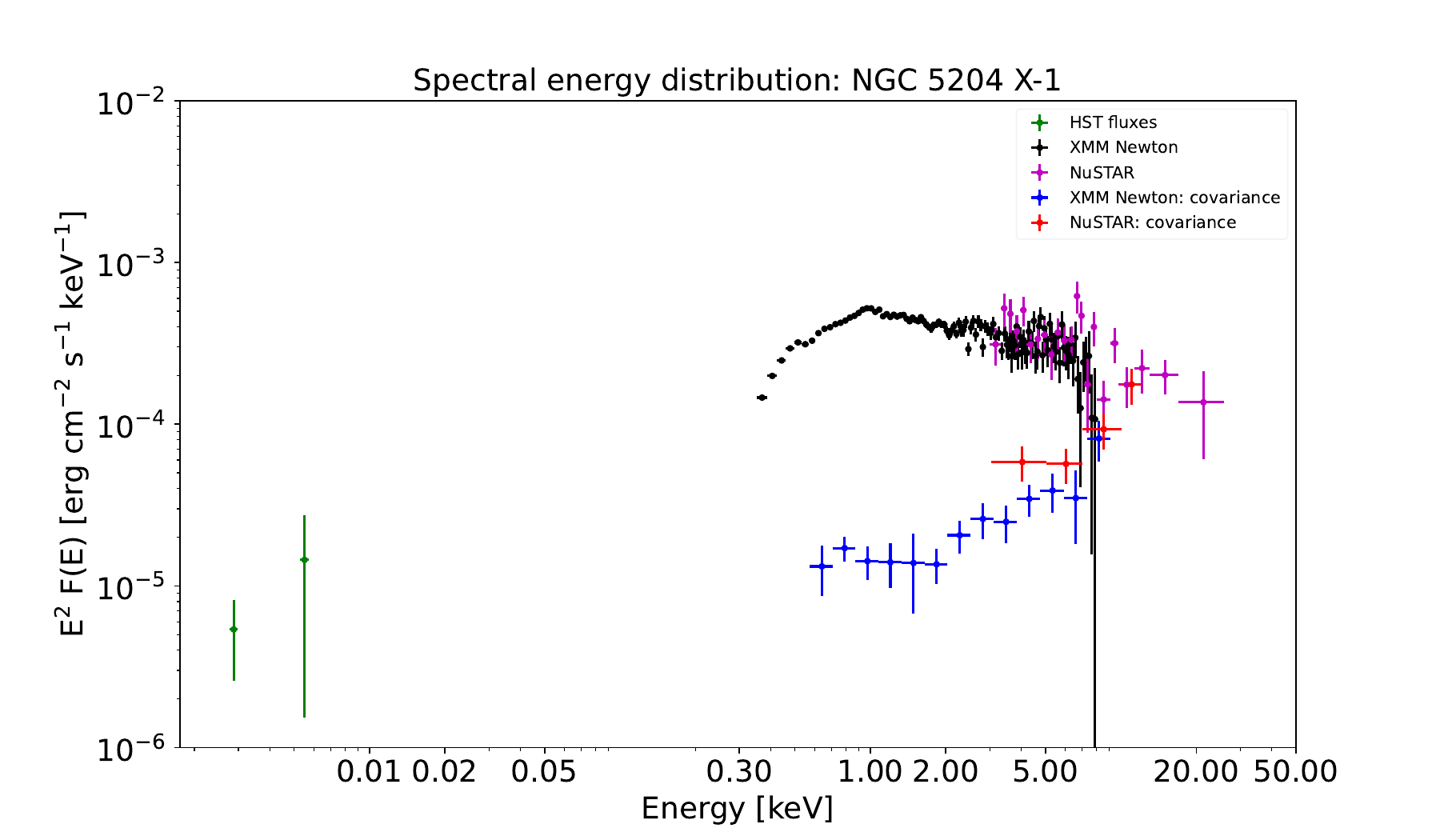}
\caption{Spectral energy distributions (SEDs) of the two ULXs, including {\it{NuSTAR}} and {\it{HST}} observations.}
\label{Fig:SED_ngc5204_ngc55}
\end{figure*}

In order to constrain the nature of the source more reliably, we searched for QPOs and pulsations in all the EPIC--pn and EPIC--MOS observations of the two ULXs. We compute the power spectral density and use the Fourier domain acceleration search technique based on the sliding window segmentation strategy of \citet{sathyaprakash_2019}. As such, we extracted background subtracted light--curves for each observation in the 0.3--10.0 keV and 2.0--10.0 keV energy ranges with the {\tt{evselect}} tool, using the smallest possible time bins (corresponding to an EPIC--pn frame time of 73 ms). Then, we performed a Fourier domain acceleration search on each light--curve using the {\tt{presto}} pulsar timing software, with the dispersion in the Fourier frequency bins $z_{\text{max}}$ set to a range between 0--120, translating to: 

\begin{equation}
\dot{\nu} = \frac{z_{\text{max}}}{T_{\text{obs}}^2} = (0-10^{-7}) \text{Hz s}^{-1}
\end{equation}
\newline
\newline
We found pulsations in a single observation of NGC 5204 X--1 (Obs ID: 0142770101) and NGC 55 ULX--1 (Obs ID: 0824570101) with significances of 6$\sigma$ and 3.5$\sigma$ respectively. The detailed properties of the detection are summarised in Table~\ref{Tab:pulsations_table}. We show the expected $\delta$--distribution function in the power--spectral--densities (PSDs) after correcting for the period derivative, and plot the evolution of the pulse phase and the integrated pulse profile in Figure~\ref{fig:pulsations_ngc5204_ngc55}. The acceleration search is performed in the Fourier domain after deconvolving the effect of red--noise by following the same strategies as mentioned in section 3.3: That is, we compute the FFT of the original waveform $S(f)$ in the following manner:
\begin{multline}
	R(f) = W(f) \circledast S(f) \\
	S(f) = R_{f}/W_{f} \\
	s(t) = \mathcal{F}^{-1} \bigg [ \frac{R_{f} (f)}{S_{f} (f)} \bigg ] \\
\end{multline}
Here, $S(f)$ represents the FFT of the true waveform, $R(f)$ represents the FFT of the observed waveform, while $R_{f} (f)$ is the FFT of $R(f)$ and similarly $W_{f}$ is the FFT of $W(f)$. We report the inferred spin periods $P$ and period derivatives $\dot{P}$ in the observations in which pulsations were detected. We then proceed to extract the time--of--arrival (TOAs) for each detection by cross--correlating the observed pulse profile with the observed time series, and fit the phase delays of the pulse profile with orbital evolution models to extract the binary parameters.
\newline
\newline
Type--C QPOs, which are thought to scale with BH mass have been found in some ULXs with typically soft X--ray spectra such as NGC 5408 X--1 (\citealt{deMarco2013}). Such QPOs typically correlate with soft lags, slope of the PSD and their variability, and as such we searched for the presence of low frequency quasi--periodic oscillations (QPOs) in the PSDs of the two ULXs. The averaged PSDs were geometrically binned in frequency space to improve the signal--to--noise, and we fit the PSD continuum with a combination of bessel and power--law functions, but we find that quasi--periodic oscillations remain undetected in the current datasets.

\subsection{Covariance spectra}

\begin{table*}
\caption{The parameters inferred from simultaneously modelling the time averaged and covariance spectra of NGC 5204 X--1 and NGC 55 ULX--$1$ in the manner described in section 3.3. $^{\text{a}}$ Hydrogen column density, $^{\text{b}}$ Temperature of the cool disc (in keV), $^{\text{c}}$ Normalisation of the cool disc (in keV), $^{\text{d}}$ Photon index $^{\text{e}}$ Temperature of the warm disc, $^{\text{f}}$ Hot electron temperature, $^{\text{g}}$ Exponent of the radial dependence of disc temperature, $^{\text{h}}$ Normalisation of the thermal Comptonisation model, $^{\text{i}}$ Normalisation of the $p$--free disc model.}
\begin{tabular}{lccccccccccc}
\hline
\hline
\centering
Source & ObsID & $N_{\text{H}}$ & $kT_{\text{diskbb}}$ & $kT_{\text{diskpbb}}$ & $p$ index & $\log_{10}$ $F_{\text{bb}}$ & $\log_{10}$ $F_{\text{pbb}}$ & $\Gamma$ & $kT_{\text{e}}$ \\
       &       & [10$^{22}$ cm$^{-2}$] & [keV] & [keV] & & [erg cm$^{-2}$ s$^{-1}$] & [erg cm$^{-2}$ s$^{-1}$] & & [keV] \\
\hline
NGC 5204 X--1 & 0405690101 & $0.06 \pm 0.03$ & $0.33 \pm 0.02$ & $2.2 \pm 1.2$ & $0.53 \pm 0.09$ & $-12.52 \pm 0.03$ & $-11.94 \pm 0.02$ & $1.0 \pm 0.2$ & $<20$ \\
NGC 5204 X--1 & 0405690201 & $0.06 \pm 0.02$ & $0.35 \pm 0.02$ & $1.9 \pm 0.3$ & $0.6 \pm 0.1$ & $-12.06 \pm 0.01$ & $-11.88 \pm 0.01$ & -- & -- \\
NGC 5204 X--1 & 0405690501 & $0.08 \pm 0.04$ & $0.25 \pm 0.04$ & $1.7 \pm 1.8$ & $0.5 \pm 0.1$ & $-12.02 \pm 0.02$ & $-11.87 \pm 0.01$ & $1.2 \pm 0.1$ & $2.6 \pm 0.9$ \\
NGC 5204 X--1 & 0693851401 & $0.05 \pm 0.03$ & $0.25 \pm 0.04$ & $2.3 \pm 0.2$ & $0.62 \pm 0.03$ & $-11.91 \pm 0.01$ & $-11.91 \pm 0.02$ & -- & -- \\
NGC 5204 X--1 & 0693850701 & $0.08 \pm 0.02$ & $0.22 \pm 0.02$ & $3.1 \pm 2.2$ & $0.58 \pm 0.05$ & $-12.00 \pm 0.01$ & $-11.89 \pm 0.02$ & -- & -- \\
NGC 5204 X--1 & 0741960101 & $0.02 \pm 0.03$ & $0.3 \pm 0.05$ & $2.3 \pm 1.3$ & $0.7 \pm 0.2$ & $-12.51 \pm 0.02$ & $-11.99 \pm 0.02$ & -- & -- \\
NGC 5204 X--1 & 0921360101 & $0.06 \pm 0.01$ & $0.25 \pm 0.02$ & $2.6 \pm 0.3$ & $0.55 \pm 0.02$ & $-12.33 \pm 0.02$ & $-11.87 \pm 0.01$ & -- & -- \\
NGC 5204 X--1 & 0921360201 & $0.09 \pm 0.02$ & $0.29 \pm 0.02$ & $2.1 \pm 0.5$ & $0.50 \pm 0.06$ & $-12.39 \pm 0.02$ & $-11.72 \pm 0.004$ & $1.0 \pm 0.8$ & $<40$ \\
NGC 5204 X--1 & 0142770101 & $0.06 \pm 0.02$ & $0.22 \pm 0.03$ & $3.4 \pm 1.3$ & $0.54 \pm 0.02$ & $-12.36 ^{+0.01}_{-0.02}$ & $-11.83 \pm 0.01$ & -- & -- \\
NGC 5204 X--1 & 0142770301 & $0.06 \pm 0.02$ & $0.29 \pm 0.03$ & $1.7 \pm 0.5$ & $0.8 \pm 0.5$ & $-12.26 \pm 0.02$ & $-12.02 \pm 0.02$ & -- & -- \\
NGC 5204 X--1 & 0150650301 & $0.06 \pm 0.02$ & $0.30 \pm 0.03$ & $2 \pm 25$ & $0.6 \pm 0.1$ & $-12.36 \pm 0.01$ & $-11.85 \pm 0.02$ & -- & -- \\
\hline
\hline
NGC 55 ULX--1 & 0028740101 & $0.29 \pm 0.04$ & $0.29 \pm 0.05$ & $1.1 \pm 0.1$ & $0.5 \pm 0.1$ & $-11.89 \pm 0.03$ & $-11.30 \pm 0.01$ & -- & -- \\
NGC 55 ULX--1 & 0028740201 & $0.30 \pm 0.08$ & $0.24 \pm 0.04$ & $0.9 \pm 0.3$ & $0.5 \pm 0.2$ & $-11.75 \pm 0.01$ & $-11.39 \pm 0.01$ & $1.5 \pm 1.6$ & $2 \pm 3$ \\
NGC 55 ULX--1 & 0655050101 & $0.34 \pm 0.04$ & $0.18 \pm 0.03$ & $0.67 \pm 0.06$ & $0.5 \pm 0.1$ & $-11.73 \pm 0.01$ & $-11.76 \pm 0.01$ & $< 1.5$ & $<70$ \\
NGC 55 ULX--1 & 0824570101 & $0.35 \pm 0.03$ & $0.18 \pm 0.02$ & $0.77 \pm 0.07$ & $0.5 \pm 0.2$ & $-11.25 \pm 0.01$ & $-11.89 \pm 0.01$ & -- & -- \\
NGC 55 ULX--1 & 0864810101 & $0.35 \pm 0.04$ & $0.17 \pm 0.02$ & $0.67 \pm 0.04$ & $0.5 \pm 0.1$ & $-12.21 \pm 0.01$ & $-11.69 \pm 0.004$ & $1.0 \pm 3.5$ & $<50$ \\
NGC 55 ULX--1 & 0883960101 & $0.31 \pm 0.04$ & $0.22 \pm 0.02$ & $0.7 \pm 0.1$ & $0.5 \pm 0.3$ & $-11.84 \pm 0.01$ & $-11.60 \pm 0.005$ & $<1.5$ & $<70$ \\
NGC 55 ULX--1 & 0852610101 & $0.30 \pm 0.03$ & $0.26 \pm 0.03$ & $0.7 \pm 0.1$ & $<1$ & $-11.54 \pm 0.03$ & $-11.75 \pm 0.03$ & -- & -- \\
NGC 55 ULX--1 & 0852610201 & $0.30 \pm 0.03$ & $0.26 \pm 0.02$ & $1.6 \pm 1.1$ & $0.5 \pm 0.2$ & $-11.49 \pm 0.03$ & $-11.60 \pm 0.03$ & -- & -- \\
NGC 55 ULX--1 & 0852610301 & $0.30 \pm 0.03$ & $0.27 \pm 0.03$ & $3.6 \pm 6.7$ & $<1$ & $-12.27 \pm 0.05$ & $-11.93 \pm 0.1$ & -- & -- \\
\hline
\hline
\label{Tab:xcovspec_fits_emcee}
\end{tabular}
\end{table*}

\begin{figure*}
\centering
\includegraphics[scale=0.5]{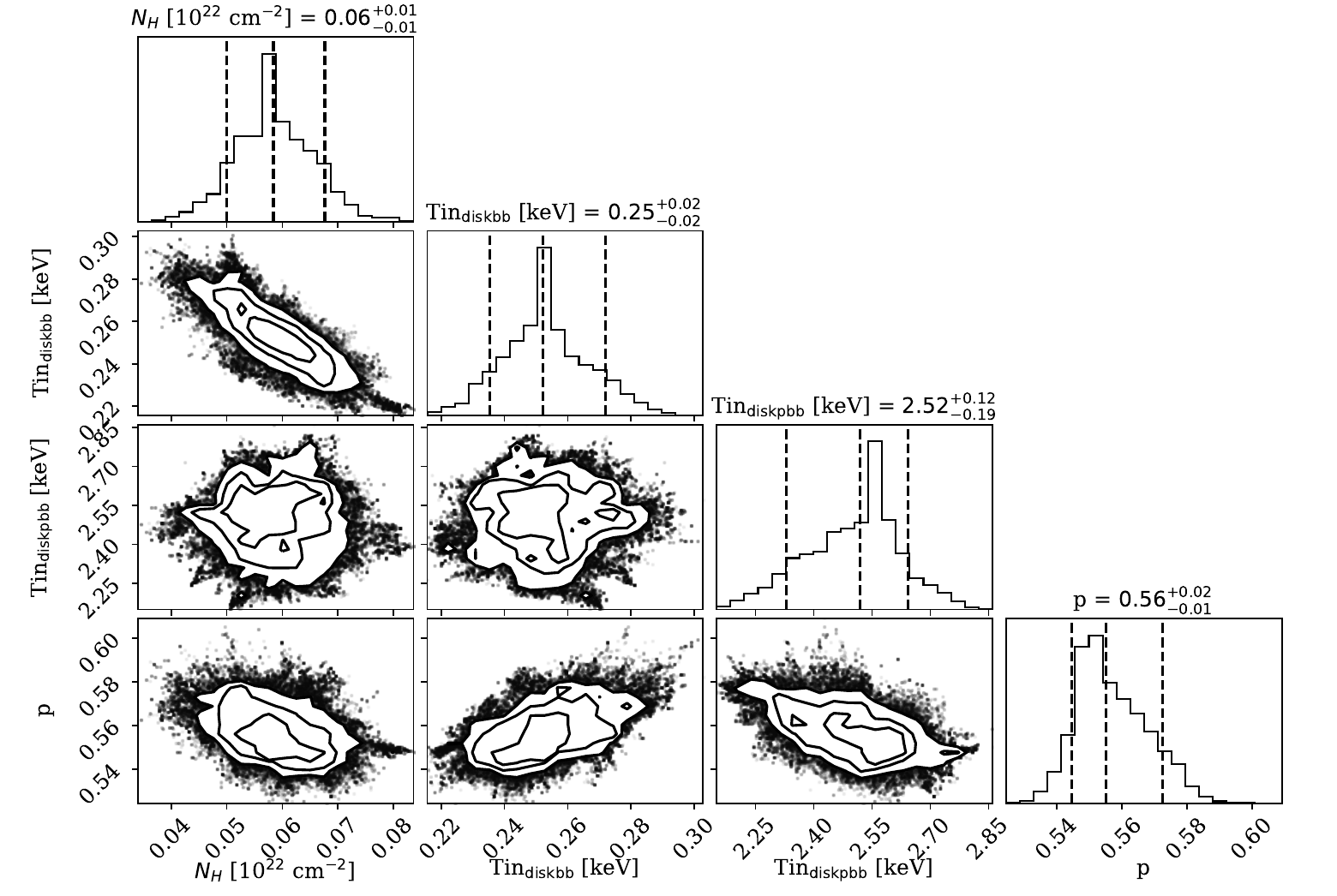}
\caption{Posterior distributions for the parameters used to fit the time-averaged and covariance spectra of NGC 5204 X--1 derived from using the {\tt{emcee}} algorithm and the following model in {\tt{xspec}}: {\tt{tbabs}} $\times$ ({\tt{diskbb}} $+$ {\tt{diskpbb}}). Here, the inner disc temperature for the soft thermal disc component (i.e. {\tt{diskbb}}) is $kT_{\text{diskbb}}$), the warm disc component is the p--free disc model {\tt{diskpbb}}, with temperature $kT_{\text{diskpbb}}$) and the hydrogen column density is $N_{\text{H}}$.}
\label{Fig:emcee_ngc55ulx1_0824570101}
\end{figure*}

We note that the time averaged spectra of both the recent observations of NGC 5204 X--1 can be best described to be in the hard ultraluminous (HUL) state based on the classification scheme of \citet{sutton_2014}, even though the two differ in X--ray flux and fractional variability $F_{\text{var}}$. In the case of NGC 55 ULX--1, the X-ray spectra appear consistent with a broadened disc (BD) in a vast majority of cases. The covariance spectrum can be used to derive more accurate parameters of the time--averaged spectral model, which is used to characterise the accretion modes of the two ULXs. The covariance spectrum is a measure of the correlated variability between the reference band and comparison band light--curves, and can be used to dis--entangle contributions from the variable and non--variable spectral components (see e.g. \citealt{middleton_2015a}; \citealt{pinto_soria_2017}; \citealt{kara_2020}). In order to compute covariance spectra for NGC 5204 X--1 and NGC 55 ULX--1, we combined light--curves from {\it{XMM-Newton}} observations in the same manner as described in section 3.1. From the amplitude of the noise subtracted and averaged CPSD $\mid \langle C_{xy}(f) \rangle$ and the comparison-band PSD $P_{Y}$, we define the covariance in a particular frequency band in accordance with the following equation:
\begin{equation}
\text{Cov} (f) = \langle y \rangle \sqrt{\frac{(\mid \langle C_{xy}(f) \rangle \mid^{2} - n_{\text{bias}}^{2}) \delta f_{b}}{\langle P_{X} (f) \rangle - P_{X,\text{noise}}}}
\end{equation}
where $\langle y \rangle$ is the mean count-rate of the comparison--band light--curve, $P_{X,\text{noise}}$ is the Poisson noise level of the averaged reference--band PSD (which is subtracted from the $P_{\text{X}}$). The uncertainty in the covariance is defined in accordance with equation 15 of \citet{uttley_2014}. Finally, due to bad time intervals (BTIs), there were gaps in the lightcurves of the {\it{XMM-Newton}} observations on timescales of a few seconds to kilo--seconds. Computing the FFTs of these lightcurves directly would lead to false artefacts in PSDs and CPSDs due to beat frequencies. We accounted for gaps in the light--curves by constructing bootstrapped PSDs using the approach described in Timmer \& Koenig (1995). The interpolated segments of the lightcurves are shown in Figures~\ref{XMM_lc_ngc5204} and~\ref{Fig:XMM_lc_ngc55}. 
\newline
\newline
To constrain degeneracies between poorly understood parameters in the spectral components, we jointly modelled the time averaged spectra and covariance spectra adopting the same geometrical configuration for super--Eddington discs as in several previous studies. This typically includes (i) an absorbed disc blackbody and a thermal comptonisation model, in addition to a simple gaussian for NGC 5204 X--1 yielding: {\tt{tbabs}} $\times$ {\tt{tbabs}} $(${\tt{diskbb}} $+$ {\tt{nthcomp}} $+$ {\tt{gauss}}$)$. We include a cool disc {\tt{diskbb}} and warm disc {\tt{diskpbb}} model and the same gaussian component for NGC 55 ULX--1: {\tt{tbabs}} $\times$ {\tt{tbabs}} $(${\tt{diskbb}} $+$ {\tt{diskpbb}} $+$ {\tt{gauss}}$)$. Here, {\tt{tbabs}} is a multiplicative model component in {\tt{xspec}} used to account for the ionisation of X--ray photons from an intervening column of hydrogen along the line-of-sight, where the hydrogen column density ($N_{H}$) is usually given in units of $10^{22}$ cm$^{-2}$. The reason for the choice of the warm disc model in the case of NGC 55 ULX--1 is because it has a much softer spectrum at higher energies relative to NGC 5204 X--1 and is hence thought to be viewed at a larger inclination, where the wind column density is larger, while the outflow velocities are slower and with a smaller ionisation parameter:
\begin{equation}
\chi = \frac{L_{\text{X}}}{n_{e} R^2}.
\end{equation}
We note that {\tt{diskbb}} is an analytical prescription used to fit the emission from a stable thin disc (usually upto 2 keV; \citealt{shakura_sunyaev_1973}) which in the case of ULXs corresponds to the outer--most regions of the inflow where the accretion rate is presumed to be $\lesssim$ Eddington. In such a case the emergent photon spectrum from a local disc region should have a standard blackbody spectrum: this spectrum is eventually convolved with similar blackbodies over several different radii. Moreover, {\tt{diskpbb}} is used to fit the hard (0.2--10 keV) portion of the X-ray continuum to model emission from the thick, advective disc where the radiative efficiency of the inflow is highly uncertain due to the possibility that some portion of the radiative energy is transferred to launch winds. Moreover, in the case of NSs the accretion flow can thermalise over scales $\sim$ magnetospheric radius $R_{\text{M}}$, bounding over an accretion column in the inner--most regions, thus Compton--scattering the observed spectrum. The covariance spectrum can help to separate time variable components (over timescales $<$ duration of the observation) from the non--variable components, which then provides an accurate measurement of the disc temperature and luminosity ($L_{\text{X}}$--$T_{\text{eff}}$ correlation), and the Eddington--ratio $\lambda = \frac{L}{L_{\text{Edd}}}$. We report the best--fit parameters of our spectral energy distribution (SED) modelling in Table~\ref{Tab:xcovspec_fits_emcee}, via Affine Invariant Markov Chain Monte Carlo Ensemble Sampler ($\tt{emcee}$), ensuring that the chains converge to reliable estimates of the interesting spectral parameters.  
\newline
\newline
Considering the above approach, we find the covariance spectra of NGC 5204 X-$1$ to be well-fit by a soft disc and an absorbed thermal Comptonisation model irrespective of whether the time averaged spectrum is classified to be in the SUL or HUL regime. For NGC 55 ULX--1, the covariance spectrum is much softer and peaks in the 0.3--2.0 keV band. A recurring theme of previous studies has been to test the nature of the soft thermal disc component, and whether the luminosity--temperature $L$--$T_{\text{eff}}$ relation corresponds to the well understood Shakura--Sunyaev disc (\citealt{shakura_sunyaev_1973}) in which the inner disc temperature is expected to correlate with the integrated emissivity as $L_{\text{diskbb}} \propto T_{\text{eff}}^{4}$. We explicitly show whether this relation holds for the particular ULXs considered in this study in Figure~\ref{fig:LT_diagrams}. It was shown by \citep{robba2021} for NGC 1313 X--2 that the two quantities are anti--correlated, and hence the disc structure we are observing in ULXs need not be a classic thermal disc. We test whether such a scenario fits for the ULXs chosen in this study using covariance spectra to break the degeneracies between the different model components, as stated above. 
\newline
\newline
\section{Discussion}

\begin{figure*}
\centering
\includegraphics[scale=0.325]{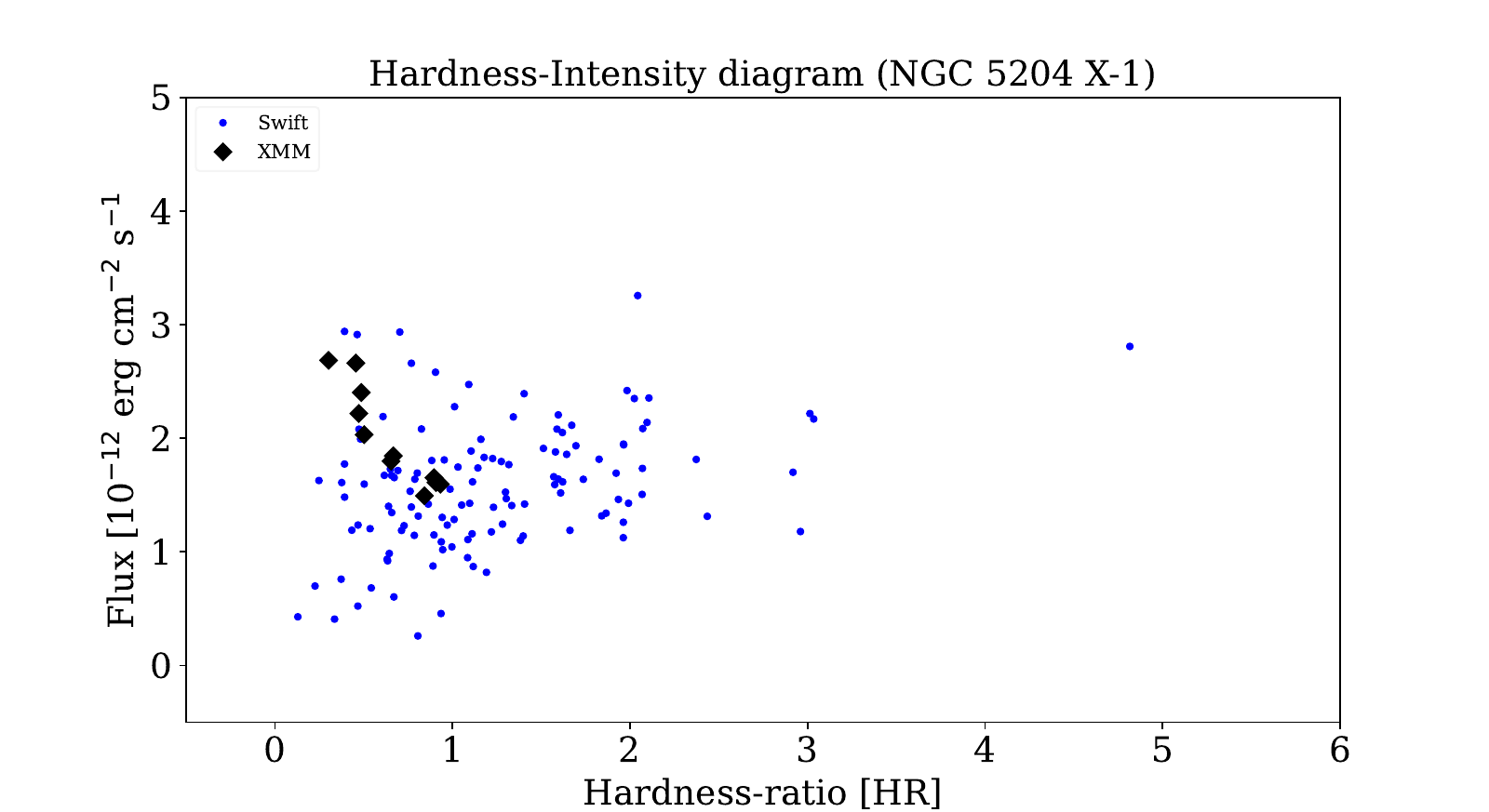}
\includegraphics[scale=0.325]{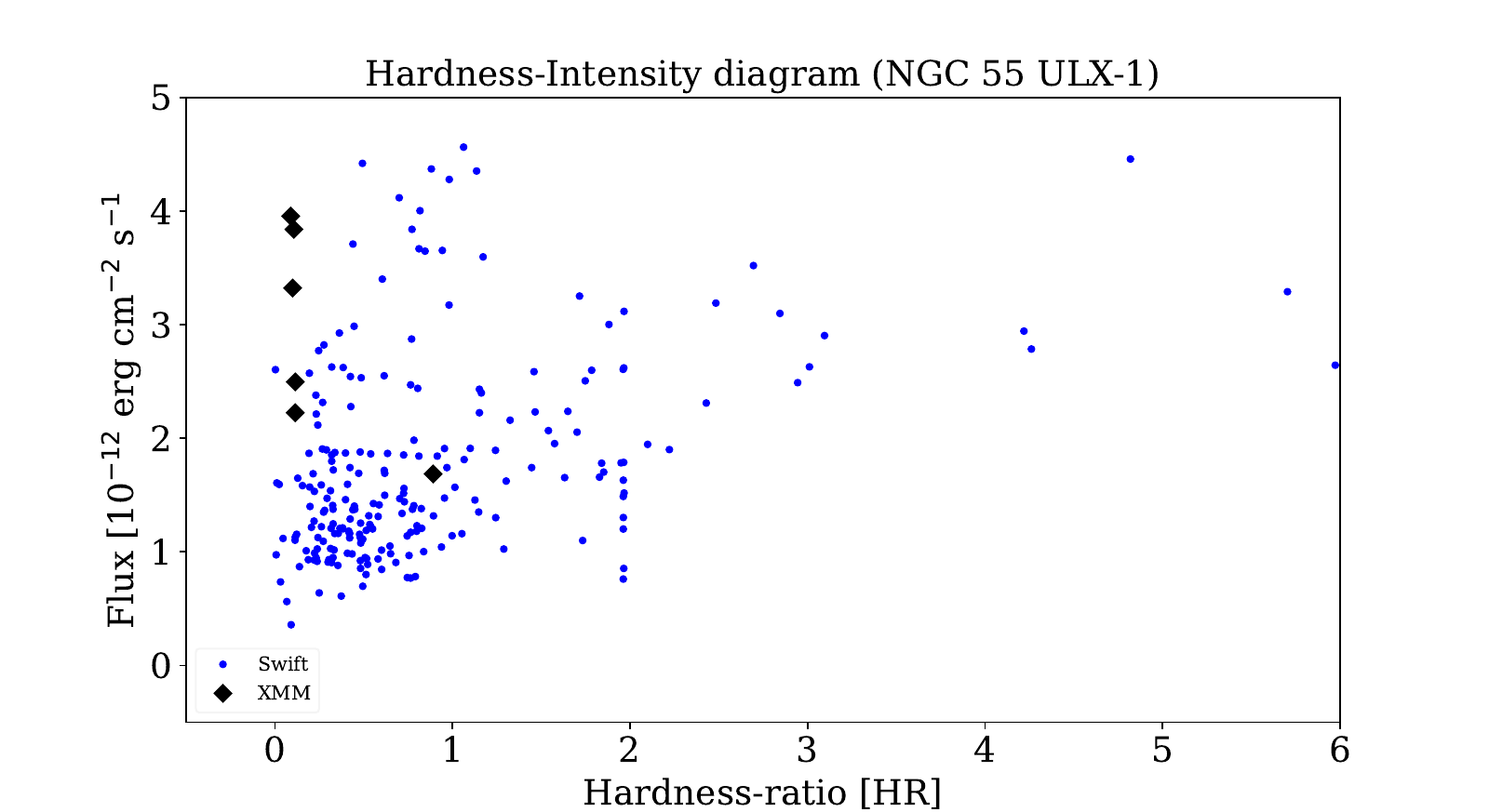}
\caption{{\it{Left}}: Hardness--intensity diagram of NGC 5204 X--1 showing the evolution of the accretion state using {\it{Swift}} and {\it{XMM--Newton}} observations. We use a hierarchical clustering algorithm to determine the optimum number of clusters, so as to group the {\it{XMM--Newton}} data into discrete states. {\it{Right}}: Same, but for NGC 55 ULX--1.}
\label{fig:hardness_intensity_diagram}
\end{figure*}

\begin{figure*}
\centering
\includegraphics[scale=0.275]{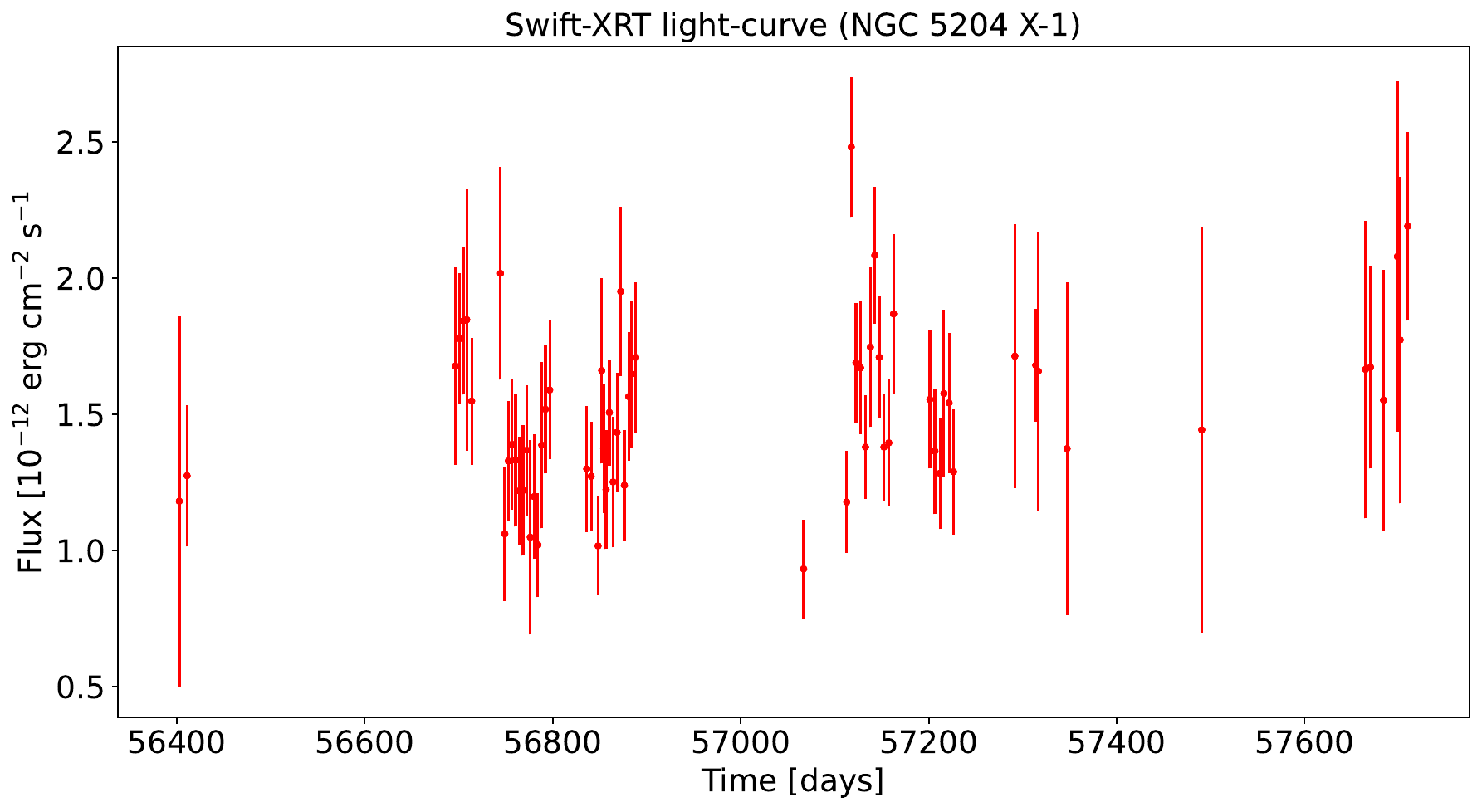}
\hspace{0.5cm} 
\vspace{0cm} 
\includegraphics[scale=0.275]{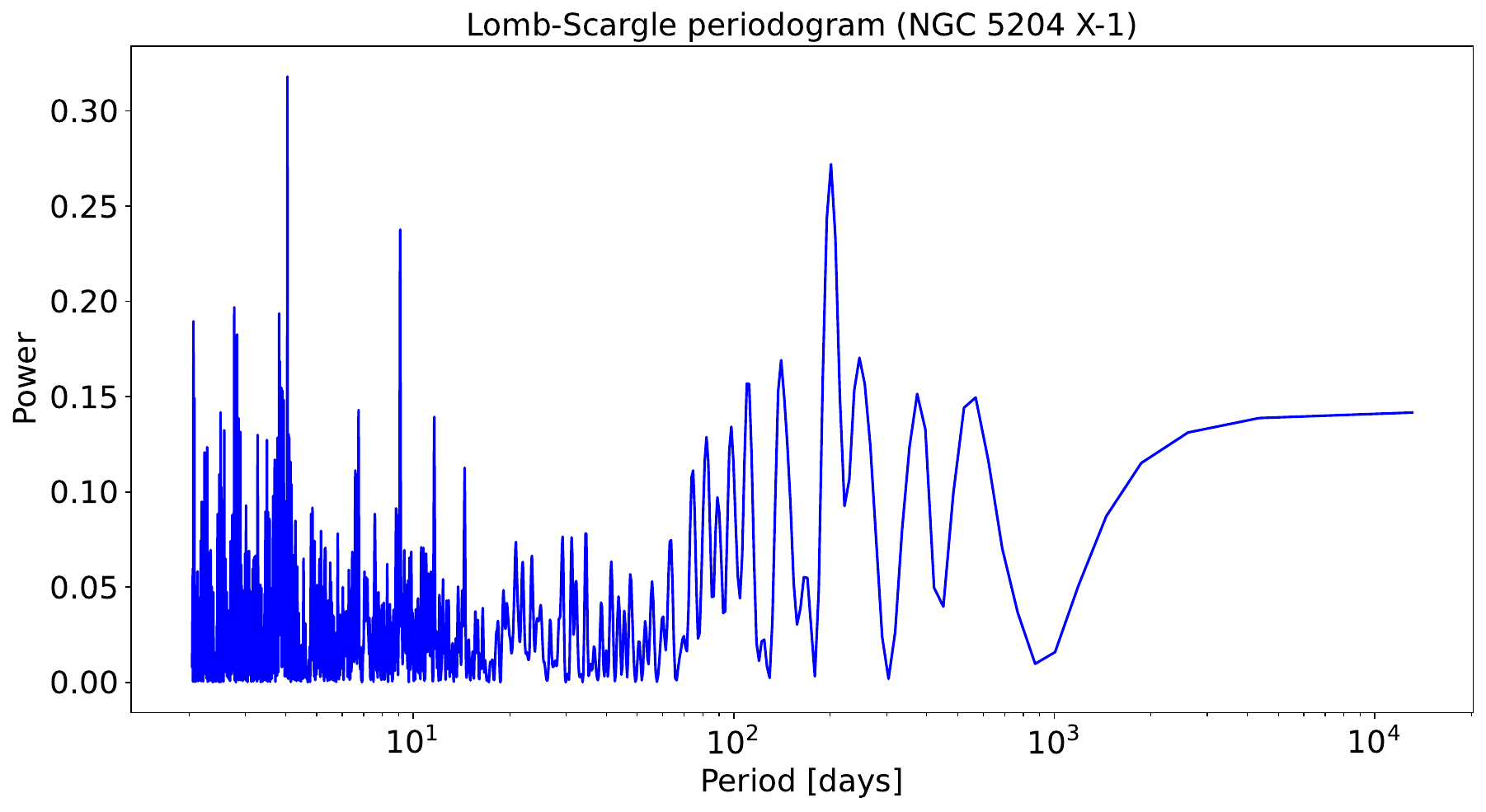}
\\[8pt]
\includegraphics[scale=0.275]{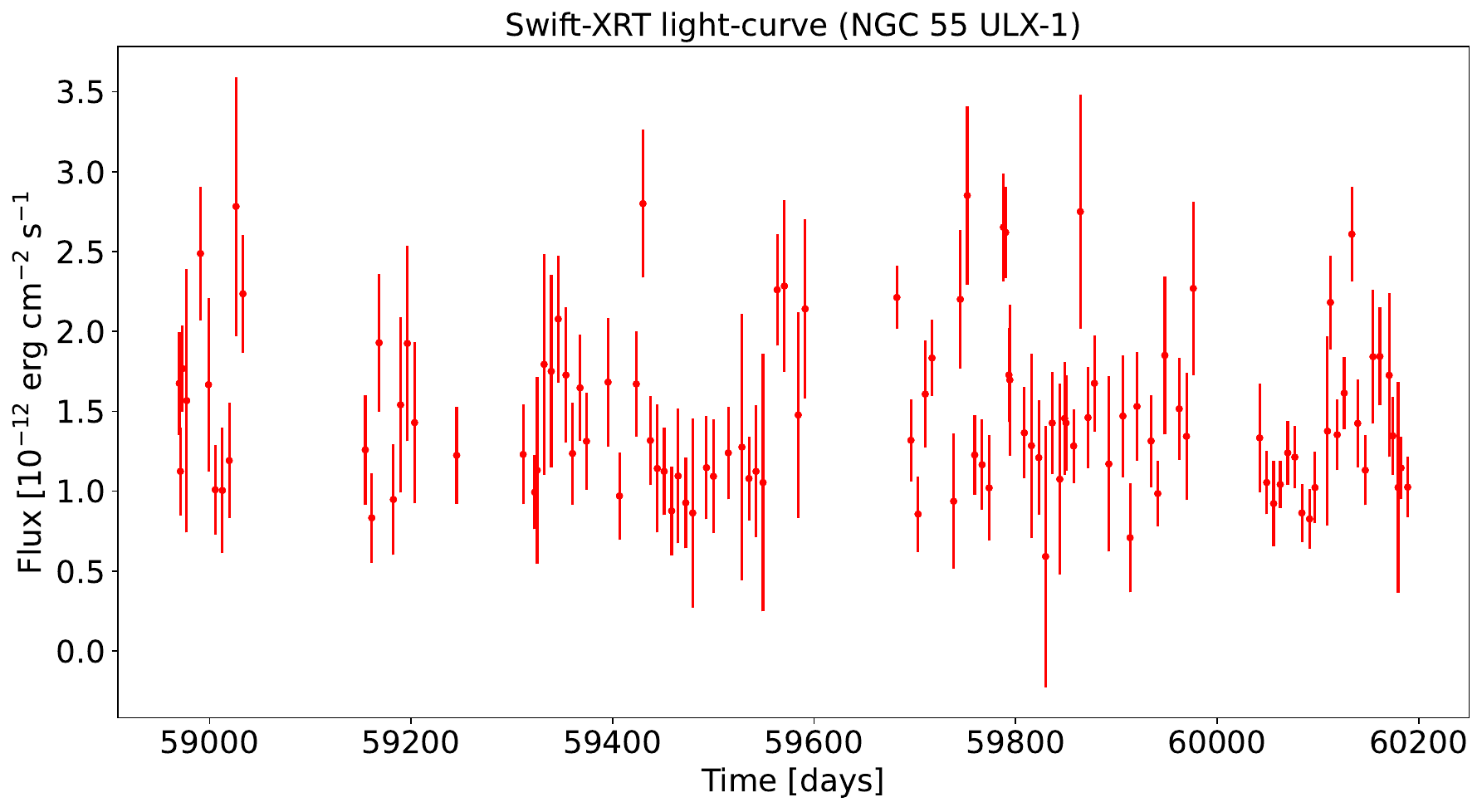}
\hspace{0.5cm}
\vspace{0cm} 
\includegraphics[scale=0.275]{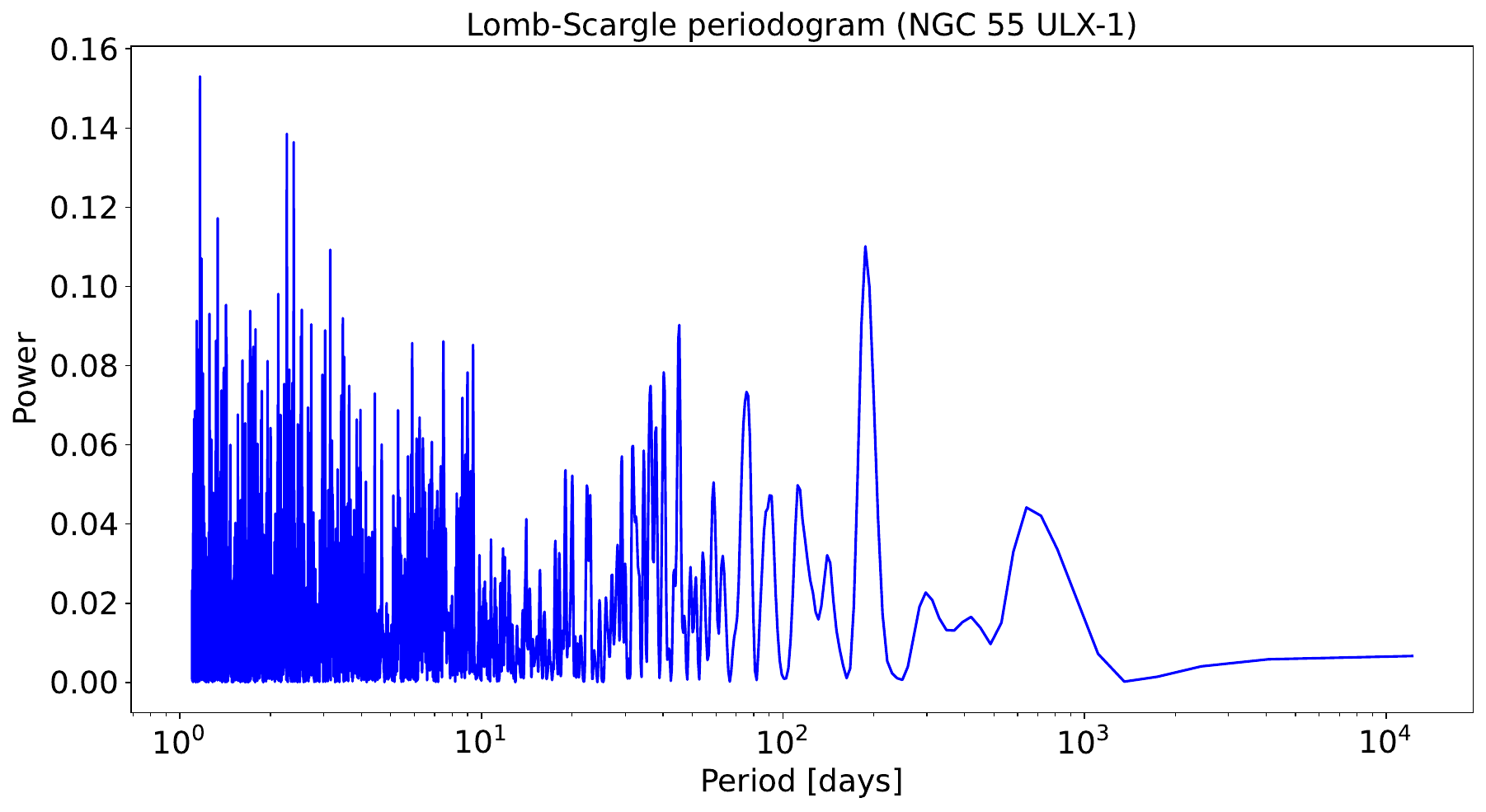}
\caption{{\it{Top panel}}: Long--term {\it{Swift}}--XRT light--curve and the associated Lomb-Scargle periodogram of the ULX NGC 5204 X--1, revealing tentatively significant super-orbital periodicities, with the significances derived from bootstrapping the non--evenly sampled light--curve. {\it{Botoom panel}}: Same as the top panel, but for NGC 55 ULX--1.}
\label{fig:superorbital_period}
\end{figure*}

The main focus of this paper was to perform a detailed analysis of the two ULXs with two distinct spectral classifications, NGC 5204 X--1 and NGC 55 ULX--1. The former is classified as a hard ultraluminous X--ray source (HUL) and is thought to be viewed at a low inclination. On the other hand, the latter has a soft X--ray spectrum that clearly dims at energies above 6 keV. Given their distinctly different viewing angles, the original aim of the study was to observe if the properties of the accretion disc wind (i.e. wind velocity $v_{\text{obs}}$, column density $N_{\text{H}}$ and the ionisation parameter $\chi$) change between the two sources. By building on the disc wind study (\citealt{barra_2022}), we add additional constraints on the accretion disc structure by searching for reverberation lags in the {\it{XMM-Newton}} observations, as would be expected in optically thick super--Eddington disc winds when viewed at a high inclination. By modelling covariance spectra (\citealt{middleton_2015a}; \citealt{kara_2020}), we study the correlation between the {\it{inner}} disc temperature and luminosity (\citealt{robba2021}). In contrast to Galactic X--ray binaries, where one usually observes $L_{\text{X}} \propto T^{4}$ (\citealt{done2010}; \citealt{czerny2019}), which suggests the presence of an inner--most stable orbit ($R_{\text{ISCO}}$), in the case of ULXs one expects an anti--correlation between $L_{\text{X}}$ and $T_{\text{eff}}$ given the presence of a thermal wind, where the wind launching radius is expected to increase in response to a rise in the mass accretion rate $\dot{m}$ and hence the integrated X-ray luminosity. We use frequency resolved covariance spectra to isolate the time variable component in the broadband X--ray spectrum, helping to break degeneracies between the model parameters when fitting multi--component disc models to the spectra. Finally, we search for pulsations and super--orbital periodicities in order to reliably comment on the nature of the source, and to constrain the orbital ephemeris of the binary system.  

By nature, ultraluminous X--ray sources are assumed to exceed the Eddington limit, which describes the limit that a radiative system can sustain its self gravity, which depends on the mass of the compact object, the disc geometry and the presence of a magnetic field if the compact object is a neutron star (\citealt{mushtukov2015}). In fact, super--Eddington discs are observed in many different classes of sources including tidal disruption events (TDEs), Narrow Line Seyfert--1 Active Galactic Nuclei (NLS--1 AGN) and high--redshift quasars, which are key to understanding the evolution and growth of black holes, and hence to resolving the formation channels of black holes. Being observed at low--redshift, ULXs provide a high resolution view of super--Eddington accretion discs and the surrounding interstellar medium (\citealt{beuchert2024}; \citealt{gurpide2024a}), which could enable constraints on the uncertainties in the disc, wind and jet models in understanding the formation of super--massive black holes from stellar mass seeds including Population III stars, which helps resolve the question of how black holes grow and evolve (\citealt{devecchi2012}; \citealt{eduardo_banados}; \citealt{gilmore2011}). The key question remains as to whether differences in the observed X--ray continuum and emission line spectra for ULXs of different spectral classifications are due to differences in the viewing angle, the various evolutionary states of a binary system, or the nature of the compact object (i.e. a neutron star or a black hole). As emphasised previously, the detection of blue--shifted emission lines with soft ultraluminous (SUL) spectra is often thought to be due to a high inclination view of the accretion disk, close to regions where optically thick winds are predicted to be launched. By using arguments for super--Eddington accretion flows, a similar accretion geometry also holds for other classes of transients, including tidal disruption events. 

\subsection{Limits on propagation or reverberation lags}

Modelling lag--frequency and lag--energy profiles (\citealt{uttley_2014}; \citealt{de_marco_2013}; \citealt{tde_zhang}) enables constraints on the underlying size scale and structure of the accretion flow as this is in--directly tied to the mass $m$ and mass accretion rate $\dot{m}$ of the compact object and hence yields clues on whether the time lags we observe are propagation or reverberation lags. In Galactic X--ray binaries, both lag mechanisms can be resolved: propagation lags arise due to turbulent fluctuations in $\dot{m}$ due to viscous diffusion of in--falling material that transports angular momentum outwards: such fluctuations can eventually be entrained by a jet that later dissipates on larger scales. The disc emission variability as a function of radius is usually modelled as a series of multiplicative convolutions of the accretion rate from the outer disc to the inner disc regions, while reverberation lags represent a smoothed and delayed response to the flux variability from the inner accretion flow due to being reprocessed by the photosphere of the super--Eddington thick disc or possibly the boundary layer of a neutron star. 
\newline
\newline
As such, lag--frequency spectra can be modelled with multiple components and their timescales can yield clues as to the spherisation radius and/or the presence or absence of a dipolar magnetic field. In order to fit the time--delay profiles, we convolve the driving (reference--band) light--curve with a top--hat filter of width $\tau_{d}$ with higher values of $\tau_{d}$ smoothing out higher frequency variability. As such, we use two distinct components to model the reverberation lag component: $H_{\text{reverb-BK}}(f)$ represents the impulse response to account for light-crossing echoes over a finite time window of width $\tau_w$, while $H_{\text{reverb-PL}}(f)$ is the transfer function used to model extended, diffuse scattering or delayed multi--zone echoes starting from a transition time $t_{0}$. Finally, we use a simple power--law to model thermal or viscous propagation lags across an extended accretion disc $H_{\text{prop}}(f)$. 
\begin{multline}
H_{\text{reverb-BK}}(f) = e^{-i \pi f \tau_w} \text{sinc}(\pi f \tau_w) \\
H_{\text{reverb-PL}}(f) = t_0^\alpha e^{t_0/\tau_d} \left(\frac{1}{\tau_d} + i 2\pi f\right)^{\alpha - 1} \Gamma\left(1-\alpha, t_0 \left(\frac{1}{\tau_d} + i 2\pi f\right)\right) \\
H_{\text{prop}}(f) = e^{-i 2\pi A f^{1-\beta}} \\
H_{\text{total}}(f) = \left[ w_1 H_{\text{reverb-BK}}(f)  + w_2 H_{\text{reverb-PL}}(f)  \right] H_{\text{prop}}(f)  \\
\end{multline}
where $\Gamma$ is the upper incomplete gamma function, $\tau_w$ and $\tau_d$ represent the smoothing timescales of the scattered emission, $\beta$ is the exponent of the power--law corresponding to the transfer function describing the propagation lags, $w_1$ and $w_2$ are amplitudes of the two separate reverberation components, $t_{0}$ represents the time at which the driving flare reverberates and $\alpha$ is the exponent corresponding to the envelope model of the reverberation lag. The time--lag model can then be defined as the argument of the impulse response in the same manner as for the real data:
\begin{multline}
\tau(f) = \frac{1}{2\pi f}\arctan \bigg( \frac{ \Re{H_{\text{total}} }}{1 + \Im{H_{\text{total}}}} \bigg ) \\
\end{multline}
We use the above analytic expression to fit the lag--frequency spectra (as shown in Figures~\ref{fig:XMM_lag_freq_ngc5204} and~\ref{fig:XMM_lag_freq_ngc55}), illustrating the best--fit model for the case of Obs ID: 0405690101 as an example, where we believe the lags are moderately significant. The best--fit model suggests a Compton scattering timescale $\tau_d$ of the disc plasma to be $\sim 1.5$ seconds, from which one might infer the photosphere of the accretion disc wind to be $\lesssim 10^{5}$ light-seconds from the following equation:
\begin{multline}
\tau_{d} = \frac{\sigma_{T} n_{e} R_{\text{sph}}^{2}}{c} \\
R_{\text{sph}} = \frac{c\tau_{d}}{\sigma_{T} N_{\text{H}}} \\
\end{multline}
We note that the spherisation radius $R_{\text{sph}}$ describes the location at which the disc breaches the Eddington limit. By constraining the latter from the lag--frequency spectral fits, we can obviously place limits on $n_{e}$, $m$ and $\dot{m}$, in addition to the presence or size of a dipolar magnetic moment $\mu$ provided the compact object is a neutron star (alongside other observations noted in the following section). We note that these quantities are degenerate and requires multi--wavelength data--sets and analyses. 
\newline
\newline
In previous studies, relativistic reverberation codes such as {\tt{reltrans}} have been successfully used to model complex lag--frequency spectra of Narrow--line Seyfert--1 (NLS1) AGN and Tidal Disruption Events (TDEs) (\citealt{ingram_reltrans}). These models use ray--tracing simulations to sketch the geodesic of photons and their transmission through multiple layers of ionised plasma, after considering turbulence and gravitational lensing effects using 3D geometries of accretion discs. We plan to develop and use these simulations in future work, while noting that our current model is online. We also note that {\tt{reltrans}} applies reflection kernels to model relativistic gravitational and Doppler--boosted emission (blue--red)shifts of the inner disc emission (see \citealt{kara_garcia_2025}), but the underlying physics in the case of ULXs is probably different since asymmetric doubled--horned Iron (Fe) emission lines are not observed in their time--averaged--spectra, despite sufficient detection sensitivities between 6--7 keV (\citealt{walton2011}). This could either imply sub--solar abundances (requiring exotic evolutionary channels) or simple reflection from highly ionised plasma, but more importantly accretion geometries that could be substantially different from AGNs or TDEs. Studies of such accretion discs are crucial to build a coherent picture of super--Eddington systems, which can influence black hole growth across cosmic time. Simple analytic functions such as a top--hat kernel or a power--law are used to derive a mean optical depth $\tau_{\text{d}}$ of the scattering layer(s) to set limits on the column density and, by implication, simultaneously constrain the presence of a dipolar magnetic field of the compact object if such quantities cannot be determined from pulsations data alone. We adopt a canonical thick disc geometry, as proposed for super--Eddington accretion flows in \cite{king_middleton_2023}, implying that the lag timescale $\tau_{\text{mean}}$ and the electron density $n_{\text{e}}$ are related as follows:
\begin{equation}
n_e = \frac{1}{c\sigma_{T}\tau_{\text{mean}}}
\end{equation}
This allows us to determine the column density of the accretion flow, yielding $N_{\text{H}} = \dot{m} R_{\text{sph}} \sim 0.8 \times 10^{21}$ cm$^{-2}$ in the case where (i) the compact object is a black hole and (ii) $N_{\text{H}} = \dot{m} R_{\text{m}}$ for a neutron star. $R_{\text{sph}}$ is derived from the observed X--ray luminosity, which can thus be compared to values obtained from fitting multi--instrument broadband (high--resolution) X--ray spectra. Therefore, we fit the lag--frequency spectra for the highest signal--to--noise {\it{XMM--Newton}} observations of both ULXs with the models described in equations 8 and 9 finding that in most cases (aside possibly from Obs ID: 0405690101 and Obs ID: 0405690501), the data show a lack of structure, presumably due to the low fractional variability of the X--ray emission. We are able to confirm that in the moderately significant cases, the time lags, related to the Compton scattering timescale in equation 9, are below $\sim 1.5$ ks, implying that the spherisation radius is $\lesssim 10^{11}$ R$_{\text{g}}$.
\begin{table*}
\caption{A table showing the information for the pulsations detected for both ULXs in this study, where we quote the binary orbital parameters inferred by modelling the Doppler shifts of the pulse periods. The pulsation significances are also included.}
\begin{tabular}{lccccccc}
\hline
\hline
\centering
Source & ObsID & Pulsation significance & $P_{\text{epoch}}$ & $P_{\text{orb}}$ & $a\sin i$ & $e$ & Pulsed fraction \\
       &       & [$\sigma$]             & [MJD]              & [days]           & [light--seconds] & \\
\hline
NGC 5204 X--1 & 0142770101 & 6.0 & 52777.05 & 1.3 $\pm 0.6$ & $20 \pm 5$ & $<0.1$ & (11 $\pm$ 1)\% \\
NGC 55 ULX--1 & 0824570101 & 3.5 & 58442.07 & 1.0 $\pm 0.2$ & $2 \pm 0.3$ & $<0.3$ & (7 $\pm$ 1)\% \\
\hline
\label{Tab:pulsations_table}
\end{tabular}
\end{table*}
In future work, we will explore different physical sources to model lag--frequency spectra (in addition from adopting an analytical approach that only considers Compton scattering of radiation from disc winds). For example, in systems such as AGN reverberation models such as {\tt{RELTRANS}} adopt a ray--tracing method to sketch the geodesic of photons and their transmission through multiple layers of ionised plasma winds, and can therefore account for turbulence, photo--ionisation and gravitational lensing physics that can be effectively analysed using 3D models of accretion discs. In such cases, the ionisation parameter $\chi$ is simultaneously derived by fitting SED models to high resolution grating spectra, provided they feature ionised iron, nitrogen and oxygen line complexes (as in {\tt{XILLVER}}), assuming the gas is in collisional ionisation equilibrium. 
\newline
\newline
As a general point, we note that the covariance spectra reveal soft disc, warm disc in addition to the hard power--law component, implying that the disc emission does vary over timescales $\sim 2$ ks, which limits the size of the cool disc to be $\lesssim 2000$ $R_{g}$ given $\alpha = 0.01$. This holds if one assumes an $\alpha$ thin--disc with a radial dependent viscosity where the viscous frequency describes fluctuations in mass accretion rate at a given radial region, with lower Fourier frequencies tied to accretion rate fluctuations far away from the inner most stable orbit, which scales as ($R_{\text{ISCO}} = \frac{aGM}{c^{2}}$), where $a \sim 6$ for a non--spinning black hole or a neutron star:
\begin{multline}
\Omega_{\text{visc}} = \frac{3}{2}\alpha\Omega_{K} \\
\Omega_{K} = \big( \frac{GM}{R^3} \big )^{1/2} \\
\end{multline}
{\it{NuSTAR}} observations could potentially reveal breaks in the power--density--spectra (PDS) at high Fourier frequencies (see Figure~\ref{fig:fig_nulc_appendix2} in the Appendix), and hence the location of the spherisation radius $R_{\text{sph}}$ or the region between the different accretion layers. As such, we the include {\it{NuSTAR}} time--averaged and covariance spectra at energy bands above 20 keV computed in the same manner as in section 3.1. The time averaged spectra are featureless, consistent with a power--law, and show a cut--off above 20 keV as seen for other ULXs, which characterises them to be in a distinct accretion state to canonical Galactic X--ray binaries, tidal disruption events or AGN. The covariance spectrum is also consistent with a power--law at these higher energies that smoothly connect with the {\it{XMM--Newton}} data. Finally, we place upper limits on any pulsations in the {\it{NuSTAR}} observations of the two ULXs.     

\subsection{Constraints from pulsations and quasi--periodic oscillations}

\begin{figure*}
\centering
\includegraphics[scale=0.15]{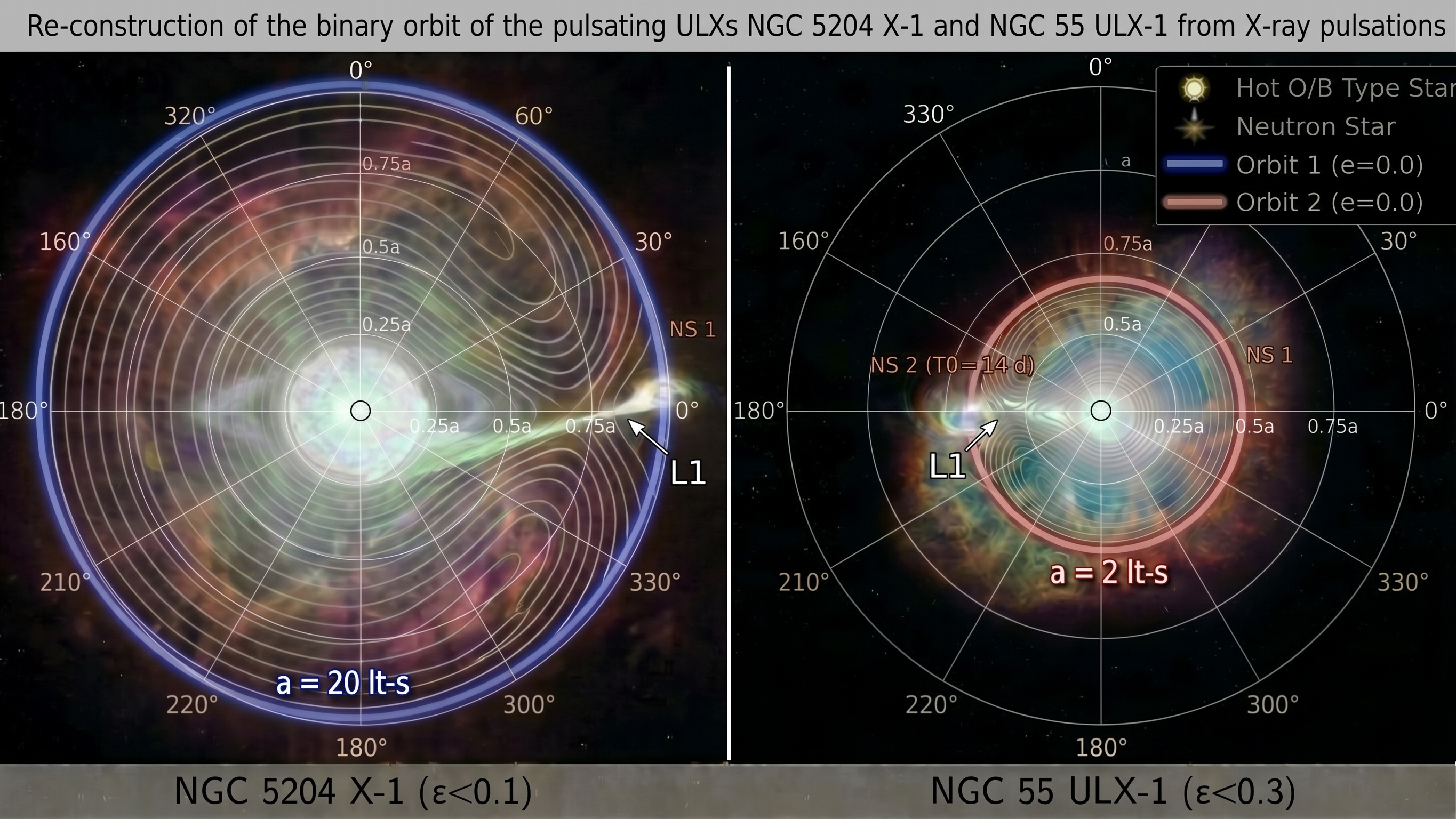}
\caption{A illustration of the probable binary orbit of the two ULXs inferred from fitting the phase delays of the pulsations detected in the {\it{XMM--Newton}} observations at reference epoch $T_{0} =$ 52645.2 MJD for NGC 5204 X--1 (i.e. Obs ID: 0142770101) and reference epoch $T_{0} = $ 58440.3 MJD for NGC 55 ULX--1 (i.e. Obs ID: 0824570101). The data reveals a larger orbit for NGC 5204 X--1 with $P_{\text{orb}} = 1.3 \pm 0.6$ days, $a\sin i = 20 \pm 5$ light--seconds, $e < 0.1$ at $T_{0}$ = 52777.05 MJD, while $P_{\text{orb}} = 1.0 \pm 0.2$ days, $a\sin i = 2.0 \pm 0.3$ light--seconds, $e<0.3$ and $T_{0} = 58585.53$ MJD for NGC 55 ULX--1. We assume that the phase delay shifts are dominated by orbital motions alone, which is a safe assumption if the the spin--up rate $\dot{nu}$ remains $\lesssim 10^{-10}$ Hz s$^{-1}$ given the short exposure time in both cases, especially if the dipolar magnetic field $B_{\text{dipolar}} < 10^{14}$ Gauss.}
\label{fig:pulsations_analysis}
\end{figure*}

In order to probe the nature of the source more reliably, we searched for pulsations in all the EPIC--pn and EPIC--MOS observations of the two ULXs studied here using the Fourier Domain Acceleration Search (FDAS) technique. We found pulsations in a single observation of NGC 5204 X--1 (i.e. ObsID: 0142770101) with a significance of 6$\sigma$, and in a single observation of NGC 55 ULX--1 (i.e. ObsID: 0824570101) at a slightly lower 3.5$\sigma$ significance. The detailed properties of the detection are summarised in Table~\ref{Tab:pulsations_table}. We show the power--spectral densities after correcting for the period derivative (or orbital acceleration) in Figures~\ref{fig:pulsations_analysis}, and in the insets, we include the time evolution of the pulse phase and the integrated pulse profile. We note that the pulse profile is well--fit by a simple sinusoidal component with the presence of harmonics $N_{\text{harm}} > 2$ found to be negligible. We also quote the pulsed fraction in Table~\ref{Tab:pulsations_table}. We note that the {\it{NuSTAR}} data do not reveal any significant periodicities despite sufficiently high time resolution. Here, we emphasise that the acceleration search is performed in the Fourier domain, to remove the effect of beat periods by de-convolving the FFT of the observed time series ($R(f)$) by the FFT of the window function ($W(f)$), using the strategy outlined in section 3.3. We note an interesting feature of the pulsating signal found in NGC 55 ULX--1, which clearly begins to brighten in luminosity towards the end of the observation confirming the transient nature of such events, and perhaps their viability in other hard ULXs such as M51 ULX--7 and NGC 1313 X--2. Importantly, we use a post--Keplerian orbital evolution model to set the first (preliminary) constraints on the orbital ephemeris of the two systems, as inferred from the evolution of the detected spin--frequencies. In Figure 19 we sketch the binary orbit in polar ($a$, $\theta$) co--ordinates, which suggests that the orbital size scale of NGC 55 ULX--1 is smaller than NGC 5204 X--1. Hence, we constrain the orbital period, semi--major axis, eccentricity and reference epoch (or orbital phase): see Table~\ref{Tab:pulsations_table} for specific values. Given that pulsations were detected only in a single observation of the two ULXs (likely because the pulsed fraction increased only during those epochs), we are unable to constrain the spin--up rate, the orbital period derivative, or any other post--Keplerian parameters such as Shapiro delays. Such studies may become viable with more observations in future missions, although we assume $\dot{\nu} \sim 10^{-10}$ Hz s$^{-1}$ by using datasets from previous results when fitting the phase evolution. 
\newline
\newline
We also searched for the presence of any quasi--periodic oscillations (QPOs) in both the hard--band (2--10 keV) and soft band (0.3--1,2 keV) PSDs for both ULXs, but such features appear to remain in--significant for the {\it{XMM--Newton}} observations considered here. We strategically binned the PSDs using geometric binning factors $\nu_{b} = 1.0, 1.05, 1.1$, as illustrated in Figures~\ref{fig:XMM_psds_ngc5204} and~\ref{fig:XMM_psds_ngc55}.    

\begin{figure*}
\centering
\includegraphics[scale=0.5]{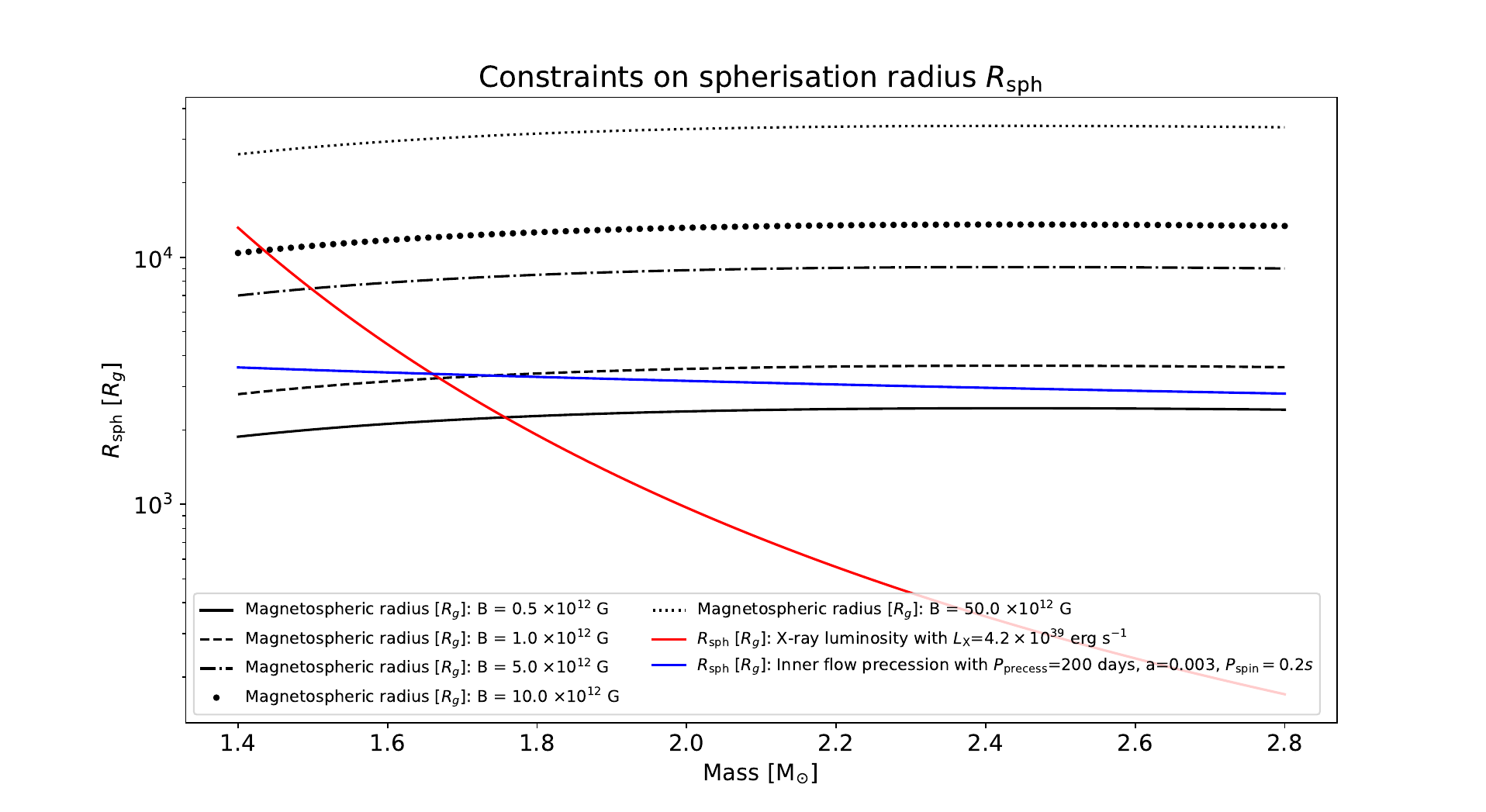}
\caption{Constraints on the spherisation radius $R_{\text{sph}}$ of the accretion flow surrounding NGC 5204 X--$1$ in units of $R_{\text{g}} = \frac{GM}{c^{2}}$ from the observed precession (super--orbital) period and X--ray luminosity (\citealt{king2008}; \citealt{chashkina2019}) for a range of NS masses. The model curves assume a spin parameter $a \sim 0.003$ given the observed spin period of $\sim 0.2$ seconds and the inflow asymmetry parameter $\Lambda \sim 0.5$ (see \citealt{ghosh_lamb_1979}). The magnetospheric radius is also plotted for a range of possible dipolar field strengths (\citealt{mushtukov_2019}), yielding the constraint that $R_{\text{m}} < R_{\text{sph}}$ in order for pulsations to be observed, which would indeed imply $B \lesssim 10^{12}$ Gauss.}
\label{fig:rsphconst}
\end{figure*}

\subsection{L--$T_{\text{eff}}$ correlation}

\begin{figure*}
\centering
\includegraphics[scale=0.5]{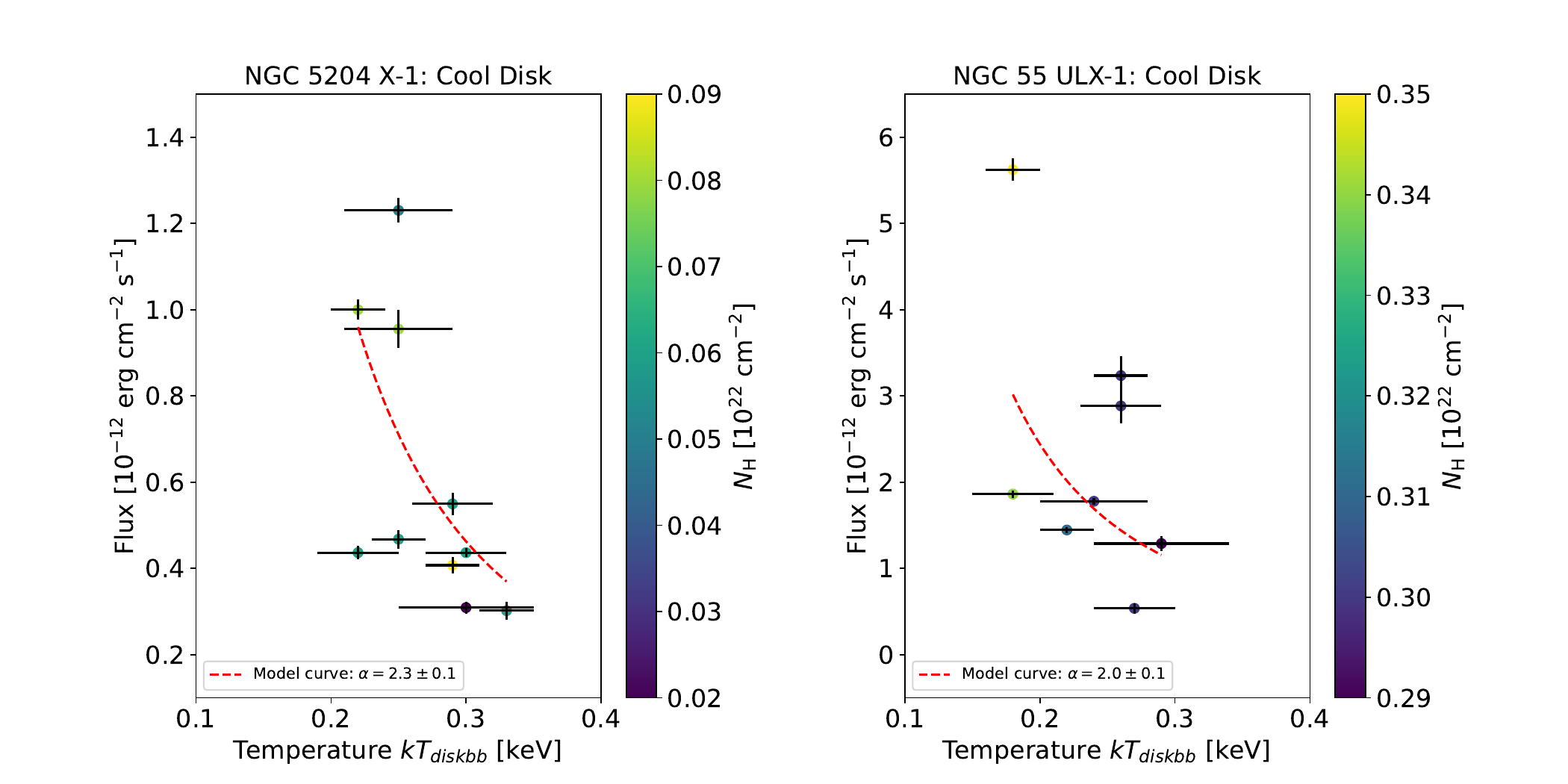}
\caption{Disc flux--temperature plane for the two ULXs assuming $L \propto T^{-\alpha}$, where we infer $\alpha=2.3 \pm 0.1$ for NGC 5204 X--1 and $\alpha=2.0 \pm 0.1$ for NGC 55 ULX--1. The negative correlation between X--ray flux and temperature suggests that the phenomenological soft thermal component is likely associated with the photosphere of a radiatively driven outflow as inferred from fitting the continua of other ULXs in the same energy band (see \citealt{robba2021}).}
\label{fig:LT_diagrams}
\end{figure*}

By modelling the covariance spectra over a carefully chosen low--frequency band (i.e. $\sim 10^{-4}$ -- $5 \times 10^{-4}$ Hz), we resolved the cool disc component. We use this to test whether an inverse relation is observed between the soft disc luminosity and temperature, which would be expected if the emitting medium is dominated by the turbulent photosphere of an ultra--fast wind, as also expected for other super--Eddington systems such as tidal disruption events (\citealt{karae_2017}). We consider the expected radiative diffusion timescales due to Compton scattering effects as determined for a small (but reliable) sample of ULXs when choosing the frequency range of interest. We also note that this choice is motivated by the fact that Poisson noise typically dominates the power density spectra (PDS) at frequencies higher than $\sim 0.1$ Hz, especially in the absence of simultaneous {\it{NuSTAR}} observations. We model the covariance spectrum and time--averaged spectrum with a multi--component accretion disc model: $\tt{diskbb}$ and $\tt{diskpbb}$ and compute the integrated flux of these components over the 0.3--12 keV energy band using {\tt{cflux}} in {\tt{xspec}}. We demonstrate the disc luminosity--temperature diagrams for the two ULXs as a function of the column density in Figure~\ref{fig:LT_diagrams}. The data reveal no obvious trends with $N_{\text{H}}$. We use the Pearson and Spearman rank correlation metrics to deduce whether a concrete relation exists between the luminosity and temperature $L$--$T_{\text{eff}}$ for the two ULXs considered in this work (see also \citealt{kajava2009}). The Pearson statistic values for NGC 5204 X--1 and NGC 55 ULX--1 are $-0.6$ and $-0.7$ with a p--values of $0.05$ and $0.02$. We find that they are moderately anti--correlated, and a simple power--law fit yields the exponent $\alpha = 2.3 \pm 0.1$ for the cool disc component of NGC 5204 X--1 and $\alpha = 2.0 \pm 0.1$ for NGC 55 ULX--1, in line with models that predict blackbody emission from the photosphere of a wind (\citealt{chashkina2019}).

\begin{figure*}
\centering
\includegraphics[scale=0.4]{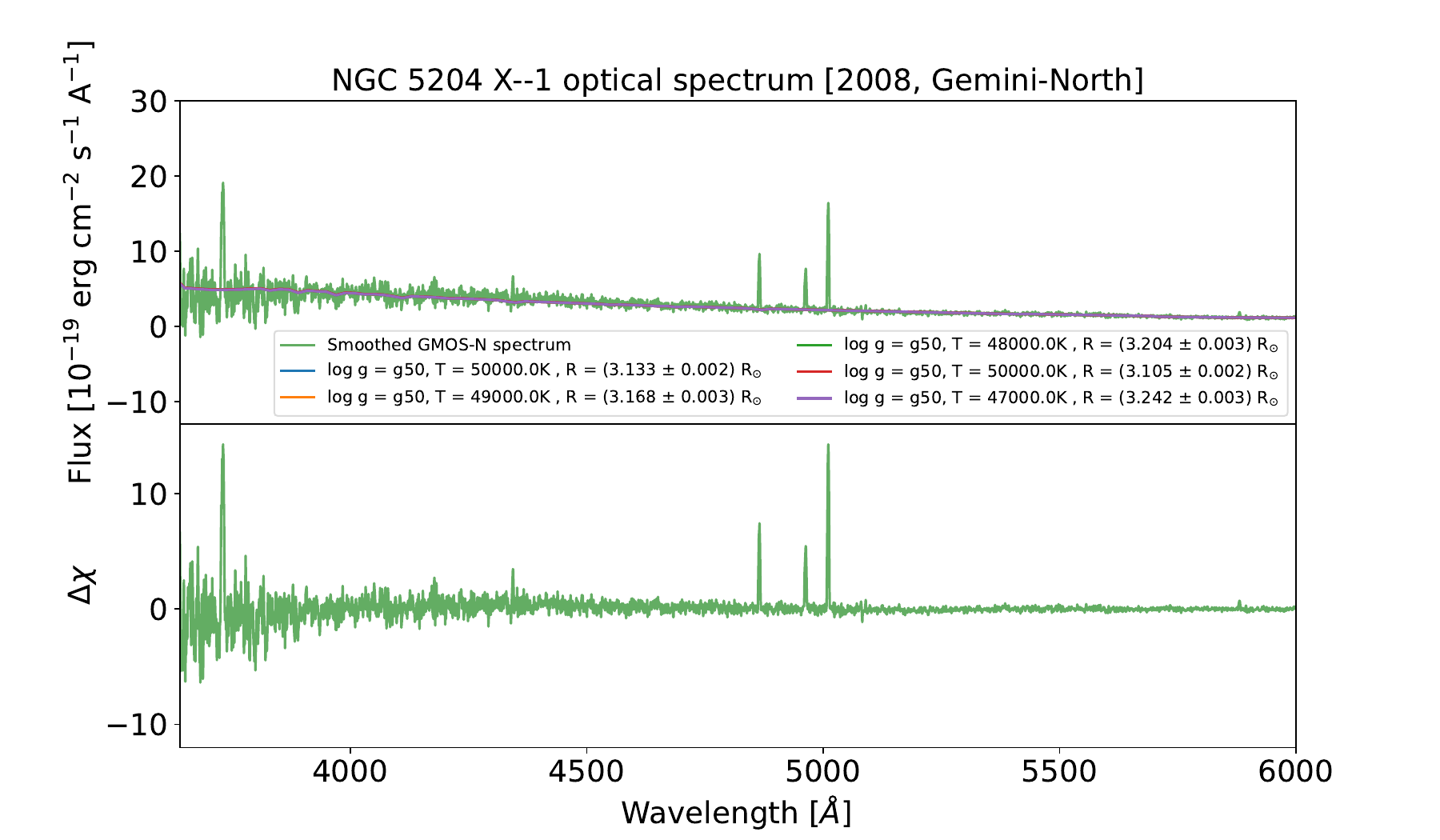}
\includegraphics[scale=0.4]{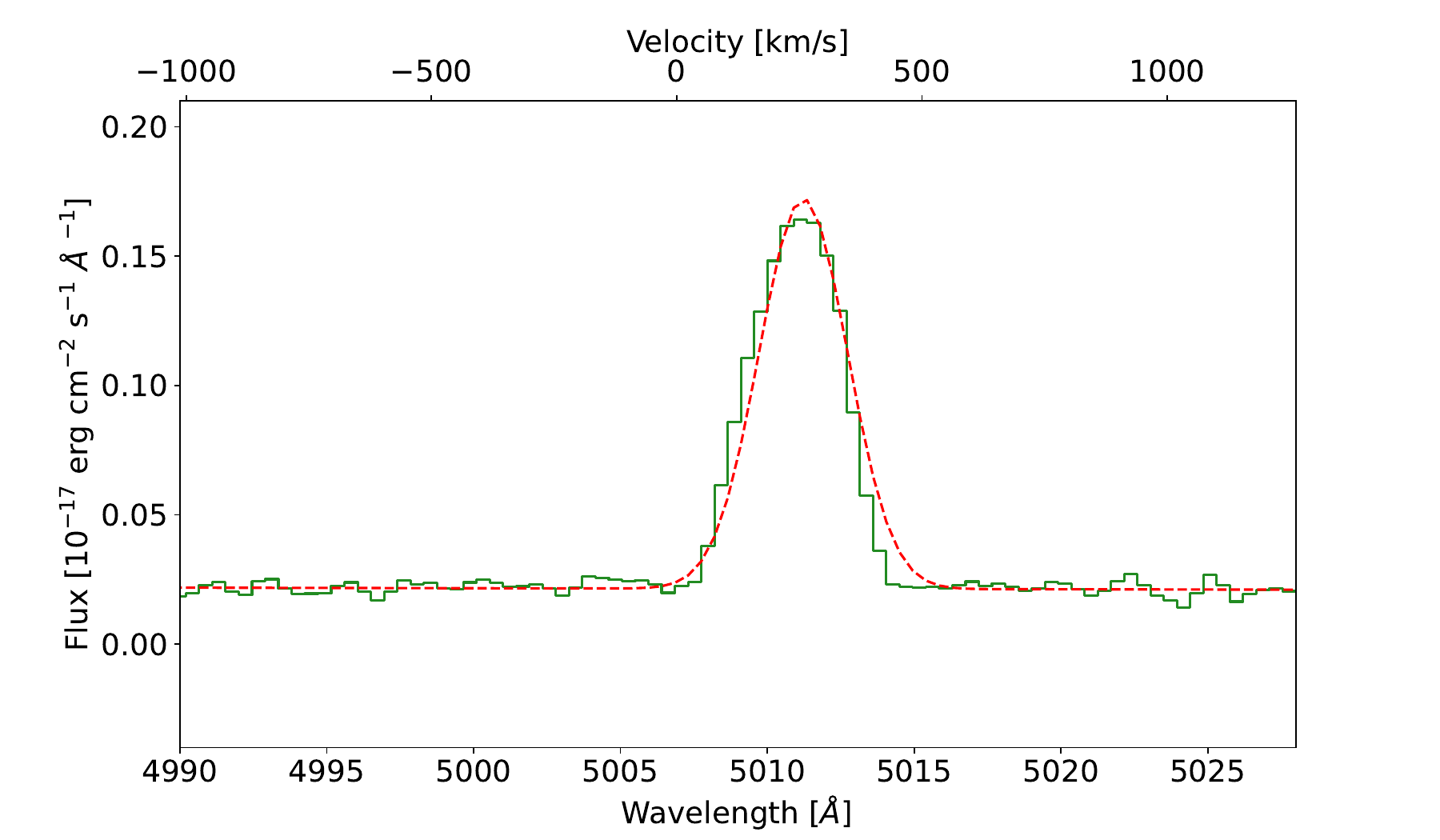}
\caption{{\it{Top}}: Gemini--North spectrum of NGC 5204 X--1 showing the optical flux between 3500--6000 \AA, featuring the well--known H$\beta$ and the forbidden [OIII,4959,5007] emission line complex from the photo--ionised nebula. We modelled the continuum with stellar templates from the Castelli--Kurucz atlas, which indicates that the stellar counterpart is likely a hot O/B star with possible signs of being stripped given the unexpectedly small stellar radius. {\it{Bottom}}: The [OIII 5007] emission line profile modelled with a simple Gaussian to infer radial velocities.} 
\label{fig:optspec_gemini_2008_ngc5204x1}
\end{figure*}

\subsection{Super--orbital periodicity in the {\it{Swift}}--XRT data}

Furthermore, we searched for super--orbital periodicities by computing the Lomb--Scargle periodogram of the time series data accumulated from several years of {\it{Swift-—XRT}} observations. Such long--timescale periodicities are usually tied to a bulk, solid--body precession of a warped inner accretion flow if driven by Lense--Thirring torques communicated over larger scales via bending waves, requiring an effective flow viscosity $\alpha < \frac{H}{R}$ and the sound--crossing time to be less than the precession timescale. Moreover, super--orbital periodicities are usually longer than the orbital period timescale and if precession is a viable mechanism then super--orbital periodicities would be useful for scheduling observations to study the evolution of the disc wind, as the ULX is viewed from different orientation angles. In accordance with \citealt{gurpide_2021}, we detect a $\sim$ 200 day super--orbital variability in the {\it{Swift}--XRT} Lomb--Scargle periodogram of both NGC 5204 X--$1$ and NGC 55 ULX--1. Such timescales may yield further limits on the spherisation radius $R_{\text{sph}}$ in accordance with the following equation:
\begin{equation}
t_{\text{LT,prec}} = \frac{GM\pi}{3c^{3}a_{\text{*}}} r_{\text{sph}}^{3} \left [ \frac{  1 - \bigg( \frac{r_{\text{in}}}{r_{\text{sph}}} \bigg)^{3} } { \log \bigg ( \frac{r_{\text{sph}}}{r_{\text{in}}} \bigg ) } \right ]
\end{equation}
Given the above constraint and the observed X--ray luminosity being $\sim 5 \times 10^{39}$ erg s$^{-1}$, we obtain a family of $R_{\text{sph}}$ curves in Figure~\ref{fig:rsphconst} using the Newton--Raphson root finding strategy (\citealt{poutanen2007}; \citealt{king2008}; \citealt{middleton_king2023}). After plotting the magnetospheric radii curves in the same figure, we infer limits on the dipolar field $B <  1 \times 10^{14}$ Gauss since $R_{\text{m}}$ is required to be $< R_{\text{sph}}$, such that the accretion flow to be funnelled to the poles of the NS that then gives rise to the X--ray pulsations. However, for magnetised accretion discs the dipolar field of the NS may also precess when mis--aligned with the spin axis, or alternatively the NS could freely precess (\citealt{vasilopoulos2020}), and it would be interesting to explore such mechanisms in a future work. The expected super--orbital timescales could be comparable to $\sim 200$ days if $B < 5 \times 10^{12}$ Gauss. However, we emphasise that it is possible to explain the super-—orbital period if the flow precession is driven by Lense—Thirring torques, which predicts a range in spherisation radii that agree with the expected values from the mass accretion rate scaled to the Eddington ratio $\dot{m}$ in the context of the canonical thick disc model. We note that one might also confirm if precession actually drives the super--orbital variability by considering whether we see phase--dependent changes in the X--ray spectra, as mapped by the hardness--intensity diagram (see Figure~\ref{fig:hardness_intensity_diagram}) or by modelling the long--term {\it{Swift}} XRT light--curves (see Figure~\ref{fig:LT_diagrams} and future work).
\newline
\newline
Our Data Reduction methods interpolate bad time intervals (BTIs) by sampling from light--curves that reside in similar regions of the hardness--intensity space. To this aim, we group together the {\it{XMM-Newton}} observations into similar, but physically unclassified, accretion states by applying a hierarchical density--based clustering algorithm: {\tt{HDBSCAN}} (\citealt{hdbscan_accelerated_clustering}: also see Galactic X--ray binaries as an analogy). This algorithm removes biases in the maximum number of hardness--intensity clusters $N_{\text{c}}$ that the data might contain, since it does not require an initial guess in $N_{\text{c}}$. It hierarchically groups together clusters based on a pre--defined local density estimate using a nearest--neighbour algorithm that requires a mutual reachability distance, while continually dis--connecting sparse data points, which builds a minimum spanning tree (\citealt{hdbscanpp}). 

\subsection{Discovery of an optical counterpart and its implications on binary evolution}

Now, we obtain some insights into the evolutionary history of the binary system by considering optical {\tt{HST}} and {\tt{Gemini--North}} observations. In order to model the optical--X--ray spectral energy distribution (SED), we extrapolated the X--ray emission using a multi--colour disc component to see whether the emission at these higher wavelengths can be explained purely by thermal emission, or if the SED requires some level of reprocessed emission, a new stellar component, dust emission or large--scale jet emission. Indeed, using optical observations in our SED allows us refine our constraints on the hydrogen column density $N_{\text{H}}$, especially since this is expected to change between the two ULXs due to differences in viewing angles. 
\newline
\newline
Moreover, the Gemini--North optical spectrum of NGC 5204 X--1 reveals that the stellar counterpart is likely a hot O/B--type star in an advanced evolutionary phase with some signs of being stripped given the un-expectedly smaller stellar radius (\citealt{PodsiadlowskiP_2003}). The spectrum includes the [H$\beta$, 4861] and the [OIII, 4959,5007] forbidden doublets suggesting that these lines are formed in a photo--ionised nebula surrounding the ULX (\citealt{pakull2003}) with radial velocities $\sim 200$ km $s^{-1}$. However, the spectra don't feature strong absorption lines, possibly due to the limited instrumental wavelength resolution or physically being diluted by the photosphere of a diffuse wind envelope or the accretion disc surrounding the compact object (\citealt{gladstone2009}; \citealt{king_pringle_2001}), or the ULX residing in regions of the galaxy where chemical enrichment is less efficient pointing to sub-solar abundances (\citealt{swartz2004}; \citealt{swartz2011}; \citealt{zezas2009}). Please refer to our recent paper for more details on the optical observations of the ULX counterpart (\citealt{sathyaprakash2026}). Further details on NGC 55 ULX--1 are withstanding given the lack of high resolution Gemini--North observations or the availability of any VLT spectra (\href{https://mast.stsci.edu/portal/Mashup/Clients/Mast/Portal.html}{see MAST portal}).
\newline
\newline
We also use the UV/optical magnitudes of the ULX counterparts to determine if the stellar counterpart can be modelled with a single emission component (\citealt{motch2014}). If time resolved photometry of the stellar companion is available, the disc contribution can be constrained if, for example, optical observations are scheduled close to the phase minimum of the ellipsoidal modulation of the star. However, the VLT archives do not provide time resolved spectra at present for either of the two sources, although integral--field {\tt{MUSE}} observations are available for NGC 5204 X--1 that may be used in future work to study the ionised large--scale plasma components in more detail. The SED derived from the {\it{HST}} UV and optical magnitudes imply a hot O/B--type star in the case of NGC 5204 X--1, while hot disc emission might dominate the UV bands for NGC 55 ULX--1 given the observed slope. For both ULXs, we note the optical counterpart appears slightly too bright with too flat an SED to be associated with a diffuse star cluster, in which case the distance to the ULX would have to be incorrect by a factor of $\sim$ 20--25. This suggests that the X-ray luminosity would not be in the ULX regime even though the observed spectral properties are entirely consistent with ULXs. 

\section{Conclusions and Summary}

In this work, we studied the X--ray timing and spectral properties of two ULXs, namely NGC 5204 X--1 and NGC 55 ULX--$1$ in distinct accretion states. We used time--averaged spectra, covariance spectra and lag--frequency profiles to monitor the evolution of the accretion states of the two ULXs, that are expected to be observed from two different viewing angles according to information derived from high--ionisation emission line dynamics from the RGS {\it{XMM--Newton}} data. We derive constraints on the presence of any reverberation lags, but we also detect pulsations in both ULXs using a Fourier domain acceleration search strategy, allowing a map of the binary orbit and first independent constraints on the orbital period, eccentricity and semi--major axis of the system. We include optical {\it{HST}} and {\it{Gemini--South}} observations to gain further insights into the emission mechanisms of the ULXs, and to place any limits on stellar or large--scale jet emission. 

\section{Acknowledgements}

R.S. acknowledges financial support from the Italian Ministry for University and Research through the grants 2022Y2T94C (SEAWIND) and from INAF through LG 2023 BLOSSOM. R.S. wishes to thank Paolo Esposito and Ciro Pinto for comments and suggestions during this paper. The figures used in this manuscript were produced using the {\tt{xcovspec}} pipeline, which is a recently developed tool to generate covariance spectra and lag--energy spectra from archival {\it{XMM-Newton}} datasets. We made use of the following {\tt{python}v3.12} libraries: {\tt{numpy}}, {\tt{scipy}}, {\tt{astropy}}, {\tt{scikit-learn}}, {\tt{scikit-image}}, {\tt{hdbscan}} and {\tt{kapteyn}}.

\section{Data Availability}

The data used in this paper can be made available upon reasonable request to the authors. Throughout the paper, we provide references and links to the appropriate data pipelines.

\section{Appendix}

Here, we outline the method used to remove beat periods that occur from irregular gaps in the {\it{NuSTAR}} light--curve due to bad--time--intervals (BTIs). Namely, we follow the strategy outlined in equation $4$ in order to deconvolve the FFT of the window function from that of the observed PSD in order to re--construct the signal PSD. We highlight this in detail in Figure~\ref{fig:fig_nulc_appendix2}.
\begin{figure*}
\centering
\includegraphics[scale=0.4]{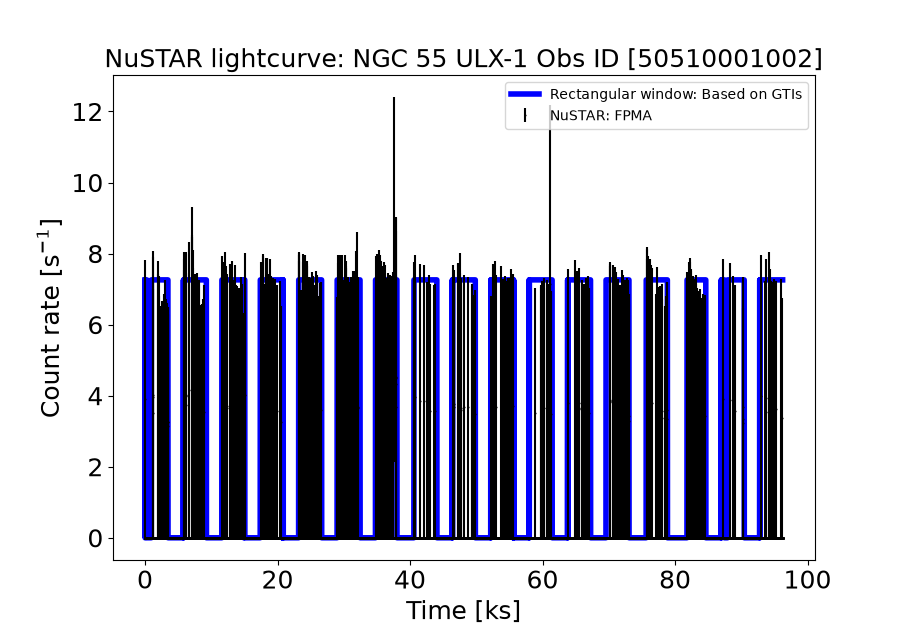}
\caption{{\it{NuSTAR}} FPMA light--curve of NGC 55 ULX--1 (Obs ID: 50510001002) highlighting gaps in the time series due to bad--time--intervals (BTIs) with the window function overlaid.}
\label{Fig:nulc_appendix1}
\end{figure*}
\begin{figure*}
\centering
\includegraphics[scale=0.4]{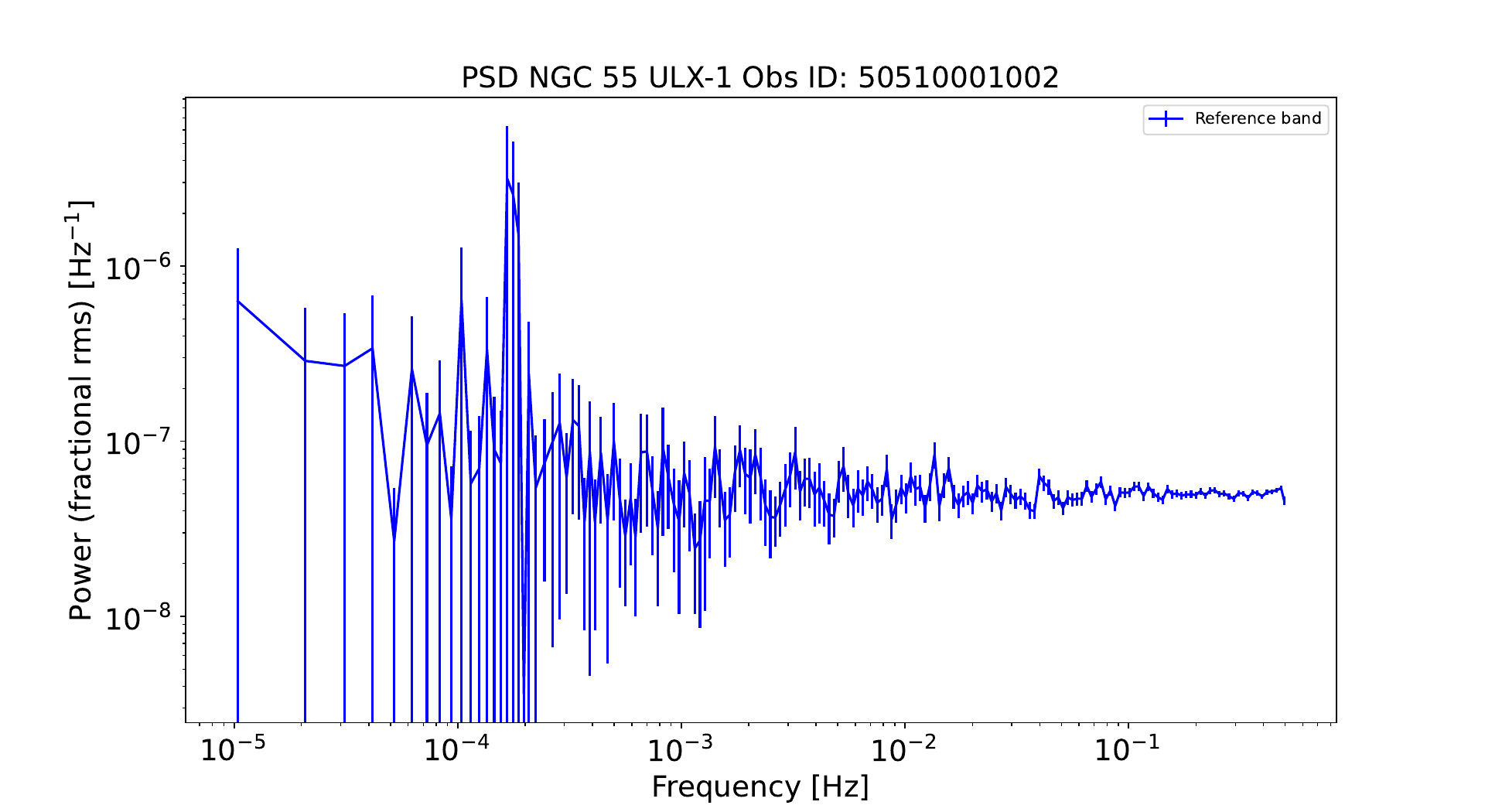}
\includegraphics[scale=0.4]{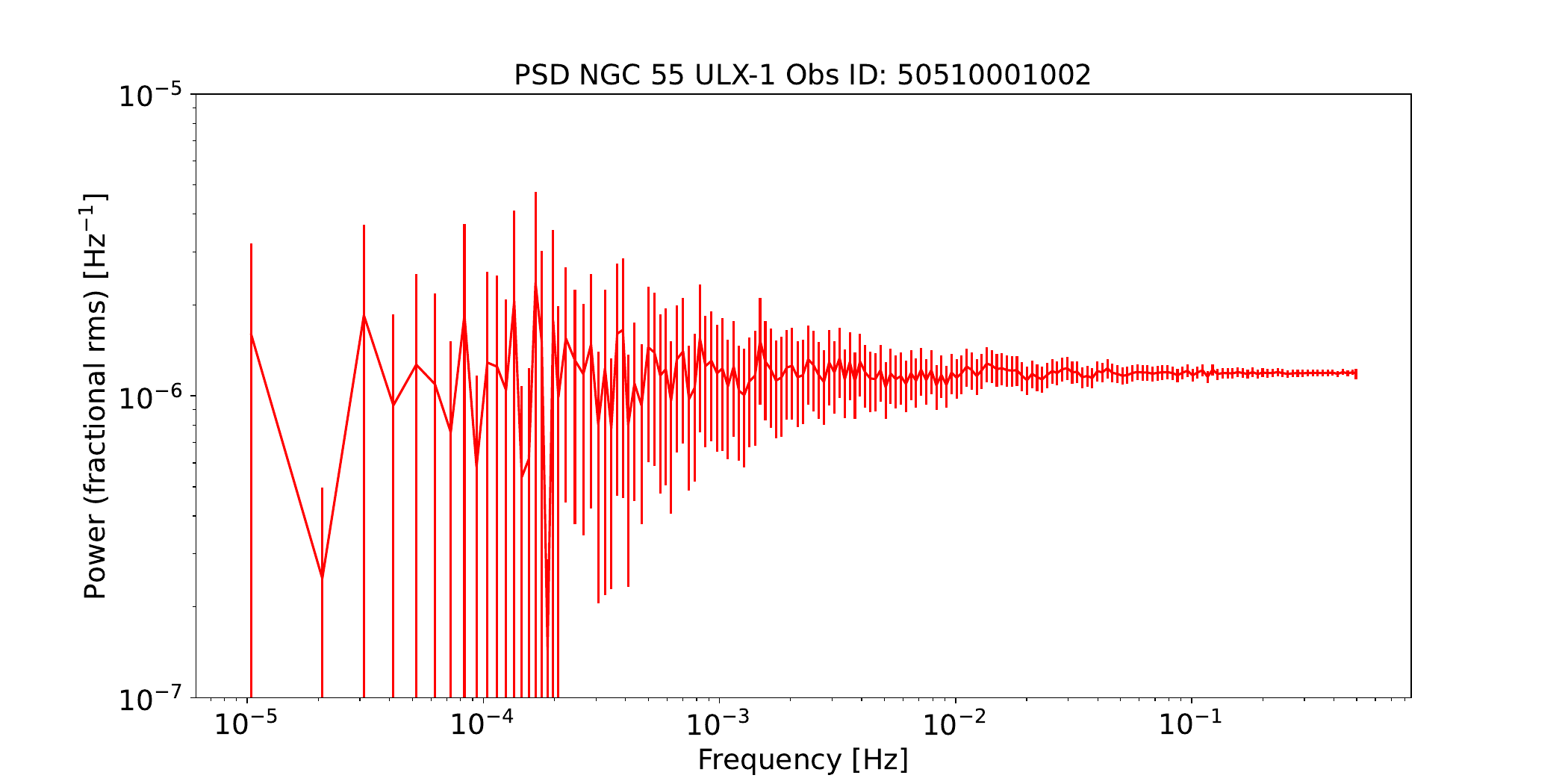}
\caption{We use the window function shown in Figure~\ref{Fig:nulc_appendix1} to remove the beat periods when computing the {\it{NuSTAR}} power--density--spectra (PSD) associated with the ULX. The top panel shows the PSD including the beat periods and the bottom panel with the beats removed via FFT deconvolution.}
\label{fig:fig_nulc_appendix2}
\end{figure*}
\bibliography{ref}
\bibliographystyle{mnras}
\appendix
\bsp	
\label{lastpage}
\end{document}